\documentclass[aps,prd,floats,floatfix,nofootinbib]{revtex4-1}
\usepackage{geometry}                % See geometry.pdf to learn the layout options. There are lots.
\usepackage{graphicx}
\usepackage{amssymb}
\usepackage{epstopdf}
\usepackage{hyperref}

\usepackage{amsmath}
\usepackage{bbold}

\newcommand{\be}{\begin{equation}}
\newcommand{\ee}{\end{equation}}
\newcommand{\ba}{\begin{align}}
\newcommand{\ea}{\end{align}}

\usepackage{color}

\begin{document}

\title{The decoupling limit in the Georgi-Machacek model}

\author{Katy Hartling}
\email{khally@physics.carleton.ca}
\author{Kunal Kumar}
\email{kkumar@physics.carleton.ca} 
\author{Heather E.~Logan}
\email{logan@physics.carleton.ca} 

\affiliation{Ottawa-Carleton Institute for Physics, Carleton University, 1125 Colonel By Drive, Ottawa, Ontario K1S 5B6, Canada}

\date{April 9, 2014}                                           % Activate to display a given date or no date

\begin{abstract}
We study the most general scalar potential of the Georgi-Machacek model, which adds isospin-triplet scalars to the Standard Model (SM) in a way that preserves custodial SU(2) symmetry.  We show that this model possesses a decoupling limit, in which the predominantly-triplet states become heavy and degenerate while the couplings of the remaining light neutral scalar approach those of the SM Higgs boson.  We find that the SM-like Higgs boson couplings to fermion pairs and gauge boson pairs can deviate from their SM values by  corrections as large as $\mathcal{O}(v^2/M_{\rm new}^2)$, where $v$ is the SM Higgs vacuum expectation value and $M_{\rm new}$ is the mass scale of the predominantly-triplet states.  In particular, the SM-like Higgs boson couplings to $W$ and $Z$ boson pairs can decouple much more slowly than in two Higgs doublet models, in which they deviate from their SM values like $\mathcal{O}(v^4/M_{\rm new}^4)$.  Furthermore, near the decoupling limit the SM-like Higgs boson couplings to $W$ and $Z$ pairs are always larger than their SM values, which cannot occur in two Higgs doublet models.  As such, a precision measurement of Higgs couplings to $W$ and $Z$ pairs may provide an effective method of distinguishing the Georgi-Machacek model from two Higgs doublet models.  Using numerical scans, we show that the coupling deviations can reach 10\% for $M_{\rm new}$ as large as 800~GeV.
\end{abstract}

\maketitle 

%%%%%%%%%%%%%%%%%%%%%%%%%%%%%%%%%%%%%%%%%%%%%%
\section{Introduction}

The recent discovery~\cite{Aad:2012tfa} of a Standard Model (SM)-like Higgs boson at the CERN Large Hadron Collider (LHC) has focused much theoretical and experimental attention on possible extensions of the SM Higgs sector.  One such extension is the Georgi-Machacek (GM) model~\cite{Georgi:1985nv}, which adds isospin triplets to the SM Higgs sector in a way that preserves the SM prediction $\rho \equiv M_W/M_Z \cos\theta_W = 1$ at tree level.  The GM model is less theoretically attractive than Higgs-sector extensions involving isospin doublets or singlets because, in the GM model, the custodial symmetry that ensures $\rho = 1$ at tree level is violated by hypercharge interactions.  This leads to divergent radiative corrections to $\rho$ at the one-loop level~\cite{Gunion:1990dt}, implying a relatively low cutoff scale (which would be needed in any case to solve the hierarchy problem).  

On the other hand, the GM model is phenomenologically attractive due to two features not present in Higgs sector extensions containing only isospin doublets or singlets.  These are the possibility that the SM-like Higgs boson has couplings to $W$ and $Z$ pairs that are larger than predicted in the SM and the presence of a doubly-charged scalar $H_5^{++}$; in fact, these features are linked insofar as the doubly-charged scalar plays an important role in the unitarization of longitudinal $WW$ scattering amplitudes when the SM-like Higgs boson coupling to $W$ pairs is enhanced relative to that in the SM~\cite{Falkowski:2012vh}.  This makes the GM model a valuable benchmark for studies of Higgs properties and searches for additional scalars beyond the SM, and its phenomenology has been extensively studied in the literature~\cite{Chanowitz:1985ug,Gunion:1989ci,HHG,Haber:1999zh,Aoki:2007ah,Godfrey:2010qb,Low:2010jp,Logan:2010en,Chang:2012gn,Chiang:2012cn,Chiang:2013rua,Kanemura:2013mc,Englert:2013zpa,Killick:2013mya,Englert:2013wga,Efrati:2014uta}. 
The GM model has also been incorporated into little Higgs~\cite{Chang:2003un,Chang:2003zn} and supersymmetric~\cite{Cort:2013foa} models, and extensions with additional isospin doublets~\cite{Hedri:2013wea} have also been considered.

Our objective in this paper is to study the approach to the \emph{decoupling limit}~\cite{Haber:1994mt} in the GM model.  In the decoupling limit, all additional particles beyond those present in the SM become heavy and the couplings of the SM-like Higgs boson approach their SM values.  This limit is of interest in the scenario that future measurements of the couplings of the SM-like Higgs boson at the LHC do not reveal large deviations from the SM expectations.

The scalar potential for the GM model first written down in Ref.~\cite{Chanowitz:1985ug} and used throughout most of the literature~\cite{Gunion:1990dt,Gunion:1989ci,Chang:2012gn,Englert:2013zpa,Efrati:2014uta} imposed a $Z_2$ symmetry on the scalar triplets in order to simplify the form of the potential.  Under this constraint, the potential depends on only two dimensionful parameters which can be eliminated in favor of the vacuum expectation values (vevs) of the doublet and triplet scalars after electroweak symmetry breaking.  This implies that all the scalar masses can be written in the form $m^2_i = \lambda_i v^2$, where $v \equiv 2 M_W/g \simeq 246$~GeV is the SM Higgs vev and $\lambda_i$ represents some linear combination of the scalar quartic couplings.  Because the sizes of the scalar quartic couplings are bounded by the requirement of tree-level perturbative unitarity to be $\mathcal{O}(1)$, the masses of all the scalars in the GM model are bounded to be less than about 700~GeV~\cite{Aoki:2007ah}; in particular, the $Z_2$-symmetric version of the GM model does not possess a decoupling limit.

In this paper we study the most general gauge- and custodial SU(2)-invariant tree level scalar potential of the GM model without imposing a $Z_2$ symmetry.  This full potential was first written down in Ref.~\cite{Aoki:2007ah} and to our knowledge has been used for phenomenology only in Refs.~\cite{Chiang:2012cn,Chiang:2013rua}.  The full potential contains two additional dimension-3 operators beyond those present in the $Z_2$-symmetric version, providing two additional dimensionful parameters that can drive a decoupling limit.  We show that such a decoupling limit indeed exists and explore its phenomenology.  In particular, we show that in the decoupling limit $M_{\rm new} \gg v$, (i) the additional scalars beyond the light SM-like Higgs boson $h$ become heavy with masses $m_H \sim \mathcal{O}(M_{\rm new})$ and increasingly degenerate with mass splittings $\Delta m_H \sim \mathcal{O}(v^2/M_{\rm new})$; and (ii) the tree-level couplings of the light SM-like Higgs boson $h$ to fermion pairs and $W$ or $Z$ boson pairs, as well as the loop-induced couplings of $h$ to photon pairs or $Z \gamma$ which receive contributions from the new charged scalars in the loop, deviate from the corresponding SM Higgs couplings by a relative correction of at most $\mathcal{O}(v^2/M_{\rm new}^2)$.  

Our most interesting result is that, depending on how the decoupling limit is taken, the deviation of the $h$ coupling to $W$ or $Z$ boson pairs can decouple as $(v^2/M_{\rm new}^2)$, in contrast to the situation in two Higgs doublet models or the Minimal Supersymmetric Standard Model in which this deviation vanishes as $(v^4/M_{\rm new}^4)$~\cite{Gunion:2002zf}.  Furthermore, near the decoupling limit the $hWW$ and $hZZ$ couplings are always larger than their SM values, a phenomenon which cannot be achieved at tree level in models containing only scalar doublets or singlets.  A precision measurement of the Higgs coupling to $W$ or $Z$ boson pairs is thus extremely interesting in the GM model, and may provide the first evidence for scalars transforming under SU(2)$_L$ as representations larger than doublets.

This paper is organized as follows.  In Sec.~\ref{sec:model} we write down the most general scalar potential and the resulting scalar mass eigenstates.  In Sec.~\ref{sec:theoryconstraints} we summarize the theoretical constraints on the model parameters from perturbative unitarity, boundedness-from-below of the scalar potential, and the avoidance of custodial SU(2)-breaking vacua.  In Sec.~\ref{sec:decoupling} we examine the approach to the decoupling limit and discuss the decoupling behavior of the couplings of the SM-like Higgs boson.  We also compare the decoupling behavior to that in the two-Higgs-doublet model and scan over the GM model parameter space in order to evaluate the allowed ranges of couplings of the SM-like Higgs boson as a function of the masses of the heavier scalars.  We conclude in Sec.~\ref{sec:conclusions}.  Feynman rules, formulas for Higgs decays to $\gamma\gamma$ and $Z \gamma$, and a translation table for the alternative parameterizations of the scalar potential used in the literature are collected in the appendices.

%%%%%%%%%%%%%%%%%%%%%%%%%%%%%%%%%%%%%%%%%%%%%%
\section{The model}
\label{sec:model}

The scalar sector of the Georgi-Machacek model consists of the usual complex doublet $(\phi^+,\phi^0)$ with hypercharge\footnote{We use $Q = T^3 + Y/2$.} $Y = 1$, a real 
triplet $(\xi^+,\xi^0,\xi^-)$ with $Y = 0$, and  a complex triplet $(\chi^{++},\chi^+,\chi^0)$ with $Y=2$.  The doublet is responsible for the fermion masses as in the SM.
In order to make the global SU(2)$_L \times$SU(2)$_R$ symmetry explicit, we write the doublet in the form of a bi-doublet $\Phi$ and combine the triplets to form a bi-triplet $X$:
\begin{eqnarray}
	\Phi &=& \left( \begin{array}{cc}
	\phi^{0*} &\phi^+  \\
	-\phi^{+*} & \phi^0  \end{array} \right), \\
	X &=&
	\left(
	\begin{array}{ccc}
	\chi^{0*} & \xi^+ & \chi^{++} \\
	 -\chi^{+*} & \xi^{0} & \chi^+ \\
	 \chi^{++*} & -\xi^{+*} & \chi^0  
	\end{array}
	\right).
	\label{eq:PX}
\end{eqnarray}
The vevs are defined by $\langle \Phi  \rangle = \frac{ v_{\phi}}{\sqrt{2}} \mathbb{1}_{2\times2}$  and $\langle X \rangle = v_{\chi} \mathbb{1}_{3 \times 3}$, where the $W$ and $Z$ boson masses constrain
\begin{equation}
	v_{\phi}^2 + 8 v_{\chi}^2 \equiv v^2 = \frac{4 M_W^2}{g^2} \approx (246~{\rm GeV})^2.
	\label{eq:vevrelation}
\end{equation} 
Note that the two triplet fields $\chi^0$ and $\xi^0$ must obtain the same vev in order to preserve custodial SU(2).
Furthermore we will decompose the neutral fields into real and imaginary parts according to
\begin{equation}
	\phi^0 \to \frac{v_{\phi}}{\sqrt{2}} + \frac{\phi^{0,r} + i \phi^{0,i}}{\sqrt{2}},
	\qquad
	\chi^0 \to v_{\chi} + \frac{\chi^{0,r} + i \chi^{0,i}}{\sqrt{2}}, 
	\qquad
	\xi^0 \to v_{\chi} + \xi^0,
\end{equation}
where we note that $\xi^0$ is already a real field.

The most general gauge-invariant scalar potential involving these fields that conserves custodial SU(2) is given by\footnote{Several different parameterizations of the scalar potential of the Georgi-Machacek model exist in the literature.  We give a translation table in Appendix~\ref{app:translations}.}
\begin{eqnarray}
	V(\Phi,X) &= & \frac{\mu_2^2}{2}  \text{Tr}(\Phi^\dagger \Phi) 
	+  \frac{\mu_3^2}{2}  \text{Tr}(X^\dagger X)  
	+ \lambda_1 [\text{Tr}(\Phi^\dagger \Phi)]^2  
	+ \lambda_2 \text{Tr}(\Phi^\dagger \Phi) \text{Tr}(X^\dagger X)   \nonumber \\
          & & + \lambda_3 \text{Tr}(X^\dagger X X^\dagger X)  
          + \lambda_4 [\text{Tr}(X^\dagger X)]^2 
           - \lambda_5 \text{Tr}( \Phi^\dagger \tau^a \Phi \tau^b) \text{Tr}( X^\dagger t^a X t^b) 
           \nonumber \\
           & & - M_1 \text{Tr}(\Phi^\dagger \tau^a \Phi \tau^b)(U X U^\dagger)_{ab}  
           -  M_2 \text{Tr}(X^\dagger t^a X t^b)(U X U^\dagger)_{ab}.
           \label{eq:potential}
\end{eqnarray} 
Here the SU(2) generators for the doublet representation are $\tau^a = \sigma^a/2$ with $\sigma^a$ being the Pauli matrices,
the generators for the triplet representation are
\begin{equation}
	t^1= \frac{1}{\sqrt{2}} \left( \begin{array}{ccc}
	 0 & 1  & 0  \\
	  1 & 0  & 1  \\
	  0 & 1  & 0 \end{array} \right), \quad  
	  t^2= \frac{1}{\sqrt{2}} \left( \begin{array}{ccc}
	 0 & -i  & 0  \\
	  i & 0  & -i  \\
	  0 & i  & 0 \end{array} \right), \quad 
	t^3= \left( \begin{array}{ccc}
	 1 & 0  & 0  \\
	  0 & 0  & 0  \\
	  0 & 0 & -1 \end{array} \right),
\end{equation}
and the matrix $U$, which rotates $X$ into the Cartesian basis, is given by~\cite{Aoki:2007ah}
\begin{equation}
	 U = \left( \begin{array}{ccc}
	- \frac{1}{\sqrt{2}} & 0 &  \frac{1}{\sqrt{2}} \\
	 - \frac{i}{\sqrt{2}} & 0  &   - \frac{i}{\sqrt{2}} \\
	   0 & 1 & 0 \end{array} \right).
	 \label{eq:U}
\end{equation}
We note that all the operators in Eq.~(\ref{eq:potential}) are manifestly Hermitian, so that the parameters in the scalar potential must all be real.  Explicit CP violation is thus not possible in the Georgi-Machacek model.  

In terms of the vevs, the scalar potential is given by\footnote{We will discuss the conditions required to avoid alternative minima in Sec.~\ref{sec:altmin}.}
\begin{equation}
	V(v_\phi,v_\chi) = \frac{\mu_2^2}{2} v_\phi^2 + 3 \frac{\mu_3^2}{2} v_\chi^2
	+ \lambda_1 v_\phi^4 
	+ \frac{3}{2} \left( 2 \lambda_2 - \lambda_5 \right) v_\phi^2 v_\chi^2
	+ 3 \left( \lambda_3 + 3 \lambda_4 \right) v_\chi^4
	- \frac{3}{4} M_1 v_\phi^2 v_\chi - 6 M_2 v_\chi^3.
\end{equation}
Minimizing this potential yields the following constraints:
\begin{eqnarray}
	0 = \frac{\partial V}{\partial v_{\phi}} &=& 
	v_{\phi} \left[ \mu_2^2 + 4 \lambda_1 v_{\phi}^2 
	+ 3 \left( 2 \lambda_2 - \lambda_5 \right) v_{\chi}^2 - \frac{3}{2} M_1 v_{\chi} \right], 
		\label{eq:phimincond} \\
	0 = \frac{\partial V}{\partial v_{\chi}} &=& 
	3 \mu_3^2 v_{\chi} + 3 \left( 2 \lambda_2 - \lambda_5 \right) v_{\phi}^2 v_{\chi}
	+ 12 \left( \lambda_3 + 3 \lambda_4 \right) v_{\chi}^3
	- \frac{3}{4} M_1 v_{\phi}^2 - 18 M_2 v_{\chi}^2.
	\label{eq:chimincond}
\end{eqnarray}
Inserting $v_{\phi}^2 = v^2 - 8 v_{\chi}^2$ [Eq.~(\ref{eq:vevrelation})] into Eq.~(\ref{eq:chimincond}) yields a cubic equation for $v_{\chi}$ in terms of $v$, $\mu_3^2$, $\lambda_2$, $\lambda_3$, $\lambda_4$, $\lambda_5$, $M_1$, and $M_2$.  With $v_{\chi}$ (and hence $v_{\phi}$) in hand, Eq.~(\ref{eq:phimincond}) can be used to eliminate $\mu_2^2$ in terms of the parameters in the previous sentence together with $\lambda_1$.  We illustrate below how $\lambda_1$ can also be eliminated in favor of one of the custodial singlet Higgs masses $m_h$ or $m_H$ [see Eq.~(\ref{eq:lambda1})].

The physical field content is as follows.
When expanded around the minimum, the scalar potential gives rise to ten real physical fields together with three Goldstone bosons.  The Goldstone bosons are given by
\begin{eqnarray}
	G^+ &=& c_H \phi^+ + s_H \frac{\left(\chi^++\xi^+\right)}{\sqrt{2}}, \nonumber\\
	G^0  &=& c_H \phi^{0,i} + s_H \chi^{0,i},
\end{eqnarray}
where
\begin{equation}
	c_H \equiv \cos\theta_H = \frac{v_{\phi}}{v}, \qquad
	s_H \equiv \sin\theta_H = \frac{2\sqrt{2}\,v_\chi}{v}.
\end{equation}
The physical fields can be organized by their transformation properties under the custodial SU(2) symmetry into a fiveplet, a triplet, and two singlets.  The fiveplet and triplet states are given by
\begin{eqnarray}
	H_5^{++}  &=&  \chi^{++}, \nonumber\\
	H_5^+ &=& \frac{\left(\chi^+ - \xi^+\right)}{\sqrt{2}}, \nonumber\\
	H_5^0 &=& \sqrt{\frac{2}{3}} \xi^0 - \sqrt{\frac{1}{3}} \chi^{0,r}, \nonumber\\
	H_3^+ &=& - s_H \phi^+ + c_H \frac{\left(\chi^++\xi^+\right)}{\sqrt{2}}, \nonumber\\
	H_3^0 &=& - s_H \phi^{0,i} + c_H \chi^{0,i}.
\end{eqnarray}
Within each custodial multiplet, the masses are degenerate at tree level.  Using Eqs.~(\ref{eq:phimincond}--\ref{eq:chimincond}) to eliminate $\mu_2^2$ and $\mu_3^2$, the fiveplet and triplet masses can be written as
\begin{eqnarray}
	m_5^2 &=& \frac{M_1}{4 v_{\chi}} v_\phi^2 + 12 M_2 v_{\chi} 
	+ \frac{3}{2} \lambda_5 v_{\phi}^2 + 8 \lambda_3 v_{\chi}^2, \nonumber \\
	m_3^2 &=&  \frac{M_1}{4 v_{\chi}} (v_\phi^2 + 8 v_{\chi}^2) 
	+ \frac{\lambda_5}{2} (v_{\phi}^2 + 8 v_{\chi}^2) 
	= \left(  \frac{M_1}{4 v_{\chi}} + \frac{\lambda_5}{2} \right) v^2.
\end{eqnarray}
Note that the ratio $M_1/v_{\chi}$ is finite in the limit $v_{\chi} \to 0$, as can be seen from Eq.~(\ref{eq:chimincond}) which yields
\begin{equation}
	\frac{M_1}{v_{\chi}} = \frac{4}{v_{\phi}^2} 
	\left[ \mu_3^2 + (2 \lambda_2 - \lambda_5) v_{\phi}^2 
	+ 4(\lambda_3 + 3 \lambda_4) v_{\chi}^2 - 6 M_2 v_{\chi} \right].
\end{equation}

The two custodial SU(2) singlets are given in the gauge basis by
\begin{eqnarray}
	H_1^0 &=& \phi^{0,r}, \nonumber \\
	H_1^{0 \prime} &=& \sqrt{\frac{1}{3}} \xi^0 + \sqrt{\frac{2}{3}} \chi^{0,r}.
\end{eqnarray}
These states mix by an angle $\alpha$ to form the two custodial-singlet mass eigenstates $h$ and $H$, defined such that $m_h < m_H$:
\begin{eqnarray}
	h &=& \cos \alpha \, H_1^0 - \sin \alpha \, H_1^{0\prime},  \\ \nonumber 
	H &=& \sin \alpha \, H_1^0 + \cos \alpha \, H_1^{0\prime}. \label{mh-mH}
\end{eqnarray}
The mixing is controlled by the $2\times 2$ mass-squared matrix
\begin{equation}
	\mathcal{M}^2 = \left( \begin{array}{cc}
			\mathcal{M}_{11}^2 & \mathcal{M}_{12}^2 \\
			\mathcal{M}_{12}^2 & \mathcal{M}_{22}^2 \end{array} \right),
\end{equation}
where
\begin{eqnarray}
	\mathcal{M}_{11}^2 &=& 8 \lambda_1 v_{\phi}^2, \nonumber \\
	\mathcal{M}_{12}^2 &=& \frac{\sqrt{3}}{2} v_{\phi} 
	\left[ - M_1 + 4 \left(2 \lambda_2 - \lambda_5 \right) v_{\chi} \right], \nonumber \\
	\mathcal{M}_{22}^2 &=& \frac{M_1 v_{\phi}^2}{4 v_{\chi}} - 6 M_2 v_{\chi} 
	+ 8 \left( \lambda_3 + 3 \lambda_4 \right) v_{\chi}^2.
\end{eqnarray}
The mixing angle is fixed by 
\begin{eqnarray}
	\sin 2 \alpha &=&  \frac{2 \mathcal{M}^2_{12}}{m_H^2 - m_h^2},    \nonumber  \\
	\cos 2 \alpha &=&  \frac{ \mathcal{M}^2_{22} - \mathcal{M}^2_{11}  }{m_H^2 - m_h^2},    
\end{eqnarray}
with the masses given by
\begin{eqnarray}
	m^2_{h,H} &=& \frac{1}{2} \left[ \mathcal{M}_{11}^2 + \mathcal{M}_{22}^2
	\mp \sqrt{\left( \mathcal{M}_{11}^2 - \mathcal{M}_{22}^2 \right)^2 
	+ 4 \left( \mathcal{M}_{12}^2 \right)^2} \right].
	\label{eq:hmass}
\end{eqnarray}

It is convenient to use the measured mass of the observed SM-like Higgs boson as an input parameter.  The coupling $\lambda_1$ can be eliminated in favor of this mass by inverting Eq.~(\ref{eq:hmass}):
\begin{equation}
	\lambda_1 = \frac{1}{8 v_{\phi}^2} \left[ m_h^2 
	+ \frac{\left( \mathcal{M}_{12}^2 \right)^2}{\mathcal{M}_{22}^2 - m_h^2} \right].
	\label{eq:lambda1}
\end{equation}
Note that in deriving this expression for $\lambda_1$, the distinction between $m_h$ and $m_H$ is lost.  This means that, depending on the values of $\mu_3^2$ and the other parameters, this (unique) solution for $\lambda_1$ will correspond to either the lighter or the heavier custodial singlet having a mass equal to the observed SM-like Higgs mass.

%%%%%%%%%%%%%%%%%%%%%%%%%%%%%%%%%%%%%%%%%%%%%%

\section{Theoretical Constraints on Lagrangian Parameters}
\label{sec:theoryconstraints}

\subsection{Perturbative unitarity of scalar field scattering amplitudes}

Perturbative unitarity of $2 \to 2$ scalar field scattering amplitudes requires that the zeroth partial wave amplitude, $a_0$, satisfy $|a_0| \leq 1$ or $|{\rm Re} \, a_0| \leq \frac{1}{2}$.  Because the $2 \to 2$ scalar field scattering amplitudes are real at tree level, we adopt the second, more stringent, constraint.  The partial wave amplitude $a_0$ is related to the matrix element $\mathcal{M}$ of the process by
\begin{equation}
	\mathcal{M} = 16\pi\sum_J (2J+1) a_J P_J(\cos\theta),
\end{equation}
where $J$ is the (orbital) angular momentum and $P_J(\cos\theta)$ are the Legendre polynomials.  We will use this to constrain the magnitudes of the scalar quartic couplings $\lambda_i$.  These unitarity bounds for the scalar quartic couplings in the GM model were previously computed in Ref.~\cite{Aoki:2007ah}.  We have recomputed them independently and agree with the results of Ref.~\cite{Aoki:2007ah}.

We work in the high energy limit, in which the only tree-level diagrams that contribute to $2 \to 2$ scalar scattering are those involving the four-point scalar couplings; all diagrams involving scalar propagators are suppressed by the square of the collision energy.  Thus the dimensionful couplings $M_1$, $M_2$, $\mu_2^2$, and $\mu_3^2$ are not constrained directly by perturbative unitarity.  In the high energy limit we can ignore electroweak symmetry breaking and include the Goldstone bosons as physical fields (this is equivalent to including scattering processes involving longitudinally polarized $W$ and $Z$ bosons).  We neglect scattering processes involving transversely polarized gauge bosons or fermions.

Under these conditions, only the zeroth partial wave amplitude contributes to $\mathcal{M}$, so that the constraint $|{\rm Re} \, a_0| < \frac{1}{2}$ corresponds to $|\mathcal{M}| < 8 \pi$.  This condition must be applied to each of the eigenvalues of the coupled-channel scattering matrix $\mathcal{M}$ including each possible combination of two scalar fields in the initial and final states.  Because the scalar potential is invariant under SU(2)$_L \times$U(1)$_Y$, the scattering processes preserve electric charge and hypercharge and can be conveniently classified by the total electric charge and hypercharge of the incoming and outgoing states.  We include a symmetry factor of $1/\sqrt{2}$ for each pair of identical particles in the initial and final states.  The basis states and resulting eigenvalues of $\mathcal{M}$ are summarized in Table~\ref{tab:evals}.  

\begin{table}
\begin{tabular}{ccll}
\hline \hline
$Q$ & $Y$ & Basis states &  Eigenvalues  \\
\hline
0  &  0  &  $[\chi^{++*}\chi^{++},\chi^{+*}\chi^{+},\xi^{+*}\xi^{+},\phi^{+*}\phi^{+}, \chi^{0*}\chi^{0},\frac{\xi^{0}\xi^{0}}{\sqrt{2}},\phi^{0*}\phi^{0}]$  &   $x_1^+, x_1^-, x_2^+, x_2^-, y_1, y_1, y_2$  \\
0  &  1  &  $[\phi^+  \xi^{+*} ,   \phi^0 \xi^0, \chi^+  \phi^{+*} , \chi^0  \phi^{0*}]$  & $y_3, y_4, y_4, y_5$ \\
0  &  2  &  $[\frac{\phi^0 \phi^0}{\sqrt{2}} , \chi^0 \xi^0 , \chi^+  \xi^{+*}]$  &  $x_2^+, x_2^-, y_2$  \\
0  &  3  &  $[\phi^0 \chi^0]$ & $y_3$ \\
0  &  4  &  $[ \frac{\chi^0 \chi^0 }{\sqrt{2}} ]$  & $y_2$ \\
\hline
1 & $-2$  & $[\xi^+ \chi^{0*}]$ & $y_2$  \\
1 & $-1$ &  $[\phi^+  \chi^{0*}  ,  \xi^+ \phi^{0*} ]$ &  $y_3, y_4$ \\
1 & 0  &  $[\xi^+ \xi^0 , \chi^{+*} \chi^{++} , \phi^+  \phi^{0*} , \chi^{0*} \chi^+]$  &  $x_2^+, x_2^-, y_1, y_2$ \\
1 & 1  &  $[\phi^0 \xi^+ , \phi^+ \xi^0 , \phi^{+*} \chi^{++}  , \phi^{0*} \chi^+  ]$ & $y_3, y_4, y_4, y_5$ \\
1 & 2  &  $[\phi^+ \phi^0 , \chi^+ \xi^0 ,  \chi^{++} \xi^{+*} , \chi^0 \xi^+ ]$ & $x_2^+, x_2^-, y_1 , y_2$\\
1 & 3  &  $[\phi^+ \chi^0 , \phi^0 \chi^+]$ &  $y_3, y_4$\\
1 & 4  &  $[ \chi^+ \chi^0 ]$ &  $y_2$ \\
\hline
2 & 0 &  $[\chi^{++} \chi^{0*} , \frac{\xi^+ \xi^+ }{\sqrt{2}}]$ & $y_1, y_2$  \\
2 & 1 &  $[\phi^+ \xi^+ , \chi^{++} \phi^{0*}]$ &   $y_3, y_4$ \\  
2 & 2 &  $[\frac{\phi^+ \phi^+}{\sqrt{2}} ,  \chi^{++} \xi^0  ,  \chi^+ \xi^+]$ & $x_2^+, x_2^-, y_2$  \\
2 & 3 &  $[\phi^+ \chi^+ , \phi^0 \chi^{++}]$ &  $y_3, y_4$ \\
2 & 4 &  $[\chi^{++}  \chi^0 , \frac{\chi^+ \chi^+}{\sqrt{2}}]$ & $y_1, y_2$  \\
\hline
3 & 2 &  $[\chi^{++} \xi^+]$ & $y_2$  \\
3 & 3 &  $[\chi^{++} \phi^+ ]$ &  $y_3$ \\
3 & 4 &  $[\chi^{++} \chi^+ ]$ &  $y_2$ \\
\hline
4 & 4 & $[\frac{\chi^{++} \chi^{++}}{\sqrt{2}}]$ & $y_2$  \\
\hline \hline
\end{tabular}
\caption{Basis states and eigenvalues of the scattering matrix $\mathcal{M}$ for $2 \to 2$ scalar scattering in the high-energy limit, classified according to the total charge $Q$ and total hypercharge $Y$ of the initial and final states.  We have included a symmetry factor of $1/\sqrt{2}$ in the matrix element for each pair of identical particles in the initial or final state.  The eigenvalues are defined in Eq.~(\ref{eq:xiyi}).  The charge-conjugates of the listed states yield the same sets of eigenvalues.}
\label{tab:evals}
\end{table}

The eigenvalues of $\mathcal{M}$ comprise the following independent combinations of $\lambda_i$ (defined in the same way as in Ref.~\cite{Aoki:2007ah}):\footnote{Our notation for the $\lambda_i$ is different from that of Ref.~\cite{Aoki:2007ah}.  This has been taken into account in the definitions of $x_i^{\pm}$ and $y_i$ in Eq.~(\ref{eq:xiyi}).  A translation between our notation and that of Ref.~\cite{Aoki:2007ah} is given in Appendix~\ref{app:translations}.}
\begin{eqnarray}
	x_1^{\pm}  &=& 12 \lambda_1 + 14 \lambda_3 + 22 \lambda_4 
	\pm \sqrt{ \left(12 \lambda_1 - 14 \lambda_3 - 22 \lambda_4 \right)^2 + 144 \lambda_2^2},     
	\nonumber \\  
	x_2^{\pm}  &=&     4 \lambda_1 - 2 \lambda_3 + 4 \lambda_4  
	\pm \sqrt{ \left( 4 \lambda_1  + 2\lambda_3 - 4 \lambda_4 \right)^2 + 4 \lambda_5^2}, 
	\nonumber \\
	y_1 &=&  16 \lambda_3 + 8 \lambda_4,     \nonumber \\
	y_2 &=&   4 \lambda_3 + 8 \lambda_4,   \nonumber \\
	y_3 &=&    4 \lambda_2 - \lambda_5,   \nonumber \\
	y_4 &=&     4 \lambda_2 + 2 \lambda_5,  \nonumber \\
	y_5 &=&     4 \lambda_2 - 4 \lambda_5.
	\label{eq:xiyi}
\end{eqnarray}
Requiring $|{\rm Re} \, a_0| < \frac{1}{2}$ imposes the conditions $|x_i^{\pm}| < 8\pi$ and $|y_i| < 8 \pi$, which must all be simultaneously satisfied.\footnote{The unitarity constraint imposed in Ref.~\cite{Chiang:2012cn} corresponds to $|x_1^{\pm}| < 16 \pi$, which is obtained by requiring $|a_0| < 1$ rather than our more stringent constraint $|{\rm Re} \, a_0| < \frac{1}{2}$.}

These conditions allow us to determine the maximum range allowed by unitarity for each of the parameters $\lambda_i$, which will be useful for setting up numerical parameter scans.  We first note that the conditions $|x_i^{\pm}| < 8 \pi$ take the general form $|z \pm \sqrt{x^2 + y^2}| < 1$, which can be rewritten without loss of generality as $\sqrt{x^2 + y^2} + |z| < 1$.  This equation describes the region bounded by a pair of cones with apices at $z = \pm 1$ that meet at a unit circle in the $x$--$y$ plane.  Clearly, then, the maximum allowed range of $y$ (i.e., $\lambda_2$ or $\lambda_5$) is obtained by setting $x = z = 0$, and the maximum allowed range in the $x$--$z$ plane is obtained by setting $y = 0$.

The coupling $\lambda_1$ is constrained by the unitarity conditions on $x_1^{\pm}$ and $x_2^{\pm}$.  The least stringent constraints come from setting $\lambda_2 = \lambda_5 = 0$ and read $|\lambda_1| < \frac{1}{3} \pi$ from $x_1^{\pm}$ and $|\lambda_1| < \pi$ from $x_2^{\pm}$.  We thus obtain the maximum range from unitarity,
\begin{equation}
	\lambda_1 \in \left( -\frac{1}{3} \pi, \frac{1}{3} \pi \right)
		\simeq \left( -1.05, 1.05 \right). 
	\label{eq:l1uni}
\end{equation}

\begin{figure}
\resizebox{\textwidth}{!}{\includegraphics{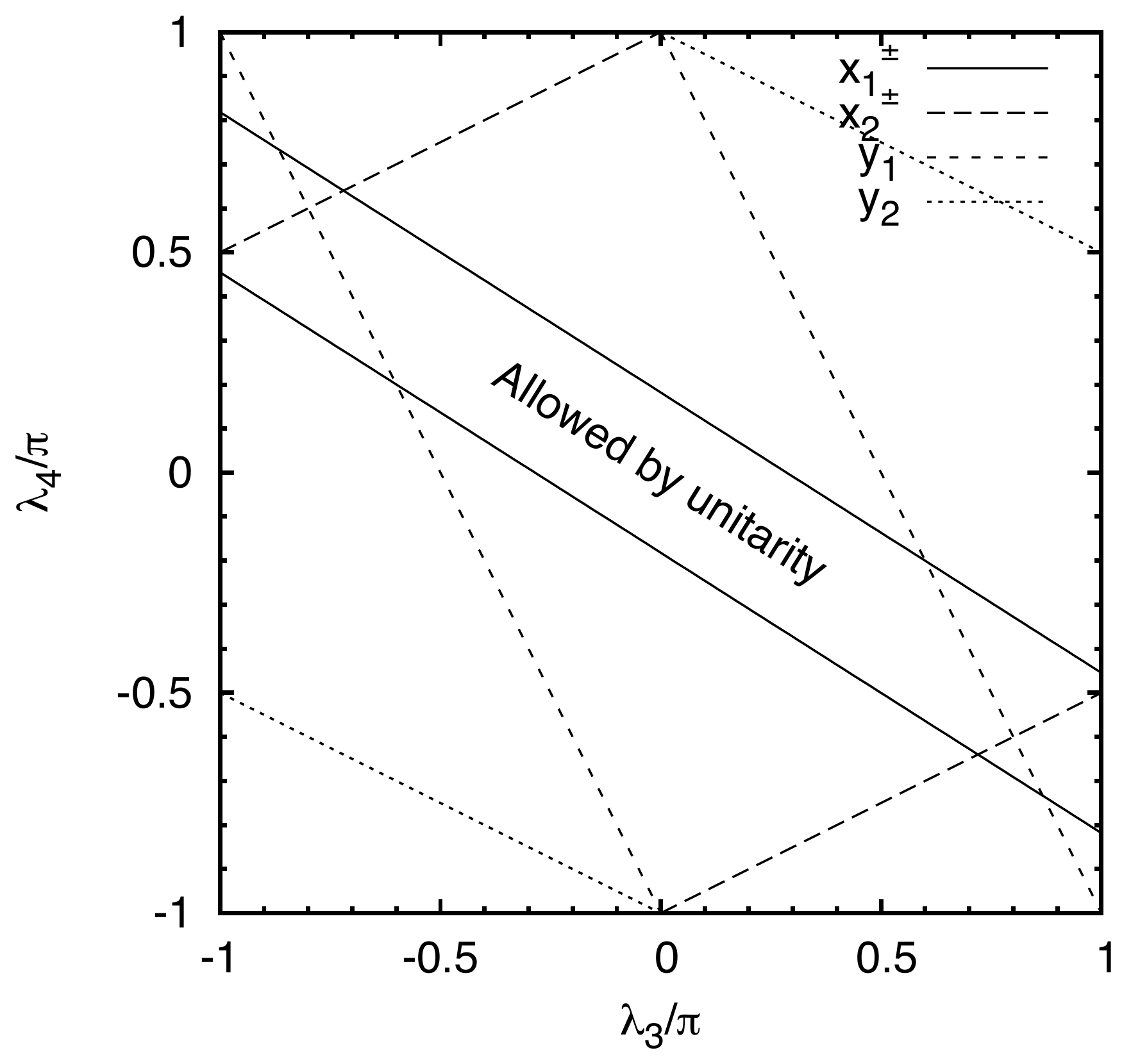}\includegraphics{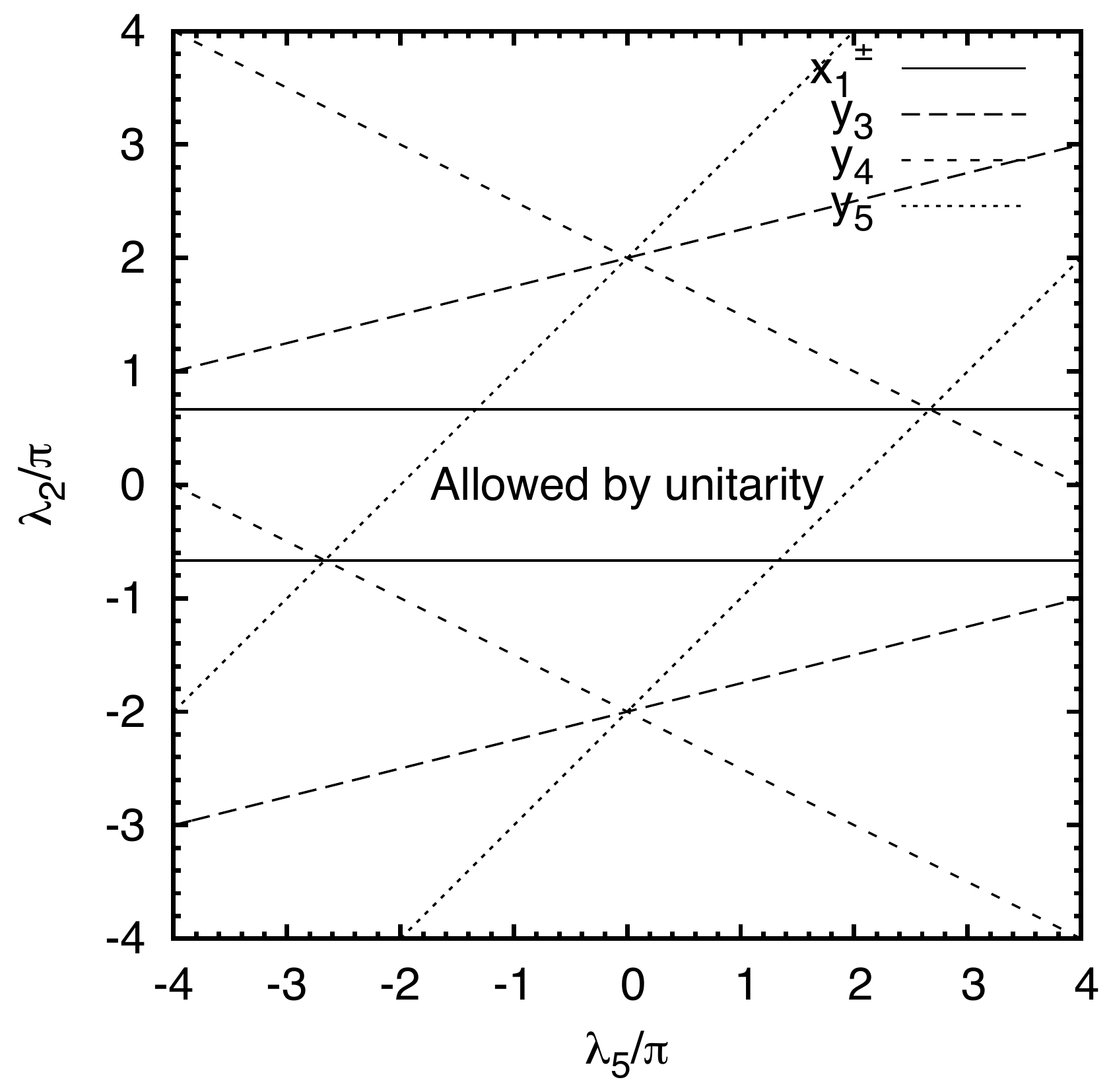}}
\caption{Constraints on the $(\lambda_3,\lambda_4)$ and $(\lambda_5, \lambda_2)$ planes from perturbative unitarity.  The constraints from $x_1^{\pm}$ and $x_2^{\pm}$ are the maximum allowed ranges obtained by setting the couplings not shown on the figure axes to zero.}
\label{fig:unitarity}
\end{figure}

Constraints on the couplings $\lambda_3$ and $\lambda_4$ come from the unitarity conditions on $x_1^{\pm}$, $x_2^{\pm}$, $y_1$, and $y_2$.  These are shown in the left panel of Fig.~\ref{fig:unitarity}, where we again take $\lambda_2 = 0$ in $x_1^{\pm}$ and $\lambda_5 = 0$ in $x_2^{\pm}$ for the least stringent constraints.  The allowed region in the $\lambda_3$--$\lambda_4$ plane is a six-sided region bounded by the constraints on $x_1^{\pm}$, $x_2^{\pm}$, and $y_1$.  The constraint on $y_2$ does not provide any additional information.  Simultaneously satisfying all constraints, we obtain the maximum ranges from unitarity,
\begin{eqnarray}
	\lambda_3 &\in& \left( -\frac{4}{5} \pi, \frac{4}{5} \pi \right)
		\simeq \left( -2.51, 2.51 \right), \nonumber \\
	\lambda_4 &\in& \left( -\frac{16}{25} \pi, \frac{16}{25} \pi \right)
		\simeq \left( -2.01, 2.01 \right).
	\label{eq:l3l4uni}
\end{eqnarray}

Constraints on the couplings $\lambda_2$ and $\lambda_5$ come from the unitarity constraints on $x_1^{\pm}$, $x_2^{\pm}$, $y_3$, $y_4$, and $y_5$.  These are shown in the right panel of Fig.~\ref{fig:unitarity}, where we take $\lambda_1 = \lambda_3 = \lambda_4 = 0$ in $x_1^{\pm}$ and $x_2^{\pm}$ for the least stringent constraints.  (The constraint from $x_2^{\pm}$ yields $|\lambda_5| < 4 \pi$, which corresponds to the left and right edges of the plot.)  The allowed region in the $\lambda_2$--$\lambda_5$ plane is a parallelogram bounded by the constraints on $x_1^{\pm}$ and $y_5$.  The constraints on $y_3$ and $y_4$ do not provide any additional information.  Simultaneously satisfying all constraints, we obtain the maximum ranges from unitarity,
\begin{eqnarray}
	\lambda_2 &\in& \left( -\frac{2}{3} \pi, \frac{2}{3} \pi \right)
		\simeq \left( -2.09, 2.09 \right), \nonumber \\
	\lambda_5 &\in& \left( -\frac{8}{3} \pi, \frac{8}{3} \pi \right)
		\simeq \left( -8.38, 8.38 \right).
	\label{eq:l2l5uni}
\end{eqnarray}

Within these maximum ranges the unitarity constraints $|x_i^{\pm}|, |y_i| < 8 \pi$ must still be imposed.  Discarding expressions that provide no additional information, we obtain the minimal set of unitarity conditions,\footnote{Imposing $|a_0| < 1$ instead of $|{\rm Re} \, a_0| < \frac{1}{2}$ would double the right-hand side of each of these expressions.}
\begin{eqnarray}
	\sqrt{ \left( 6 \lambda_1 - 7 \lambda_3 - 11 \lambda_4 \right)^2 + 36 \lambda_2^2}
	+ \left| 6 \lambda_1 + 7 \lambda_3 + 11 \lambda_4 \right| &<& 4 \pi, \nonumber \\
	\sqrt{ \left( 2 \lambda_1 + \lambda_3 - 2 \lambda_4 \right)^2 + \lambda_5^2}
	+ \left| 2 \lambda_1 - \lambda_3 + 2 \lambda_4 \right| &<& 4 \pi, \nonumber \\
	\left| 2 \lambda_3 + \lambda_4 \right| &<& \pi, \nonumber \\
	\left| \lambda_2 - \lambda_5 \right| &<& 2 \pi.
\end{eqnarray}

%%%%%%%%%%%%%%%%%%%%%%%%%%%%%%%%%%%%%%%%%%%
\subsection{Bounded-from-below requirement on the scalar potential}

The constraints that must be satisfied at tree level for the scalar potential to be bounded from below can be determined by considering only the terms in the scalar potential [Eq.~(\ref{eq:potential})] that are quartic in the fields, because these terms dominate at large field values.
Following the approach of Ref.~\cite{Arhrib:2011uy}, we parametrize the potential using the following definitions:
\begin{eqnarray}
	r &\equiv&  \sqrt{\text{Tr}(\Phi^\dagger \Phi) + \text{Tr}(X^\dagger X)} ,  \nonumber \\
	r^2 \cos^2 \gamma  &\equiv&  \text{Tr}(\Phi^\dagger \Phi)   ,  \nonumber \\   
	r^2 \sin^2 \gamma &\equiv&   \text{Tr}(X^\dagger X)   ,    \nonumber \\    
	\zeta &\equiv&   \frac{ \textrm{Tr}(X^\dagger X X^\dagger X)}{[{\rm Tr}(X^\dagger X)]^2},  \nonumber \\
	\omega &\equiv&   \frac{\text{Tr}( \Phi^\dagger \tau^a \Phi \tau^b) \text{Tr}( X^\dagger t^a X t^b)}{\textrm{Tr}(\Phi^\dagger \Phi)\textrm{Tr}(X^\dagger X)}.
	\label{eq:bfbdefs}
\end{eqnarray}
Scanning all possible field values yields the parameter ranges
\begin{equation}
 	r \in [0,\infty)  ,  \quad   
	\gamma \in \left[ 0,\frac{\pi}{2} \right]   ,    \quad  
	\zeta \in \left[ \frac{1}{3},1 \right]  \quad  \textrm{and}  \quad  
	\omega \in \left[ -\frac{1}{4}, \frac{1}{2} \right].
 \end{equation}  
The ranges of $\zeta$ and $\omega$ will be discussed in more detail below.
 
The quartic terms in the potential are given in this parametrization by,
 \begin{equation}
	V^{(4)}(r, \tan \gamma, \zeta , \omega ) = \frac{r^4}{(1+\tan^2 \gamma)^2} 
	\left[ \lambda_1 + (\lambda_2 - \omega \lambda_5) \tan^2 \gamma 
	+ (\zeta \lambda_3 + \lambda_4 ) \tan^4 \gamma \right].
	\label{eq:bfb}
\end{equation}
The potential will be bounded from below if the expression multiplying $r^4$ in Eq.~(\ref{eq:bfb}) is always positive.  The expression in the square brackets in Eq.~(\ref{eq:bfb}) is a bi-quadratic in $\tan \gamma \equiv y$ of the form $[a + b y^2 + c y^4]$.  Such an expression is positive for all values of $y \in [0,\infty)$ when
 \begin{equation}
	a > 0, \qquad
 	c > 0, \qquad {\rm and} \quad
 	b + 2\sqrt{ac} > 0.
 \end{equation}
We thus obtain the bounded-from-below conditions,
\begin{equation}
	\lambda_1 > 0,  \qquad  
	\zeta \lambda_3 + \lambda_4  > 0, 
	\qquad \textrm{and}  \quad 
	\lambda_2 - \omega \lambda_5  + 2 \sqrt{\lambda_1 (\zeta \lambda_3 + \lambda_4 )} > 0.
	\label{eq:bfbcond}
\end{equation}
These conditions must be satisfied for all allowed values of $\zeta$ and $\omega$.

The field combination $\zeta$ is given explicitly by
\begin{eqnarray}
	\zeta &=& \frac{1}{\left[ {\rm Tr}(X^{\dagger} X) \right]^2}
	\left\{ 2 \left( |\chi^0|^2 + |\chi^+|^2 + |\chi^{++}|^2 \right)^2
	+ \left[ 2 |\xi^+|^2 + (\xi^0)^2 \right]^2 \right. \nonumber \\
	&& \left. + 2 | \chi^+ \chi^+ - 2 \chi^0 \chi^{++}|^2 
	+ 4 | \xi^+ \chi^0 - \xi^0 \chi^+ - \xi^{+*} \chi^{++} |^2 \right\},
\end{eqnarray}
where 
\begin{equation}
	{\rm Tr}(X^{\dagger} X) = 2 |\chi^0|^2 + 2 |\chi^+|^2 + 2 |\chi^{++}|^2
  	+ 2 |\xi^+|^2 + (\xi^0)^2.
	\label{eq:TrXdX}
\end{equation}
To derive the allowed range of $\zeta$, we can work in a basis where the Hermitian matrix $X^\dagger X$ is diagonalized with positive real eigenvalues $x_1$, $x_2$ and $x_3$. 
In this basis,
\begin{equation}
	\zeta = 
	\frac{x_1^2 + x_2^2 + x_3^2}{x_1^2 + x_2^2 + x_3^2 + 2(x_1 x_2 + x_2 x_3 + x_3 x_1)}, 
\end{equation}
from which it follows (using $a^2 + b^2 \geq 2 ab$) that $\zeta \in \left[ \frac{1}{3},1 \right]$.\footnote{Our desired vacuum, with $\langle \chi^0 \rangle = \langle \xi^0 \rangle = v_{\chi}$, corresponds to $\zeta = 1/3$.}
 
To derive the allowed range of $\omega$, we can start by choosing the SU(2)$_L$ basis so that the field value of $\Phi$ lies entirely in the real neutral component, $\Phi = \frac{v_{\phi}}{\sqrt{2}} \mathbb{1}_{2\times2}$.  Then,
\begin{equation}
	 \frac{{\rm Tr}( \Phi^{\dagger} \tau^a \Phi \tau^b)}{{\rm Tr}(\Phi^{\dagger} \Phi)} 
	 = \frac{1}{4} \delta^{ab}.
	 \label{eq:phicombo}
\end{equation}
Inserting this into the expression for $\omega$ in Eq.~(\ref{eq:bfbdefs}) yields 
\begin{equation}
	\omega = \frac{1}{2 {\rm Tr}(X^{\dagger} X)}
	\left[ |\chi^0|^2 - |\chi^{++}|^2 + 2 \xi^0 {\rm Re} \chi^0 
	+ 2 {\rm Re}(\xi^+ \chi^{+*}) \right].
\end{equation}
Because $\omega$ is invariant under custodial SU(2), this expression can be rewritten in terms of custodial SU(2) eigenstates as follows.  We first define the custodial singlet, triplet, and fiveplet contained in $X$ according to
\begin{eqnarray}
	X_1 &=& \frac{1}{\sqrt{3}} (\xi^0 + 2 {\rm Re} \chi^0), \nonumber \\
	X_3 &=& \left( \begin{array}{c} 
		\frac{1}{\sqrt{2}} (\chi^+ + \xi^+) \\
		\sqrt{2} \, {\rm Im} \chi^0 \\
		- \frac{1}{\sqrt{2}} (\chi^{+*} + \xi^{+*}) \end{array} \right), \nonumber \\
	X_5 &=& \left( \begin{array}{c}
		\chi^{++} \\
		\frac{1}{\sqrt{2}} (\chi^+ - \xi^+) \\
		\sqrt{\frac{2}{3}} ( \xi^0 - {\rm Re}\chi^0) \\
		-  \frac{1}{\sqrt{2}} (\chi^{+*} - \xi^{+*}) \\
		\chi^{++*} \end{array} \right).
	\label{eq:custodialmults}
\end{eqnarray}
In terms of the custodial symmetry eigenstates, we have
\begin{equation}
	{\rm Tr}(X^{\dagger} X) = (X_1)^2 + |X_3|^2 + |X_5|^2,
\end{equation}
where $|X_3|^2 \equiv X_3^{\dagger} X_3$ and $|X_5|^2 \equiv X_5^{\dagger} X_5$, and
\begin{equation}
	\omega = \frac{1}{4} \frac{2 (X_1)^2 + |X_3|^2 - |X_5|^2}
	{(X_1)^2 + |X_3|^2 + |X_5|^2}.
\end{equation}
From this form it can be easily seen that $\omega \in \left[ -\frac{1}{4}, \frac{1}{2} \right]$.\footnote{Our desired vacuum, with $\langle \chi^0 \rangle = \langle \xi^0 \rangle = v_{\chi}$, corresponds to $\omega = 1/2$.}

\begin{figure}
\begin{center}
\resizebox{0.65\textwidth}{!}{\includegraphics{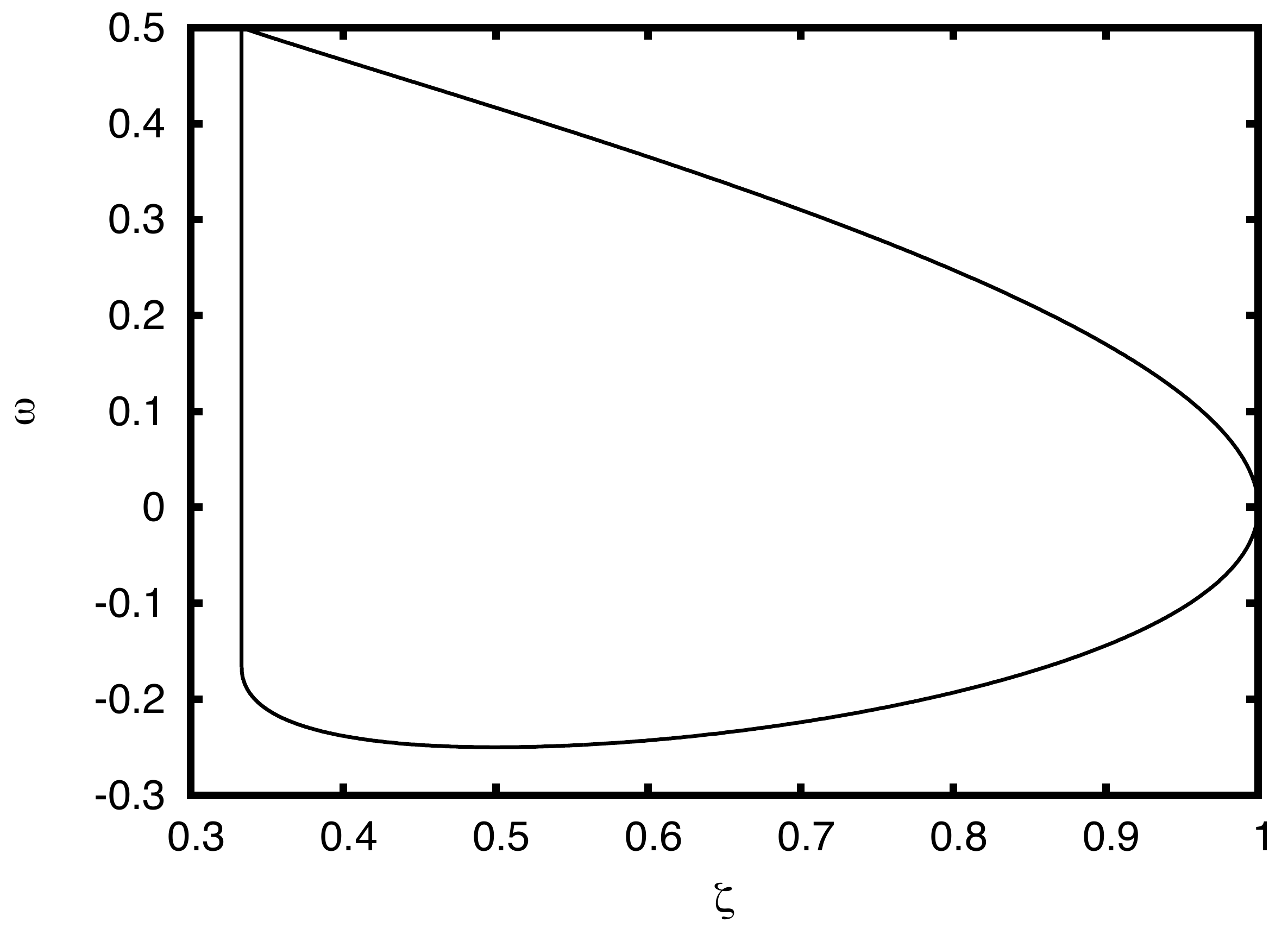}}
\end{center}
\caption{The boundary of the region in the $(\zeta,\omega)$ plane that is populated by taking all possible combinations of field values.  The region enclosed by the curve is populated.}
\label{fig:zetaomega}
\end{figure}

The region in the $(\zeta,\omega)$ plane populated by taking all possible combinations of field values is shown in Fig.~\ref{fig:zetaomega}.  For a given $\zeta$, the region encompasses $\omega \in [\omega_-, \omega_+]$, where\footnote{These bounds on $\omega$ are obtained by noting that the curved part of the boundary in Fig.~\ref{fig:zetaomega} is traced out by field combinations in which only ${\rm Re} \, \chi^0$ and $\xi^0$ are nonzero.  Taking into account the normalization by ${\rm Tr}(X^{\dagger}X)$, the formulas for $\zeta$ and $\omega$ along this boundary can then be expressed as functions of a single variable, which can in turn be expressed in terms of $\zeta$.}
\begin{equation}
	\omega_{\pm}(\zeta) = \frac{1}{6}(1 - B) \pm \frac{\sqrt{2}}{3} \left[ (1 - B) \left(\frac{1}{2} + B\right)\right]^{1/2},
\end{equation}
with
\begin{equation}
	B \equiv \sqrt{\frac{3}{2}\left(\zeta - \frac{1}{3}\right)} \in [0,1].
\end{equation}

Following Ref.~\cite{Arhrib:2011uy}, the monotonic dependence on $\zeta$ and $\omega$ in Eq.~(\ref{eq:bfbcond}) can be used to obtain the following bounded-from-below constraints:\footnote{Reference~\cite{Chiang:2012cn} computed the bounded-from-below constraints taking into account all combinations of two nonzero scalar fields.  Because our treatment allows any number of the scalar fields to be nonzero, our bounded-from-below constraints are more stringent than those of Ref.~\cite{Chiang:2012cn}.}
\begin{eqnarray}
	\lambda_1 &>& 0, \nonumber \\
	\lambda_4 &>& \left\{ \begin{array}{l l}
		- \frac{1}{3} \lambda_3 & {\rm for} \ \lambda_3 \geq 0, \\
		- \lambda_3 & {\rm for} \ \lambda_3 < 0, \end{array} \right. \nonumber \\
	\lambda_2 &>& \left\{ \begin{array}{l l}
		\frac{1}{2} \lambda_5 - 2 \sqrt{\lambda_1 \left( \frac{1}{3} \lambda_3 + \lambda_4 \right)} &
			{\rm for} \ \lambda_5 \geq 0 \ {\rm and} \ \lambda_3 \geq 0, \\
		\omega_+(\zeta) \lambda_5 - 2 \sqrt{\lambda_1 ( \zeta \lambda_3 + \lambda_4)} &
			{\rm for} \ \lambda_5 \geq 0 \ {\rm and} \ \lambda_3 < 0, \\
		\omega_-(\zeta) \lambda_5 - 2 \sqrt{\lambda_1 (\zeta \lambda_3 + \lambda_4)} &
			{\rm for} \ \lambda_5 < 0.
			\end{array} \right.
	\label{eq:bfbcond2}
\end{eqnarray}
The last two conditions for $\lambda_2$ must be satisfied for all values of $\zeta \in \left[ \frac{1}{3}, 1 \right]$.

The bounded-from-below constraints in Eq.~(\ref{eq:bfbcond2}) reduce the maximum accessible ranges of the scalar quartic couplings compared to those obtained from perturbative unitarity constraints in Eqs.~(\ref{eq:l1uni}--\ref{eq:l2l5uni}).  The bounded-from-below constraint on $\lambda_1$ trivially restricts its maximum accessible range to be
\begin{equation}
	\lambda_1 \in \left( 0, \frac{1}{3} \pi \right) \simeq \left( 0, 1.05 \right).
\end{equation}

\begin{figure}
\begin{center}
\resizebox{0.5\textwidth}{!}{\includegraphics{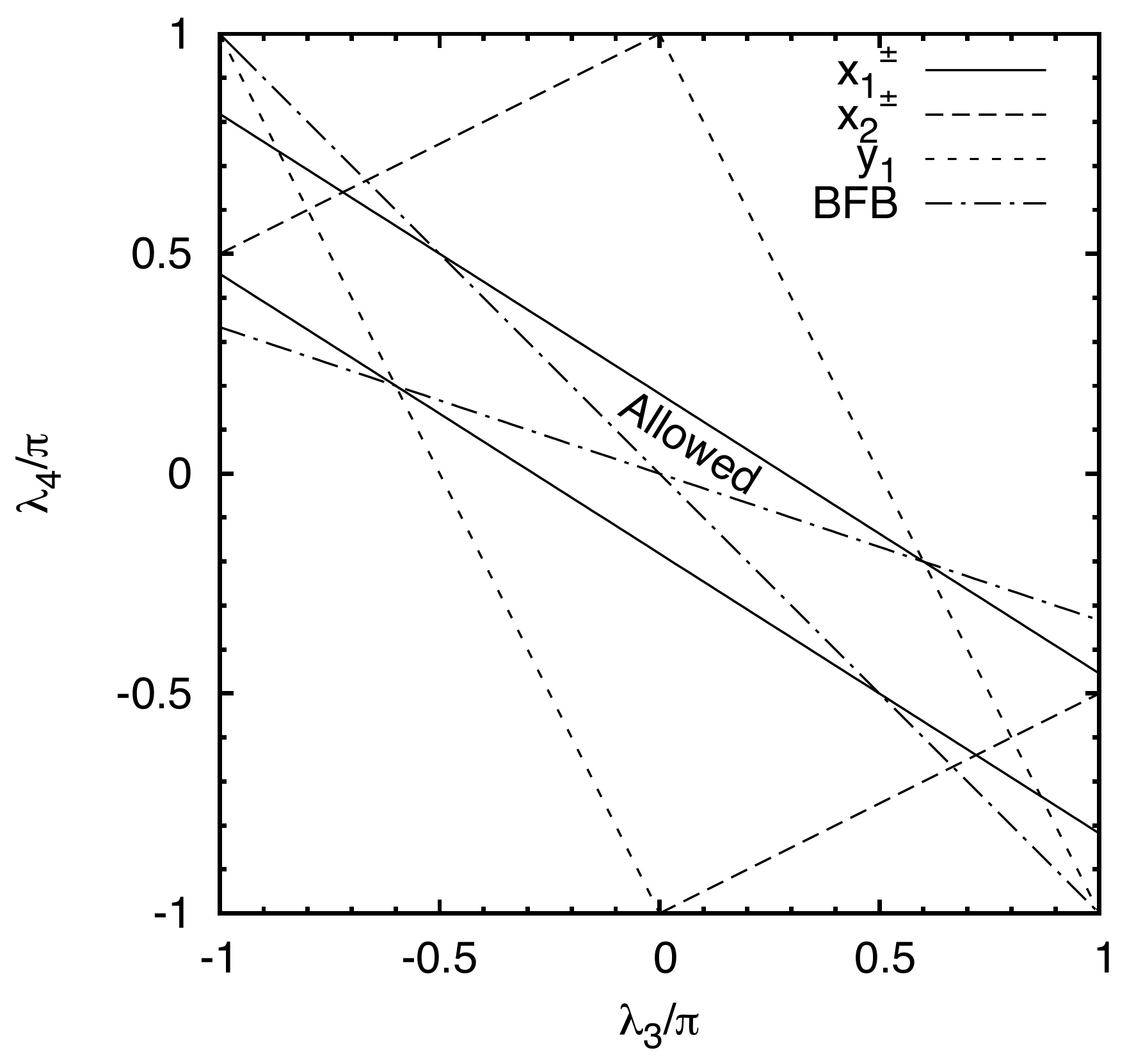}}
\end{center}
\caption{Constraints on the $(\lambda_3,\lambda_4)$ plane from perturbative unitarity, as in Fig.~\ref{fig:unitarity}, together with the bounded-from-below (BFB) constraints $\lambda_4 > - \frac{1}{3}\lambda_3$ and $\lambda_4 > -\lambda_3$.}
\label{fig:bfb}
\end{figure}

The bounded-from-below constraint on $\lambda_4$ restricts the maximum accessible ranges of $\lambda_3$ and $\lambda_4$, as shown in Fig.~\ref{fig:bfb}.  The bounded-from-below constraint excludes the regions below the dot-dashed lines, while the unitarity constraint from $|x_1^{\pm}| < 8 \pi$ restricts $\lambda_3$ and $\lambda_4$ to lie between the two solid lines (we again set $\lambda_2 = 0$ for the least restrictive constraint on $\lambda_3$ and $\lambda_4$ from $x_1^{\pm}$).  The allowed region is a triangle with vertices at $(\lambda_3/\pi,\lambda_4/\pi) = (0,0)$, $\left(-\frac{1}{2}, \frac{1}{2}\right)$, and $\left(\frac{3}{5}, -\frac{1}{5}\right)$.  The unitarity constraint from $y_1$ becomes superfluous; however, the unitarity constraint from $x_2^{\pm}$ can still be important for large enough values of $|\lambda_5|$ (in Fig.~\ref{fig:bfb} we set $\lambda_5 = 0$ for the least restrictive constraint on $\lambda_3$ and $\lambda_4$ from $x_2^{\pm}$).  The maximum accessible ranges of $\lambda_3$ and $\lambda_4$ are therefore reduced compared to those given in Eq.~(\ref{eq:l3l4uni}) to read
\begin{eqnarray}
	\lambda_3 &\in& \left( -\frac{1}{2} \pi, \frac{3}{5} \pi \right) \simeq \left( -1.57, 1.88 \right), \nonumber \\
	\lambda_4 &\in& \left( -\frac{1}{5} \pi, \frac{1}{2} \pi \right) \simeq \left( -0.628, 1.57 \right).
\end{eqnarray}

Finally, the bounded-from-below constraint on $\lambda_2$ restricts the accessible range of $\lambda_2$ as follows.  The least restrictive lower bound on $\lambda_2$ from unitarity is obtained by taking $\lambda_1 = 0$ and $7 \lambda_3 + 11 \lambda_4 = 0$ in $x_1^{\pm}$.  However, when $\lambda_1 = 0$, the bounded-from-below constraint on $\lambda_2$ forces $\lambda_2 > 0$, with the least restrictive constraint obtained for $\lambda_5 = 0$.  The least restrictive lower bound on $\lambda_2$ will occur for nonzero values of $\lambda_1$, $\lambda_3$ and $\lambda_4$ and could be obtained through a numerical scan.  However, because these maximum accessible parameter ranges are primarily useful for setting up numerical scans in the first place, we do not compute a numerical lower bound on $\lambda_2$ here.

%%%%%%%%%%%%%%%%%%%%%%%%%%%%%%%%%%%%%%%%%%%%%%
\subsection{Conditions to avoid alternative minima}
\label{sec:altmin}

We now consider the conditions on the parameters of the scalar potential that are required in order to ensure that the desired electroweak-breaking and custodial SU(2)-preserving minimum is the true global minimum.

In the notation of Eq.~(\ref{eq:bfb}), the full scalar potential can be written as
\begin{eqnarray}
	V &=& \frac{r^2}{(1 + \tan^2 \gamma)} \frac{1}{2} \left[ \mu_2^2 + \mu_3^2 \tan^2 \gamma \right]
	\nonumber \\
	&& + \frac{r^4}{(1+\tan^2 \gamma)^2} 
	\left[ \lambda_1 + (\lambda_2 - \omega \lambda_5) \tan^2 \gamma + 
	(\zeta \lambda_3 + \lambda_4 ) \tan^4 \gamma \right]
	\nonumber \\
	&& + \frac{r^3}{(1 + \tan^2\gamma)^{3/2}}  \tan\gamma \left[- \sigma M_1
	- \rho M_2 \tan^2 \gamma \right],
	\label{eq:Valtmin}
\end{eqnarray}
where $r$, $\tan\gamma$, $\zeta$ and $\omega$ were defined in Eq.~(\ref{eq:bfbdefs}) and we define two new dimensionless field combinations $\sigma$ and $\rho$ according to
\begin{eqnarray}
	\sigma &\equiv& \frac{{\rm Tr}(\Phi^{\dagger} \tau^a \Phi \tau^b) (U X U^{\dagger})_{ab}}
	{{\rm Tr}(\Phi^{\dagger} \Phi) [ {\rm Tr}(X^{\dagger} X)]^{1/2}}, \nonumber \\
	\rho &\equiv& \frac{{\rm Tr}(X^{\dagger} t^a X t^b) (U X U^{\dagger})_{ab}}
	{[{\rm Tr}(X^{\dagger} X)]^{3/2}}.
\end{eqnarray}

To derive the allowed range of $\sigma$, we start by again choosing the SU(2)$_L$ basis so that the field value of $\Phi$ lies entirely in the real neutral component, $\Phi = \frac{v_{\phi}}{\sqrt{2}} \mathbb{1}_{2\times2}$.  We can then apply Eq.~(\ref{eq:phicombo}) to reduce $\sigma$ to the simple form
\begin{equation}
	\sigma = \frac{1}{4} \frac{(2 {\rm Re} \, \chi^0 + \xi^0)}
	{[{\rm Tr}(X^{\dagger} X)]^{1/2}},
\end{equation}
where ${\rm Tr}(X^{\dagger} X)$ is given in Eq.~(\ref{eq:TrXdX}).
Because $\sigma$ is invariant under custodial SU(2), this expression can be rewritten in a very simple form in terms of the custodial SU(2) eigenstates given in Eq.~(\ref{eq:custodialmults}):
\begin{equation}
	\sigma = \frac{\sqrt{3}}{4} \frac{X_1}{\left[ (X_1)^2 + |X_3|^2 + |X_5|^2 \right]^{1/2}}.
\end{equation}
From this form it can be easily seen that $\sigma \in \left[ -\frac{\sqrt{3}}{4}, \frac{\sqrt{3}}{4} \right]$.\footnote{Our desired vacuum, with $\langle \chi^0 \rangle = \langle \xi^0 \rangle = v_{\chi}$, corresponds to $\sigma = \sqrt{3}/4$.  The vacuum with $\sigma = -\sqrt{3}/4$ is also acceptable; it corresponds to negative $v_{\chi}$.}

The field combination $\rho$ is given explicitly by 
\begin{equation}
	\rho = \frac{6}{\left[ {\rm Tr}(X^{\dagger} X) \right]^{3/2}} 
	\left\{ \xi^0 (|\chi^0|^2 - |\chi^{++}|^2)
	+ 2 {\rm Re}[\xi^+ (\chi^0 \chi^{+*} + \chi^+ \chi^{++*})] \right\}.
\end{equation}
To derive the allowed range of $\rho$, we note that $\rho$ can be rewritten as
\begin{equation}
	\rho = \frac{6 \det(X)}{\left[ {\rm Tr}(X^{\dagger}X) \right]^{3/2}}.
\end{equation}
Because $\rho$ is invariant under unitary rotations of $X$, we can choose to work in the Cartesian basis $X \to U X U^{\dagger}$ with $U$ defined in Eq.~(\ref{eq:U}).  In this basis, $X$ is a real $3 \times 3$ matrix with eigenvalues $x$, $y$, and $z$ (two of which may be complex), so that
\begin{equation}
	\rho = \frac{6 xyz}{(|x|^2 + |y|^2 + |z|^2)^{3/2}}.
	\label{eq:rhoxyz}
\end{equation}
This expression is maximized when $x = y = z$, yielding $\rho = 2/\sqrt{3}$ (this result holds regardless of whether the eigenvalues are all real).
Furthermore, taking $X \to -X$ flips the sign of $\rho$.  We thus see that $\rho \in \left[ -\frac{2}{\sqrt{3}}, \frac{2}{\sqrt{3}} \right]$.\footnote{Our desired vacuum, with $\langle \chi^0 \rangle = \langle \xi^0 \rangle = v_{\chi}$, corresponds to $\rho = 2/\sqrt{3}$.  The vacuum with $\rho = -2/\sqrt{3}$ is also acceptable; it corresponds to negative $v_{\chi}$.}

In order to check that the desired electroweak-breaking vacuum is the global minimum of the scalar potential, one minimizes the potential in Eq.~(\ref{eq:Valtmin})---i.e., determines the values of $r$ and $\tan\gamma$ by applying the minimization conditions, and computes the corresponding value of $V$---after setting the values of $\zeta$, $\omega$, $\sigma$ and $\rho$ to correspond to our desired vacuum.  One then repeats the process for every possible combination of $\zeta$, $\omega$, $\sigma$, and $\rho$ that can be obtained by varying the triplet field values in the model.  If the value of $V$ at our desired vacuum is lower than that at any other field configuration, then we are assured that the desired vacuum is the true global minimum of the potential.

\begin{figure}
\resizebox{\textwidth}{!}{
\includegraphics{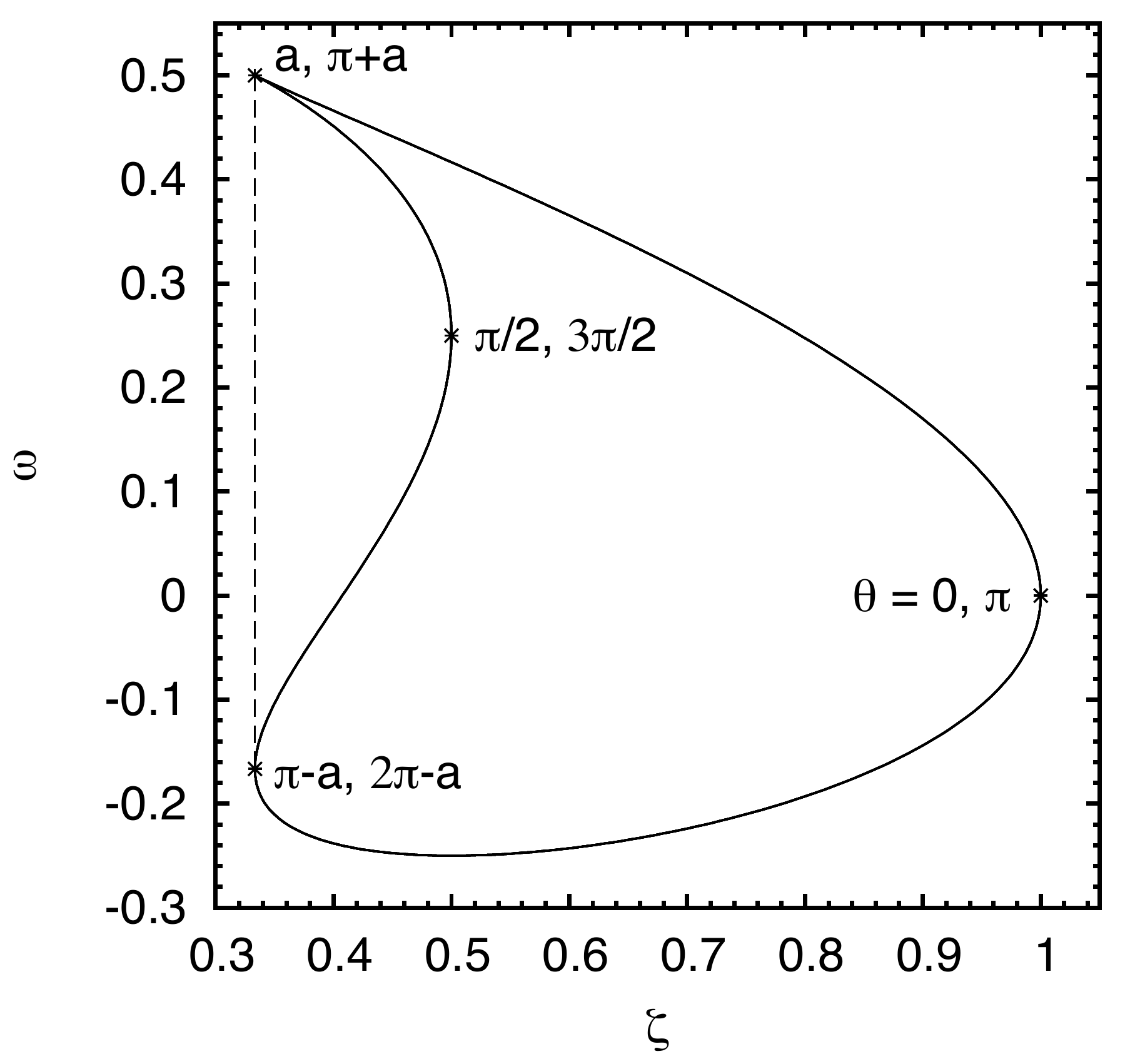}\includegraphics{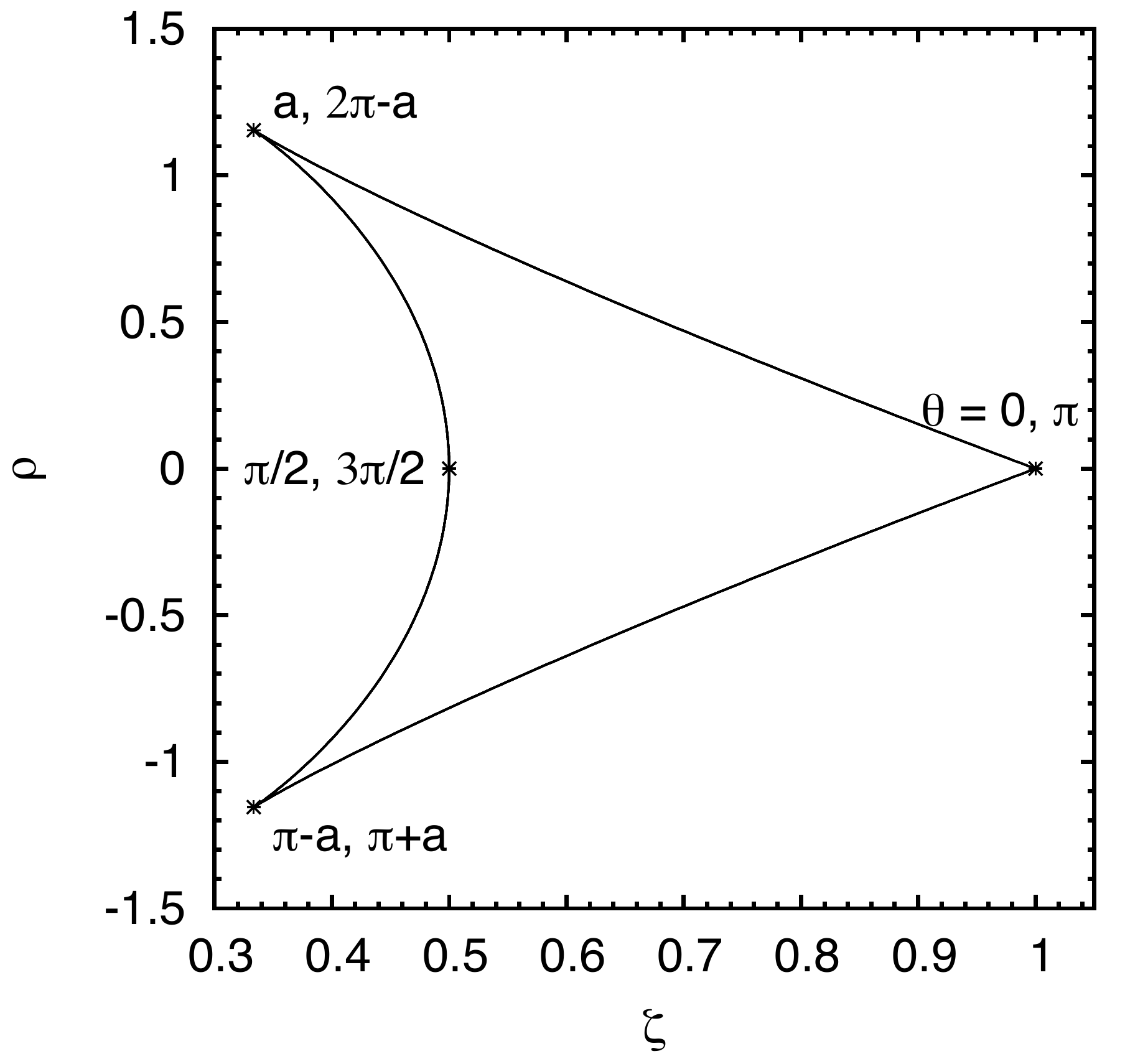}\includegraphics{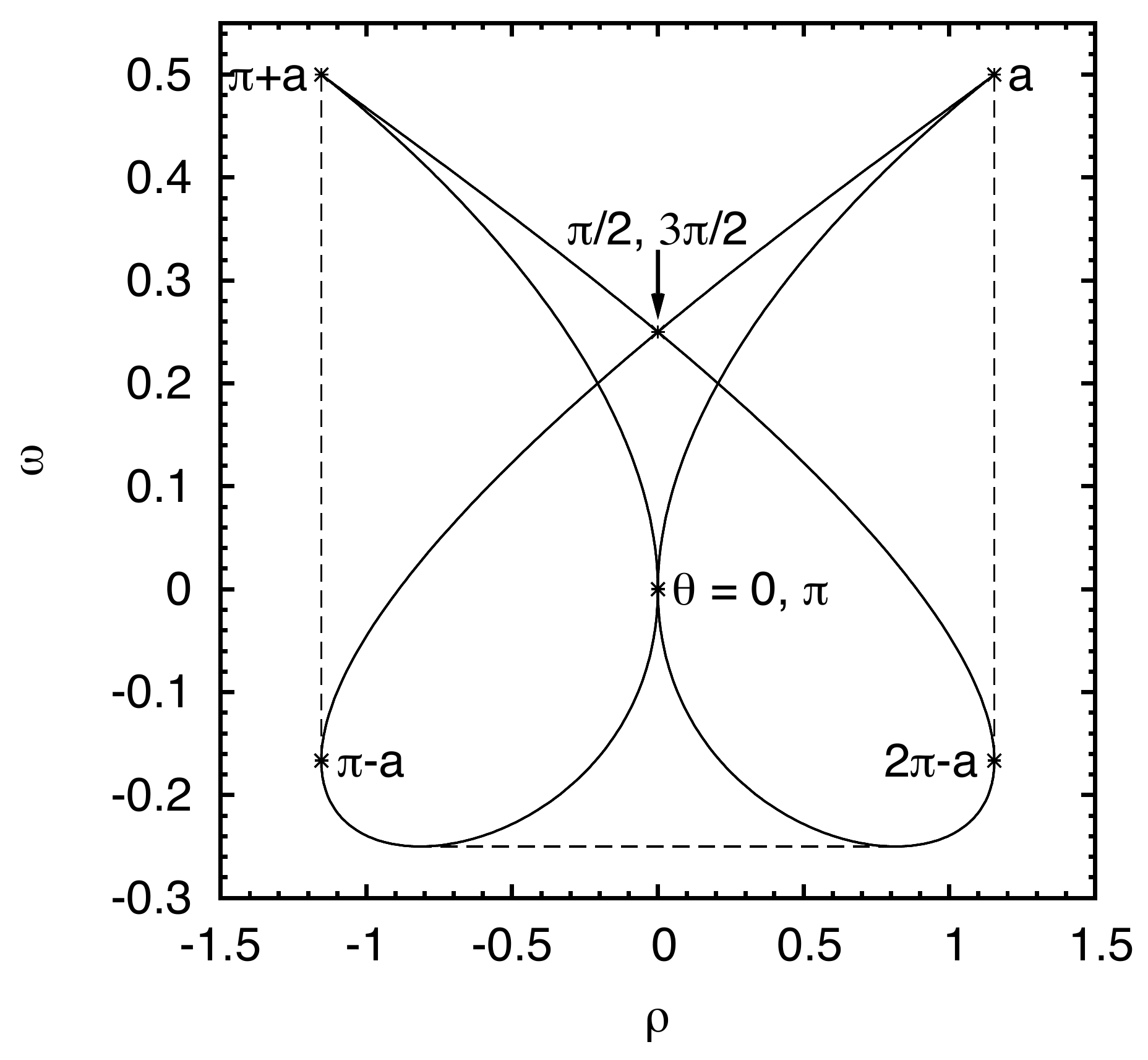}}
\resizebox{\textwidth}{!}{
\includegraphics{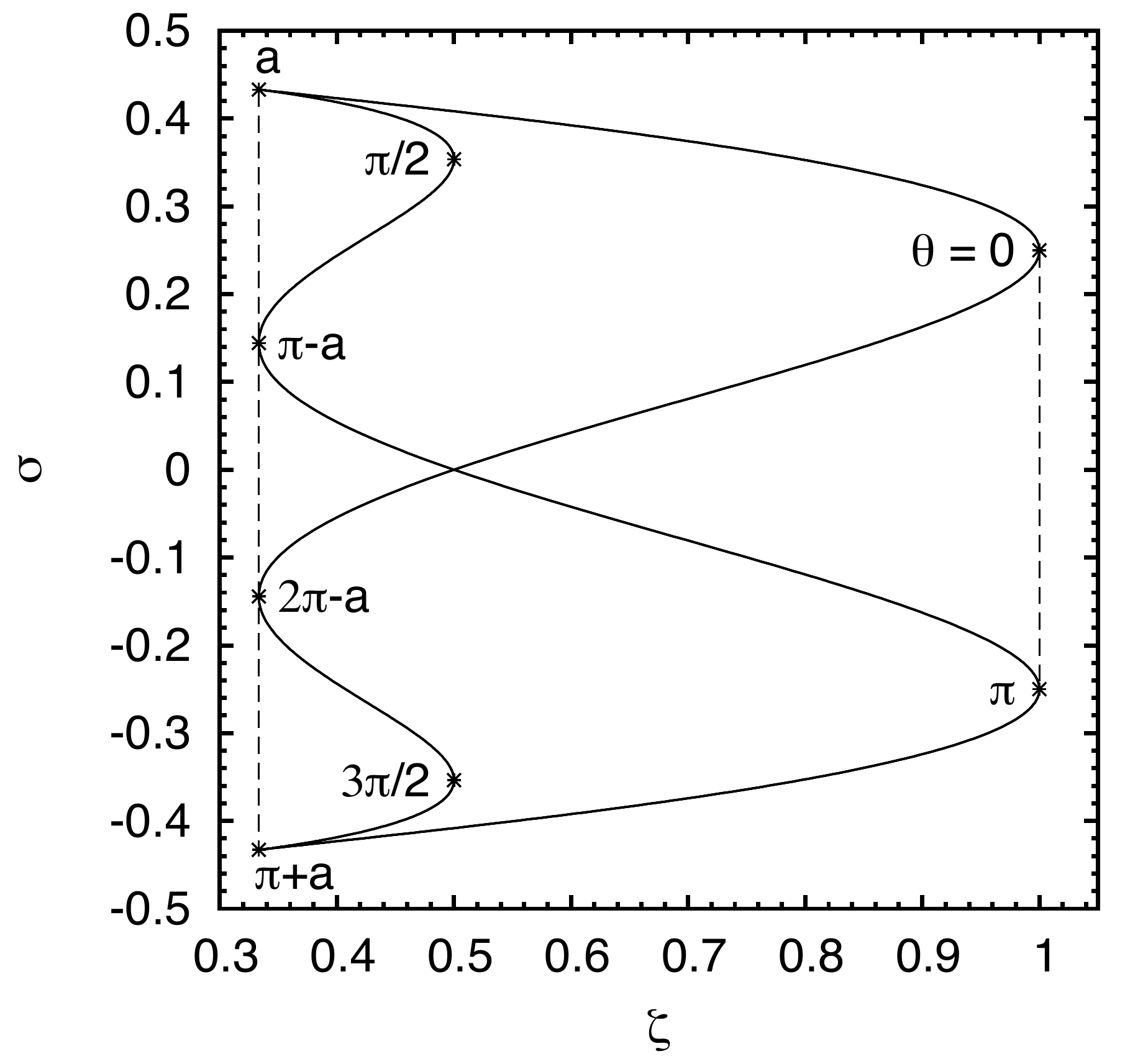}\includegraphics{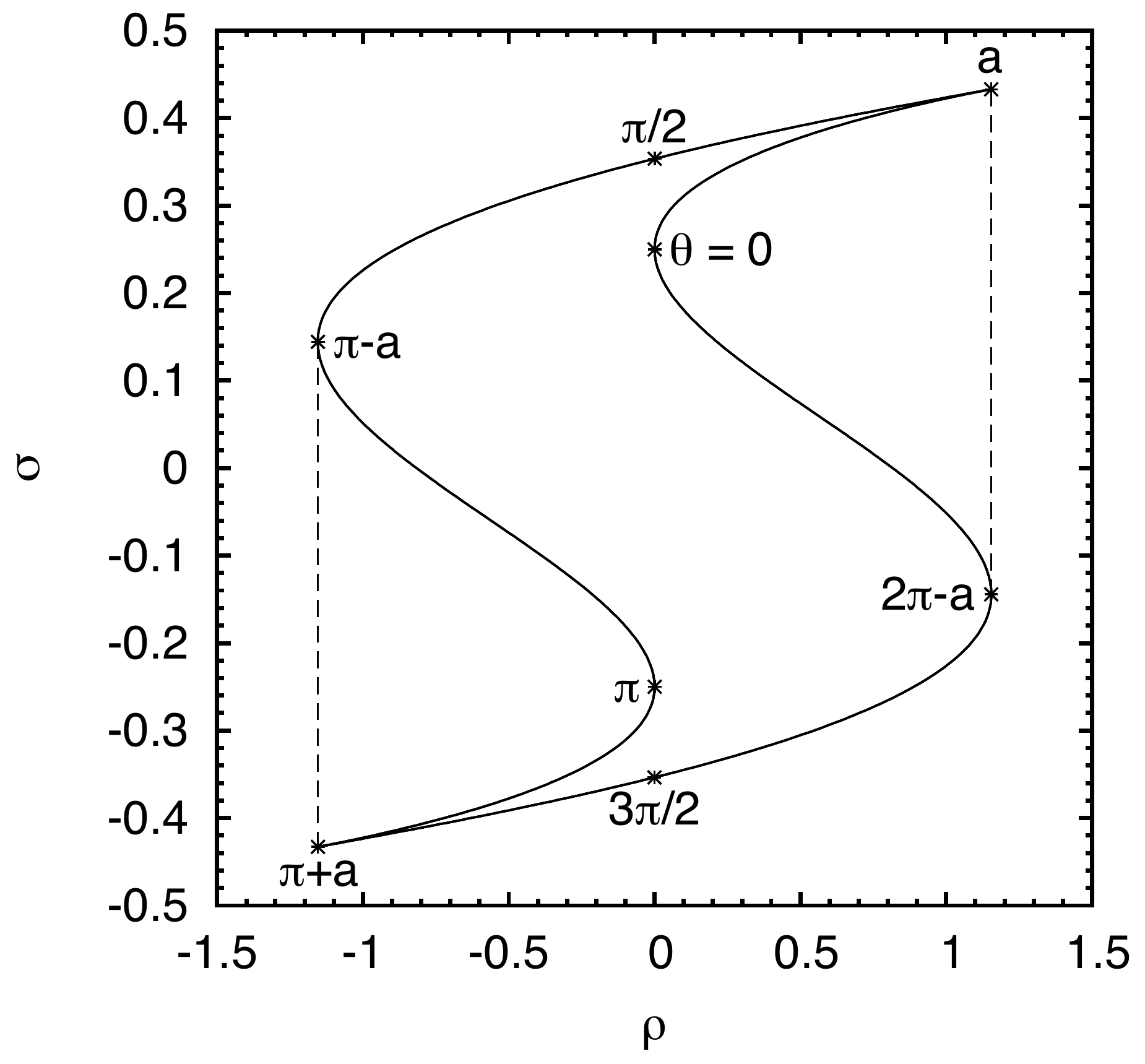}\includegraphics{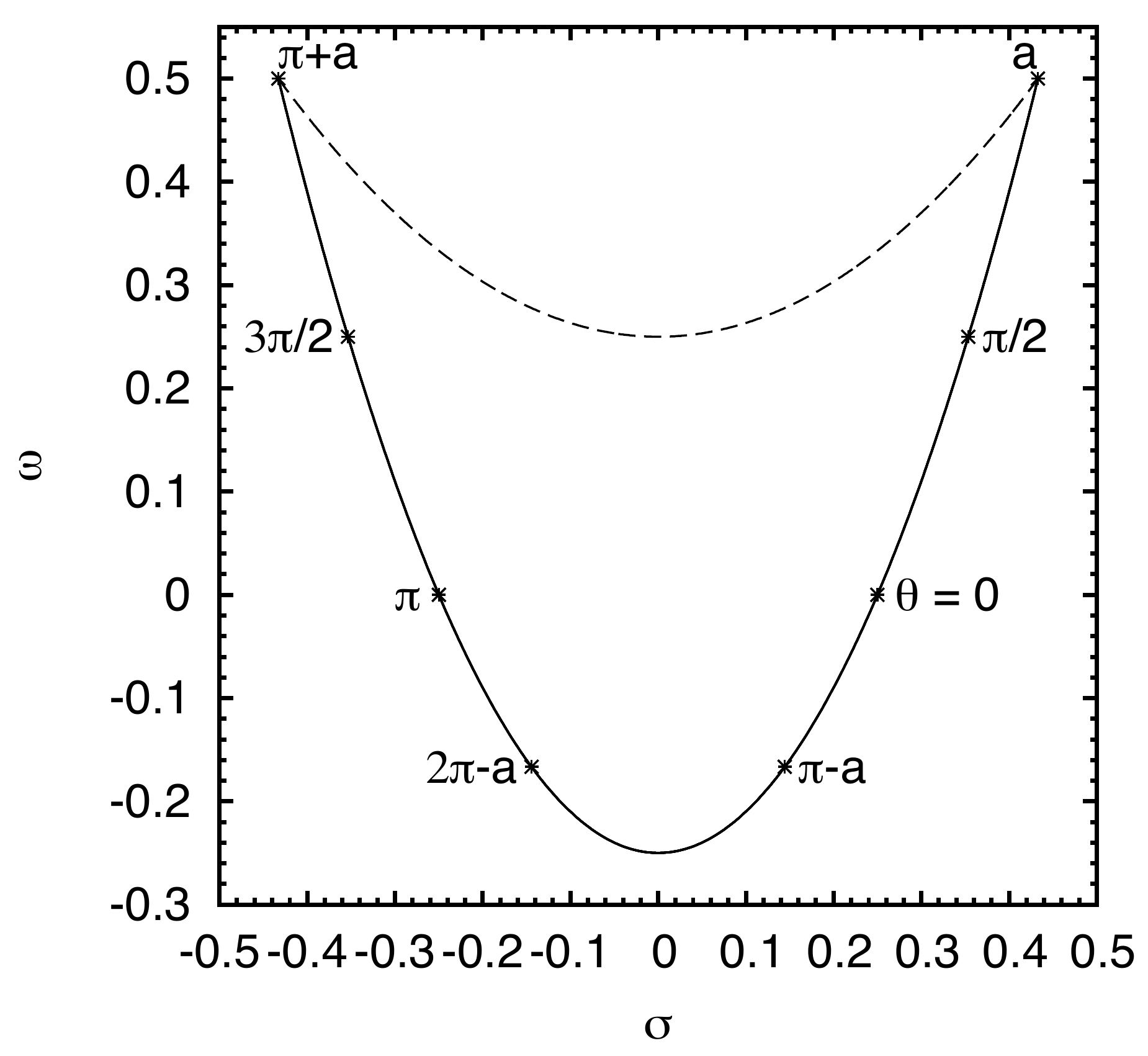}}
\caption{Six projections of the accessible region in the space of $\zeta, \omega, \sigma, \rho$.  The accessible region is that enclosed within the outermost lines (including dashed lines).  The solid line is the path traced out by the parameterization of Eq.~(\ref{eq:thetaparam}), with values of $\theta \in [0, 2\pi)$ as marked.  Our desired vacuum, with $\langle \chi^0 \rangle = \langle \xi^0 \rangle = v_{\chi}$, corresponds to $\theta = a$ for positive $v_{\chi}$ and $\theta = \pi + a$ for negative $v_{\chi}$.}
\label{fig:6views}
\end{figure}

As just described, this procedure involves a scan over an eight-dimensional parameter space of $X$ field values (the parameterization in Eq.~(\ref{eq:Valtmin}) pulls out an overall field normalization into $r$ and $\tan\gamma$).  Such a scan is numerically burdensome and calls for a more analytic approach.  In fact, because $V$ is linear in $\zeta$, $\omega$, $\sigma$ and $\rho$, the scan can be reduced to the three-dimensional parameter space that constitutes the surface of the four-dimensional volume in $(\zeta, \omega, \sigma, \rho)$ that is populated by the model.  Projections of this volume onto the six planes comprising pairs of parameters are shown in Fig.~\ref{fig:6views}.

The parameter scan is further reduced to an easy-to-handle one-dimensional space by the geometrical observation that, for any orientation of the four-dimensional volume populated by the model (achieved in the scalar potential by choosing various values and signs for $\lambda_3$, $\lambda_5$, $M_1$, and $M_2$), the ``lowest point'' always lies upon the trajectory traced out by the following simple parameterization:
\begin{equation}
	{\rm Re} \, \chi^0 = \frac{1}{\sqrt{2}} \sin \theta, 
	\qquad \xi^0 = \cos \theta,
\end{equation}
and all other triplet field values equal to zero.  The projections of this curve onto the six parameter planes are shown as solid lines in Fig.~\ref{fig:6views}, with reference values of $\theta$ marked.  

\begin{figure}
\begin{center}
\resizebox{0.5\textwidth}{!}{\includegraphics{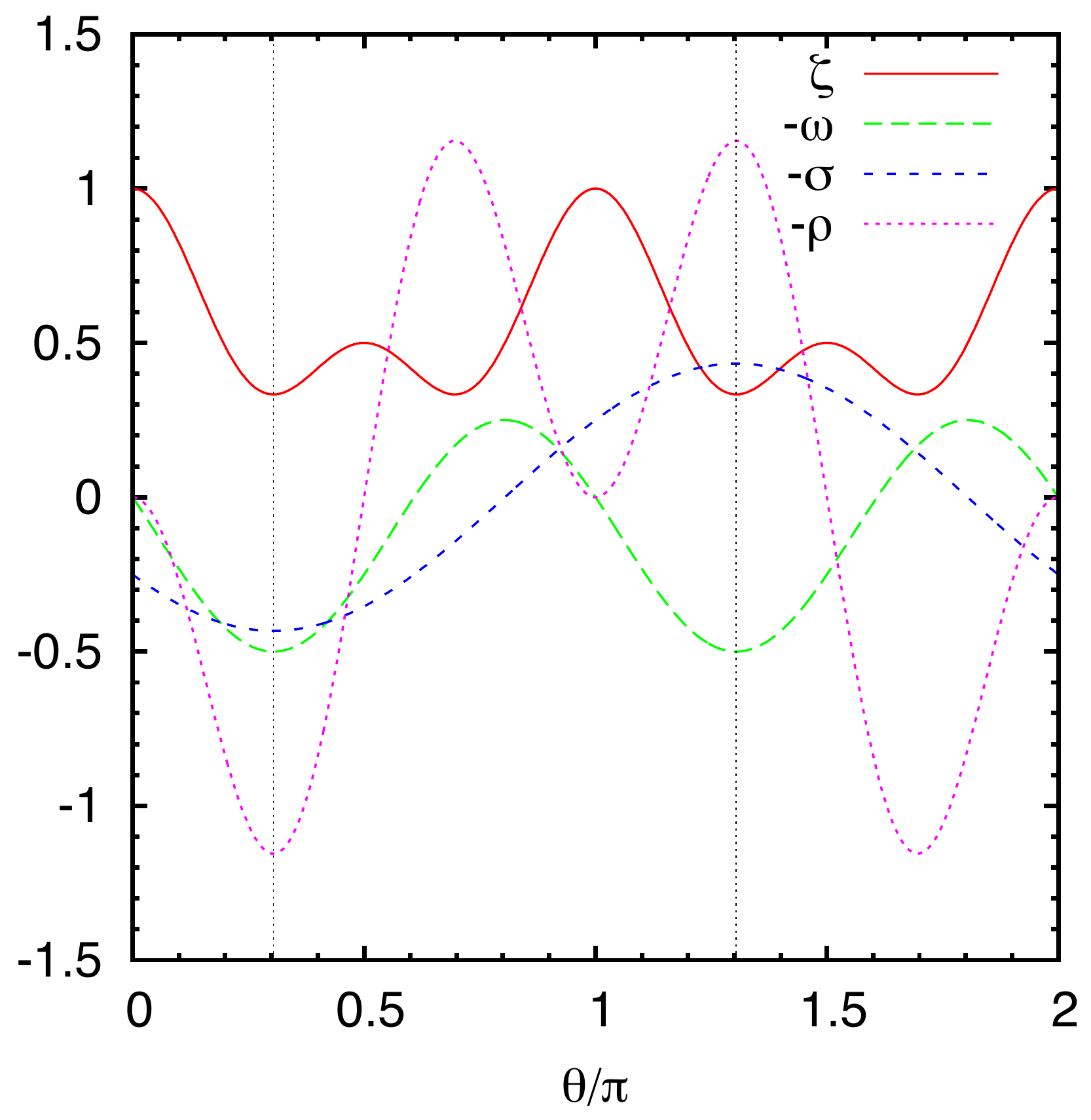}}
\end{center}
\caption{The values of $\zeta$, $-\omega$, $-\sigma$, and $-\rho$ as a function of $\theta$, in the parameterization of Eq.~(\ref{eq:thetaparam}).  The vertical dotted lines correspond to the desired minima at $\theta = a$ and $\theta = \pi + a$.}
\label{fig:thetafunc}
\end{figure}

In this parameterization, we have
\begin{eqnarray}
	\zeta &=& \frac{1}{2} \sin^4 \theta + \cos^4 \theta, \nonumber \\
	\omega &=& \frac{1}{4} \sin^2 \theta + \frac{1}{\sqrt{2}} \sin \theta \cos \theta, \nonumber \\
	\sigma &=& \frac{1}{2\sqrt{2}} \sin \theta + \frac{1}{4} \cos \theta, \nonumber \\
	\rho &=& 3 \sin^2 \theta \cos \theta.
	\label{eq:thetaparam}
\end{eqnarray}
These functions are plotted in Fig.~\ref{fig:thetafunc}.  Our desired electroweak-breaking and custodial SU(2)-preserving vacuum corresponds to $\theta = a \equiv \cos^{-1} (\frac{1}{\sqrt{3}})$.  The vacuum $\theta = \pi + a$ is also acceptable; it corresponds to negative $v_{\chi}$.  Other values of $\theta$ correspond to vacua that spontaneously break custodial SU(2).

When $\lambda_3$, $\lambda_5$, $M_1$, and $M_2$ are all positive, the desired vacuum at $\theta = a$ is always the true global minimum of the potential.\footnote{One must of course check that this minimum is deeper than the electroweak-preserving vacuum at $V=0$.}  Simultaneously flipping the signs of $M_1$, $M_2$, and $X$ leaves the scalar potential invariant, so the alternative acceptable vacuum at $\theta = \pi + a$ is always the true global minimum when $\lambda_3$ and $\lambda_5$ are positive and $M_1$ and $M_2$ are negative.  For all other sign combinations of these four parameters, the depth of alternative minima must be checked numerically as described above by scanning over $\theta$ and minimizing $V$ at each point.

%%%%%%%%%%%%%%%%%%%%%%%%%%%%%%%%%%%%%%%%%%%%%%
\section{The decoupling limit}
\label{sec:decoupling}

\subsection{Decoupling behavior of masses and couplings}

After fixing $\mu_2^2$ using the $W$ boson mass constraint, the scalar potential of the GM model contains three dimensionful parameters: $\mu_3^2$, $M_1$, and $M_2$.  Decoupling occurs when appropriate combinations of these parameters are taken large compared to the weak scale $v$.  In fact, we find that decoupling is controlled primarily by $\mu_3^2$, and the maximum allowed values of $M_1$ and $M_2$ scale with this parameter.
The upper bound on $|M_1|$ for large $\mu_3^2 \gg \lambda_i v^2$ can be derived straightforwardly as a consequence of Eq.~(\ref{eq:lambda1}) for $\lambda_1$ in terms of $m_h$ and the unitarity bound on $\lambda_1$ given in Eq.~(\ref{eq:l1uni}).  In the limit $\mu_3^2 \gg \lambda_i v^2$, we find that $M_1$ can scale at most linearly with $\sqrt{\mu_3^2}$ and that its value is constrained by $|M_1|/\sqrt{\mu_3^2} \lesssim 3.3$.
The upper bound on $|M_2|$ is less easily derived, but comes from the requirement that there be a sensible minimum of the potential with $8 v_{\chi}^2 < v^2$.  Numerically we find again that $M_2$ can scale at most linearly with $\sqrt{\mu_3^2}$ and that its value is constrained by $|M_2|/\sqrt{\mu_3^2} \lesssim 1.2$.

We will consider the behavior of the scalar mass spectrum, the vevs, the custodial-singlet mixing angle, and the couplings of the light Higgs to SM particles in the approach to decoupling. We derive explicit expansions for each of these observables in the decoupling limit, keeping terms up to next-to-leading order in inverse powers of $\mu_3^2$.  In the expansions we treat $M_1$ and $M_2$ as being of order $\sqrt{\mu_3^2}$ or smaller.  As we will show, all of the low-energy observables reduce to their appropriate SM limits as $\mu_3\rightarrow \infty$.

We also make a numerical comparison between our expansion formulas and the exact expressions for each observable.  To illustrate the approach to the decoupling limit, we consider two explicit parameter scenarios.  In the first scenario (case A) we let $\mu_3 \equiv \sqrt{\mu_3^2}$ become large while holding $M_1$ and $M_2$ constant.  In the second scenario (case B) we let $\mu_3$ become large while scaling $M_1$ and $M_2$ proportionally to $\mu_3$. The specific parameter choices in each case are given in Table~\ref{tab:cases}.  These parameter choices satisfy all of the theoretical constraints described in Sec.~\ref{sec:theoryconstraints}.  

\begin{table}
\begin{center}
\begin{tabular}{c c ccccc cc}
\hline\hline 
 Case & $\mu_3 \equiv \sqrt{|\mu_3^2|}$ & $\lambda_1$ & $\lambda_2$ & $\lambda_3$ & $\lambda_4$ & $\lambda_5$& $M_1$ &  $M_2$  \\
\hline  
 A & 300--1000~GeV & derived & 0.1 & 0.1 & 0.1 & 0.1 &100~GeV & 100~GeV \\
 B & 300--1000~GeV & derived & 0.1 & 0.1 & 0.1 & 0.1 & $\mu_3/3$ & $\mu_3/3$ \\
\hline\hline
\end{tabular}
\end{center}
\caption{Values of coupling parameters for the two decoupling scenarios considered.  We set $m_h = 125$~GeV and use this to fix $\lambda_1$ in terms of the other parameters. $\mu_2^2$ is eliminated in terms of the known SM Higgs vev $v$.}
\label{tab:cases}
\end{table}

We do not consider cases in which only one of the $M_i$ parameters scales with $\mu_3$ because the overall decoupling behavior is much more strongly influenced by $M_1$ than by $M_2$. In the expansion formulas that we derive below, $M_2$ always appears multiplied by $M_1$ in terms that are suppressed by larger powers of $\mu_3$. As a result, in the case that $M_1\propto \mu_3$ while $M_2$ is constant, the overall decoupling behavior will be very similar to that in case B. Similarly, in the case that $M_2\propto \mu_3$ while $M_1$ is constant the decoupling behavior would resemble that of case A.

The overall power law dependence of each observable on $\mu_3$ in the decoupling limit is tabulated in Table~\ref{tab:powerlaw}.  In general the convergence to the SM is more rapid in case A, where the observables approach the decoupling limit at rates proportional to $\mu_3^{-2}$ or $\mu_3^{-4}$. In comparison, in case B the decoupling rates are proportional to $\mu_3^{-1}$ or $\mu_3^{-2}$.

%%%%% TABLE 3 %%%%%%%%
\begin{table}
\begin{center}
\begin{tabular}{ccc}
\hline\hline 
Quantity & Case A &  Case B  \\
\hline  
$\frac{m_{H,3,5}}{\mu_3}-1$  & $\mu_3^{-2}$ & $\mu_3^{-2}$  \\
$v_\chi$ & $\mu_3^{-2}$ & $\mu_3^{-1}$\\
$\sin\alpha$ & $\mu_3^{-2}$ & $\mu_3^{-1}$ \\
$\kappa_V - 1$ & $\mu_3^{-4}$ & $\mu_3^{-2}$\\
$\kappa_f - 1$ & $\mu_3^{-4}$ & $\mu_3^{-2}$ \\
$g_{hhVV}/g_{hhVV}^{\rm SM} - 1$ & $\mu_3^{-4}$ & $\mu_3^{-2}$ \\
$g_{hhh}/g_{hhh}^{\rm SM} - 1$ & $\mu_3^{-4}$ & $\mu_3^{-2}$ \\
$\Delta \kappa_{\gamma}$ & $\mu_3^{-2}$ & $\mu_3^{-2}$\\
$\Delta \kappa_{Z\gamma}$ & $\mu_3^{-2}$ & $\mu_3^{-2}$\\
\hline\hline
\end{tabular}
\end{center}
\caption{The power law behavior of the heavy scalar masses, triplet vev, custodial singlet mixing angle, and light Higgs couplings for parameter cases A and B.  See text for definitions.}
\label{tab:powerlaw}
\end{table}
%%%%%%%%%%%%%%%%%%

As a first step, it is relevant to examine the expansion formula for $\lambda_1$ near the decoupling limit,
\begin{equation}
	\lambda_1 \simeq \frac{1}{8} \left[ \frac{m_h^2}{v^2}  
	+  \frac{3}{4} \frac{M_1^2}{\mu_3^2} 
	\left( 1 - 3(2\lambda_2 - \lambda_5) \frac{v^2}{\mu_3^2} 
	+ \frac{3 M_1 M_2 v^2}{\mu_3^4}   + \frac{5 m_h^2}{3 \mu_3^2} \right) \right]. 
	\label{L1deceqnALT}
\end{equation}
The first term of this formula coincides with the value of the SM quartic coupling, $\lambda_1 = m_h^2 / 8 v^2$; $\lambda_1$ approaches this value in the $\mu_3 \to \infty$ limit in case A.  In case B, however, the $\mu_3 \to \infty$ limit of $\lambda_1$ is $(m_h^2/8 v^2 + 3 M_1^2/32 \mu_3^2)$.  This expression reminds us that $M_1$ can scale at most linearly with $\mu_3$ if $\lambda_1$ is to remain consistent with the constraint from perturbative unitarity.  We also note that $\lambda_1$ does not correspond directly to the SM Higgs quartic coupling; we will compute the triple-Higgs coupling $g_{hhh}$ below and show that it exhibits decoupling even in case B.

\begin{figure}
\begin{center}
\resizebox{0.49\textwidth}{!}{\includegraphics{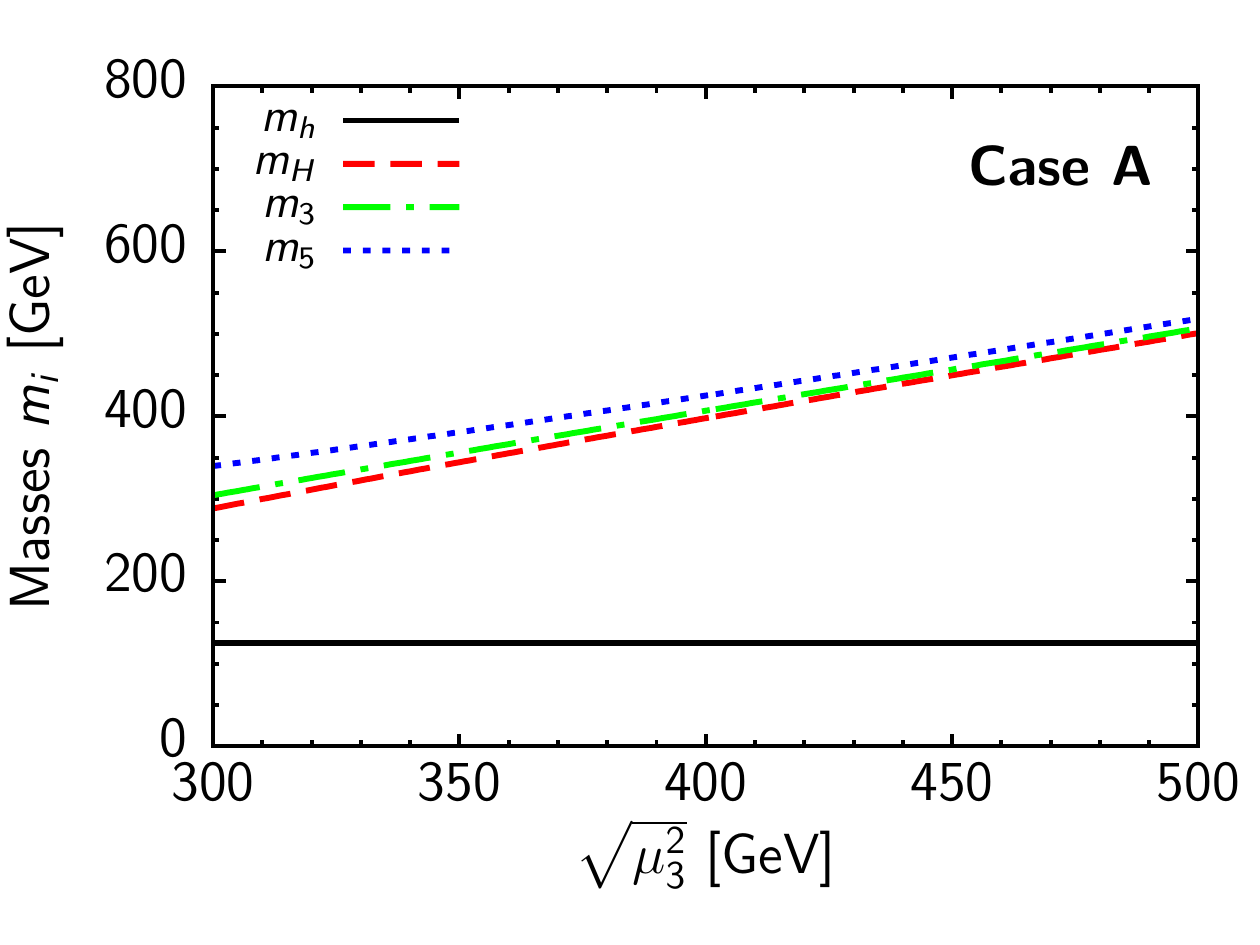}}
\resizebox{0.49\textwidth}{!}{\includegraphics{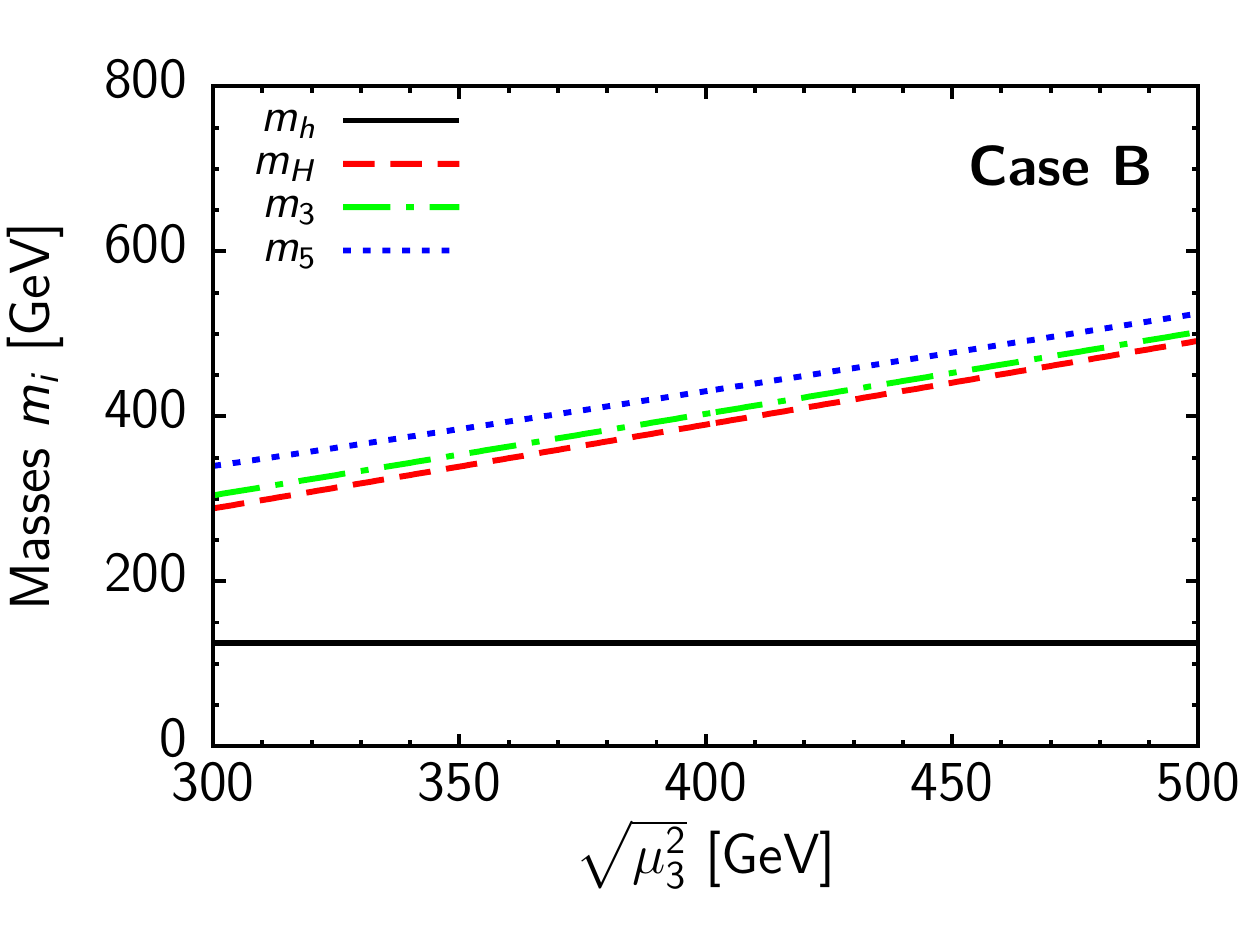}}
\resizebox{0.49\textwidth}{!}{\includegraphics{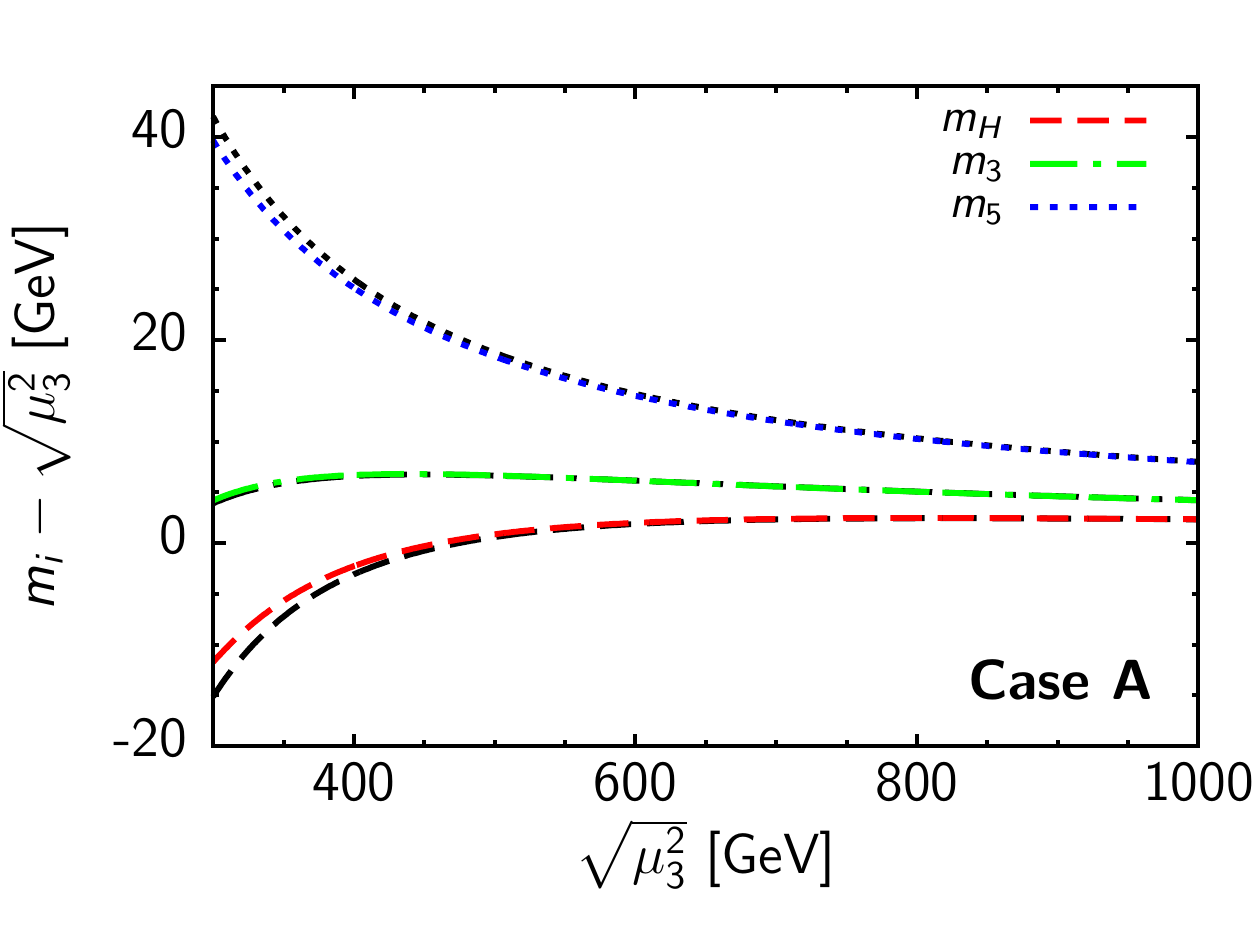}}
\resizebox{0.49\textwidth}{!}{\includegraphics{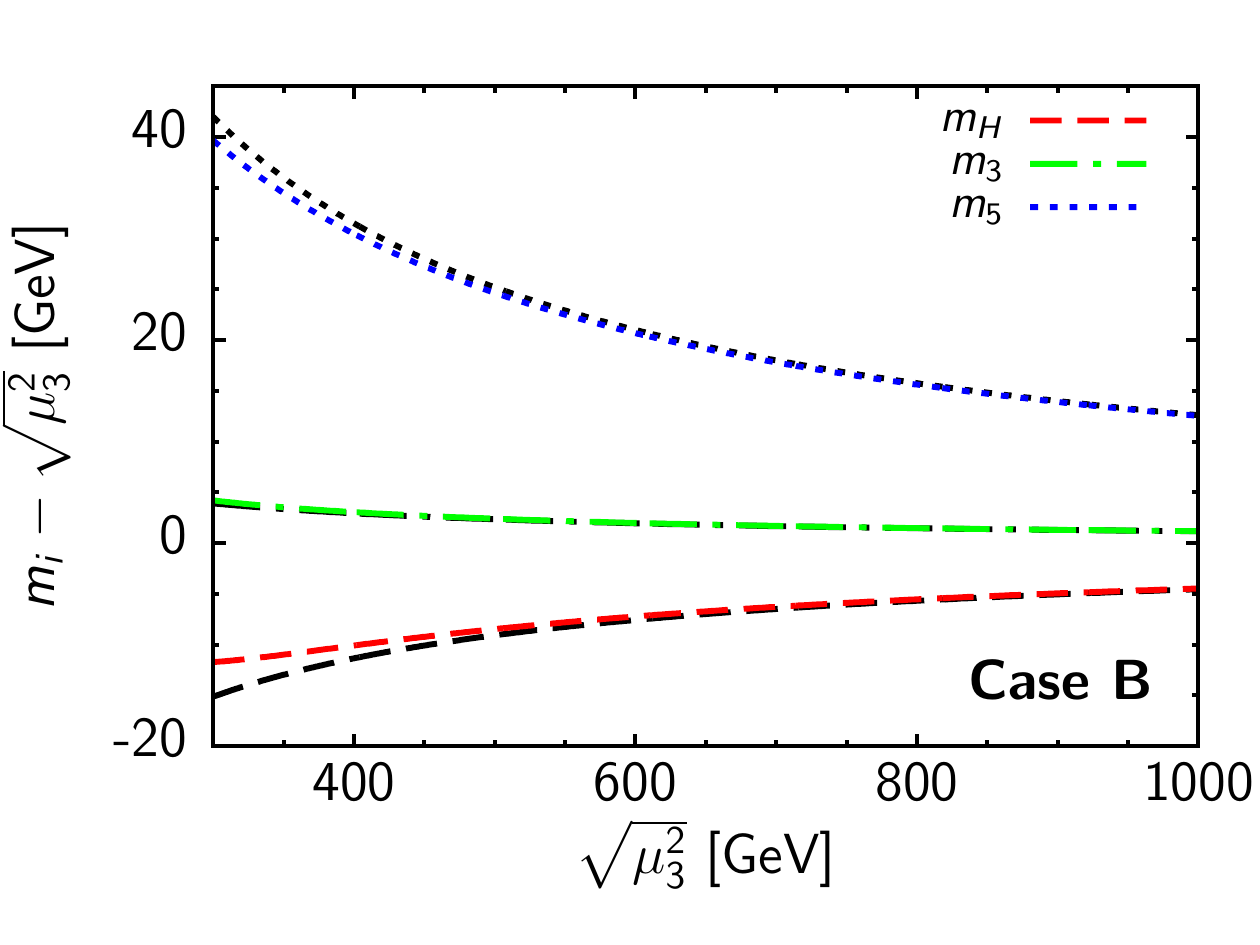}}
\caption{Top: The mass spectrum of the model as a function of $\sqrt{\mu_3^2}$ for cases A (left) and B (right). Bottom: The mass splittings $m_i - \sqrt{\mu_3^2}$ for the heavy scalars as a function of $\mu_3$ for cases A (left) and B (right).  In the bottom plots the colored (light) curves show the exact tree-level masses while the black curves are the associated expansion formulas from Eq.~(\ref{GM:mH}). For $m_3$ and $m_5$, the expansion formula curves are almost identical to the exact curves.}
\label{fig:masses}
\end{center}
\end{figure}

In the decoupling limit, the masses of the heavy scalars are given by the following expansion formulas,\footnote{Note that these are consistent with the mass spectrum in the limit that $M_1=M_2=0$. If $M_1 = M_2 = 0$ and $\mu_3^2 + (2 \lambda_2 - \lambda_5)v^2 > 0$, the scalar potential possesses an \emph{unbroken} $Z_2$ symmetry under which the triplet scalars are odd.  In this case $v_{\chi} = 0$ (so that $s_H = 0$), $H_1^0$ and $H_1^{0\prime}$ do not mix, and the lightest triplet state is stable.  The triplet masses are given by
\begin{eqnarray}
	m_{H_1^{0\prime}}^2 &=& \mu_3^2 + \left( 2 \lambda_2 - \lambda_5 \right) v^2, \nonumber \\
	m_3^2 &=& \mu_3^2 + \left( 2 \lambda_2 - \frac{\lambda_5}{2} \right) v^2, \nonumber \\
	m_5^2 &=& \mu_3^2 + \left( 2 \lambda_2 + \frac{\lambda_5}{2} \right) v^2,
\end{eqnarray}
while the mass of the physical scalar from the doublet is $m_{H_1^0}^2 = 8 \lambda_1 v^2$.  The triplets affect the couplings of the SM-like Higgs $H_1^0$ only through their loop contributions (e.g., in $H_1^0 \to \gamma\gamma$, $Z\gamma$); their loop effects decouple as $\mu_3^2$ is taken large.  This case is analogous to the Inert Doublet Model~\cite{IDM}.  We will not consider it further in this paper.}
\begin{eqnarray}
	m_H &\simeq&  \mu_3 \left[  1 + \left( 2 \lambda_2 - \lambda_5 \right) \frac{v^2}{2\mu_3^2}  
	+ \frac {3 M_1 (M_1 - 4 M_2 ) v^2}{8 \mu_3^4}     \right], \nonumber \\
	m_3 &\simeq&  \mu_3  \left[  1 
	+ \left( 2 \lambda_2 - \frac{\lambda_5}{2} \right) \frac{v^2}{2\mu_3^2}  
	+ \frac{M_1(M_1-3 M_2) v^2}{4 \mu_3^4}  \right], \nonumber \\
	m_5 &\simeq&  \mu_3  \left[ 1 
	+ \left( 2 \lambda_2 + \frac{\lambda_5}{2} \right) \frac{v^2}{2\mu_3^2}  
	+ \frac{3 M_1 M_2 v^2}{4 \mu_3^4}  \right]. 
	\label{GM:mH}
\end{eqnarray}
The fractional difference between each scalar mass and $\mu_3$ scales with $\mu_3^{-2}$ in both case~A and case~B. The behavior of the scalar masses and the difference between the masses and $\mu_3$ are illustrated as functions of $\mu_3$ in Fig.~\ref{fig:masses}.\footnote{In our numerical calculations we use $v = 246$~GeV, $m_t = 172$~GeV, $M_W = 80.399$~GeV, $M_Z = 91.1876$~GeV, and $c_W = M_W/M_Z$.}  In each case we show the exact tree-level mass values; in the lower panels of Fig.~\ref{fig:masses} we also show the expansion formulas of Eq.~(\ref{GM:mH}) in black.   As expected from Eq.~(\ref{GM:mH}), the overall decoupling behavior is similar in cases A and B; the mass splittings are larger in case B due to the numerical size of the term involving $M_1$ and $M_2$.

\begin{figure}
\begin{center}
\resizebox{0.49\textwidth}{!}{\includegraphics{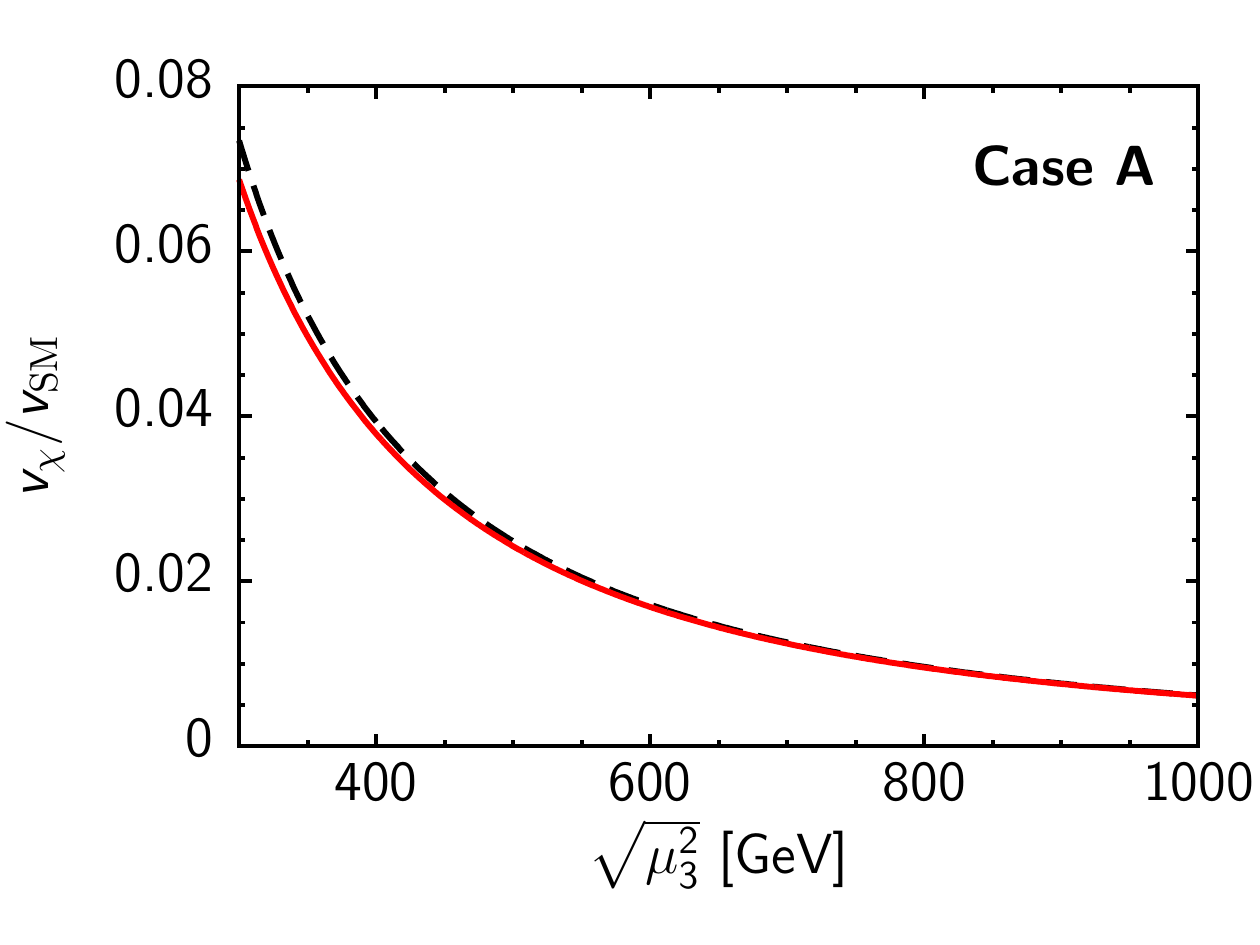}}
\resizebox{0.49\textwidth}{!}{\includegraphics{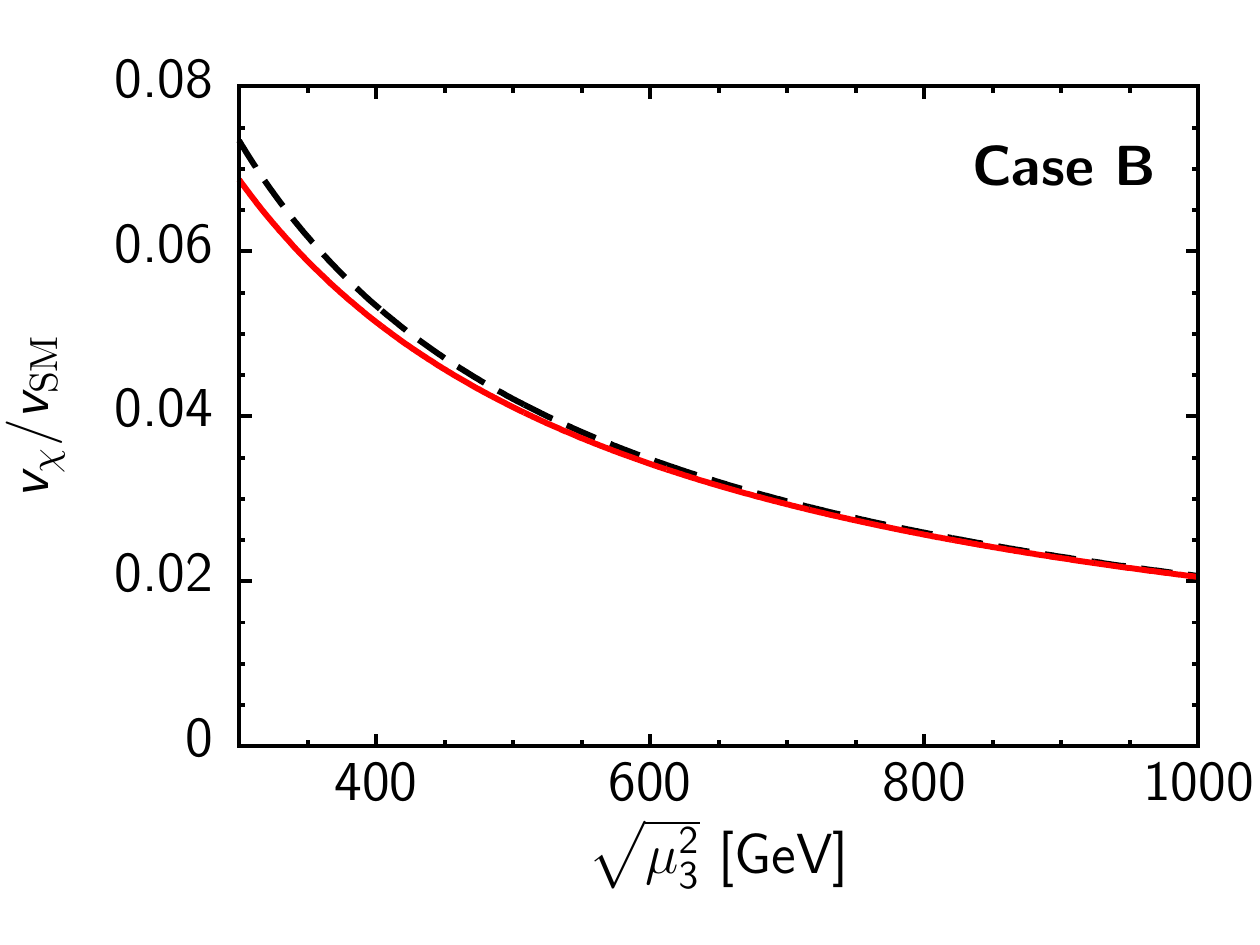}}
\resizebox{0.49\textwidth}{!}{\includegraphics{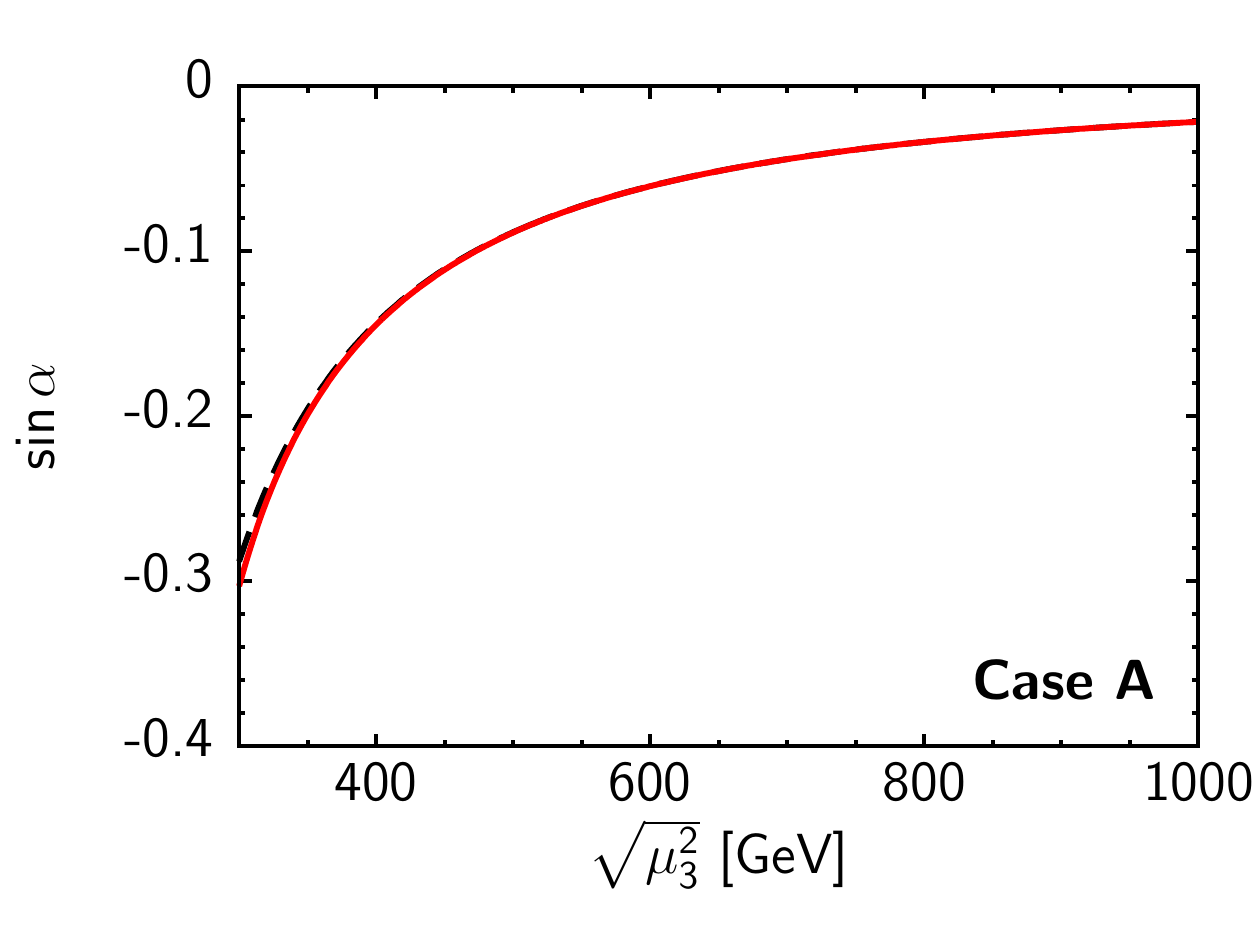}}
\resizebox{0.49\textwidth}{!}{\includegraphics{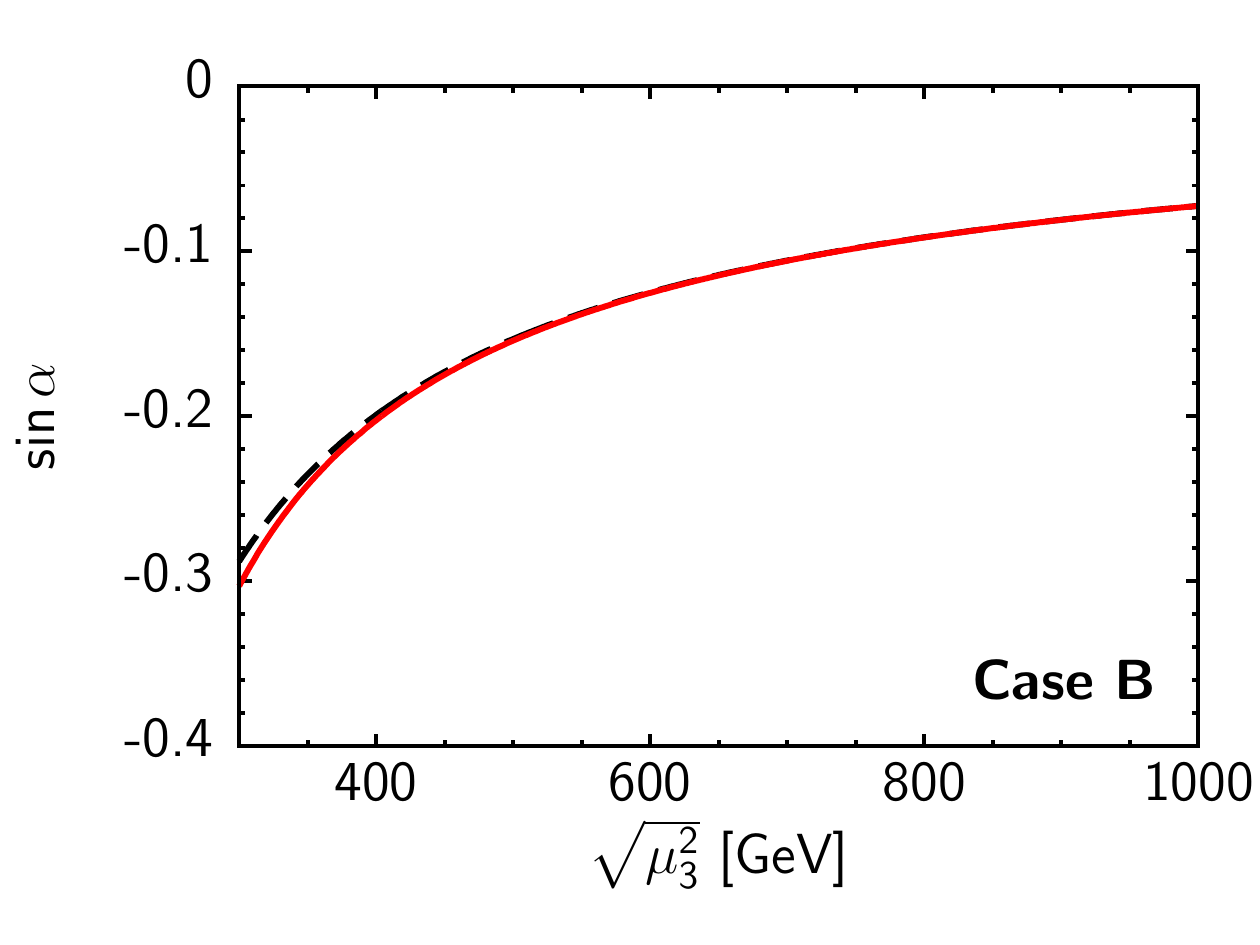}}
\caption{Top: The dependence of the triplet vev $v_\chi$ (shown normalized to $v = 246$~GeV) as a function of $\mu_3$ in cases A (left) and B (right).  Bottom: The mixing angle $\sin \alpha$ that controls the composition of the light Higgs boson $h = \phi^{0,r} \cos\alpha - H_1^{0\prime} \sin\alpha$, shown as a function of $\mu_3$ in cases A (left) and B (right).  In all plots the solid red (light) line is the exact curve, while the dashed black (dark) line is the corresponding expansion formula from Eqs.~(\ref{GM:vx}) and (\ref{GM:sin}).}
\label{fig:vevs-sin}
\end{center}
\end{figure}

Expansion formulas for the decoupling behavior of the vevs $v_\chi$ and $v_\phi$ (related by $v_{\phi}^2 + 8 v_{\chi}^2 = v^2$) are given by,
\begin{eqnarray}
	v_\chi &\simeq& \frac{M_1 v^2}{4 \mu_3^2} 
	\left[ 1 - (2 \lambda_2 - \lambda_5) \frac{v^2}{\mu_3^2} 
	+ \frac{ M_1(3 M_2-M_1) v^2}{2 \mu_3^4}   \right], \nonumber \\
	v_\phi &\simeq& v \left( 1- \frac{M_1^2 v^2}{4 \mu_3^4} \right).
	\label{GM:vx}
\end{eqnarray}
The doublet vev $v_\phi$ approaches the SM value of $v$ in the decoupling limit, as one would expect.  Likewise, the triplet vev $v_\chi$ goes to zero with its value falling like $\mu_3^{-2}$ ($\mu_3^{-1}$) in case A (case B). The decoupling behavior of $v_\chi$ is plotted for cases A and B in the top panels of Fig.~(\ref{fig:vevs-sin}).

The decoupling behavior of the mixing angle $\alpha$ is given by the expansion formula
\begin{equation}
	\sin\alpha \simeq  - \frac{\sqrt{3} M_1 v}{2 \mu_3^2} 
	\left[ 1 - 2 (2 \lambda_2 - \lambda_5) \frac{v^2}{\mu_3^2} 
	+ \frac{m_h^2}{\mu_3^2} + \frac{M_1 (24 M_2 - 5M_1) v^2}{8 \mu_3^4}  \right]. 
	\label{GM:sin}
\end{equation}
We can see that $\sin\alpha$ approaches zero as $\mu_3\rightarrow\infty$. This is to be expected, as $\sin\alpha$ controls the amount of triplet in the mass eigenstate $h$, and $\sin\alpha=0$ corresponds to a SM-like Higgs boson $h$ composed entirely of the SU(2)$_L$ doublet [recall Eq.~(\ref{mh-mH})]. The rate of the decoupling is proportional to $\mu_3^{-2}$ in case A and $\mu_3^{-1}$ in case B, similar to the decoupling pattern for $v_\chi$; we note that $\sin\alpha$ and $v_{\chi}$ are the only quantities that may decouple as slowly as $\mu_3^{-1}$, and all others decouple at a rate of $\mu_3^{-2}$ or faster. The exact expression for $\sin\alpha$ is plotted along with the expansion formula in the bottom panels of Fig.~\ref{fig:vevs-sin}.

We now consider the decoupling behavior of the couplings of the light Higgs boson $h$ to SM particles.  The relevant tree-level couplings of $h$ to vector bosons and fermions, as well as the triple-$h$ self-coupling,\footnote{The exact tree-level formula for the triple-$h$ coupling is given in Eq.~(\ref{eq:hhh}) in Appendix A.} are given by
\begin{eqnarray}
	\kappa_V &=& \cos\alpha \frac{v_\phi}{v}-\frac{8}{\sqrt{3}}\sin\alpha\frac{v_\chi}{v} 
	\simeq 1 + \frac{3}{8} \frac{M_1^2 v^2}{\mu_3^4},
	\nonumber \\
	\kappa_f &=& \cos\alpha\frac{v}{v_\phi} 
	\simeq 1 - \frac{1}{8} \frac{M_1^2 v^2}{\mu_3^4},
	\nonumber \\
	g_{hhVV} &=&   \frac{2 M_V^2}{v^2} \left( \cos^2 \alpha + \frac{8}{3}\sin^2\alpha \right)
	\simeq \frac{2 M_V^2}{v^2} \left(  1 + \frac{5}{4} \frac{M_1^2 v^2}{\mu_3^4 }\right),
	\nonumber \\
	g_{hhh} &\simeq& \frac{3 m_h^2}{v}\left\{ 1 - \frac{M_1^2v^2}{\mu_3^4} 
	\left[ \frac{7}{8} - \frac{3}{2} \frac{v^2}{m_h^2} \left( (2 \lambda_2 - \lambda_5) 
	+ \frac{M_1M_2}{\mu_3^2} \right) \right] \right\}, 
	\label{eq:hcoups} 
\end{eqnarray}
where $\kappa_V$ and $\kappa_f$ are defined as the ratios of the couplings $g_{hVV}$ and $g_{hf{\bar{f}}}$ to those of the SM Higgs boson as in Ref.~\cite{LHCHiggsCrossSectionWorkingGroup:2012nn}, and $M_V$ is the appropriate massive gauge boson mass.  Each of these couplings becomes equal to the corresponding coupling of the SM Higgs boson in the limit $\mu_3 \to \infty$.  Furthermore, each of these tree-level couplings of $h$ decouples at a rate proportional to $\mu_3^{-4}$ ($\mu_3^{-2}$) in case A (case B).  We also notice that, near the decoupling limit, $\kappa_V > 1$ and $\kappa_f < 1$, and that $|\kappa_V - 1| = 3 |\kappa_f - 1|$.

\begin{figure}
\begin{center}
\resizebox{0.49\textwidth}{!}{\includegraphics{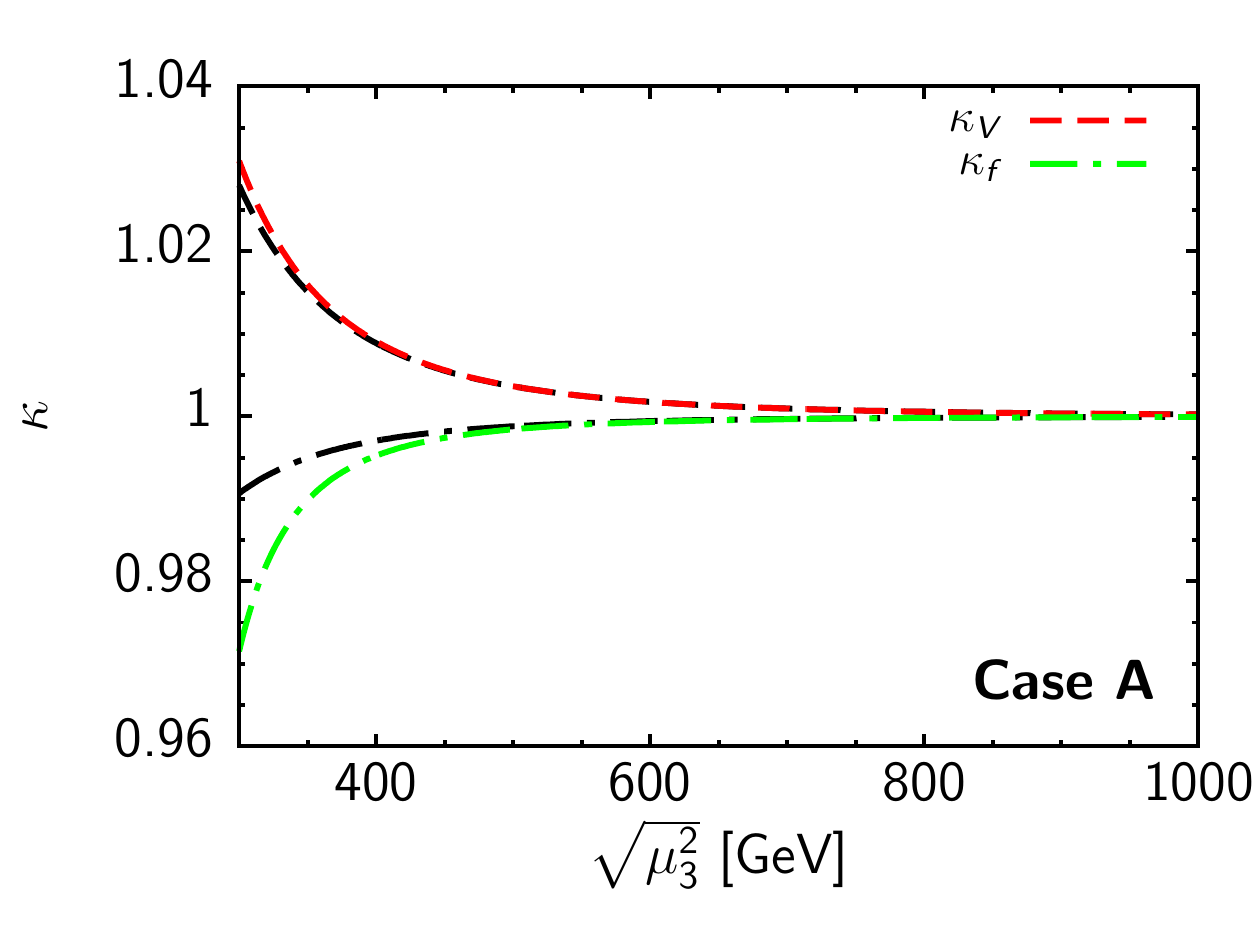}}
\resizebox{0.49\textwidth}{!}{\includegraphics{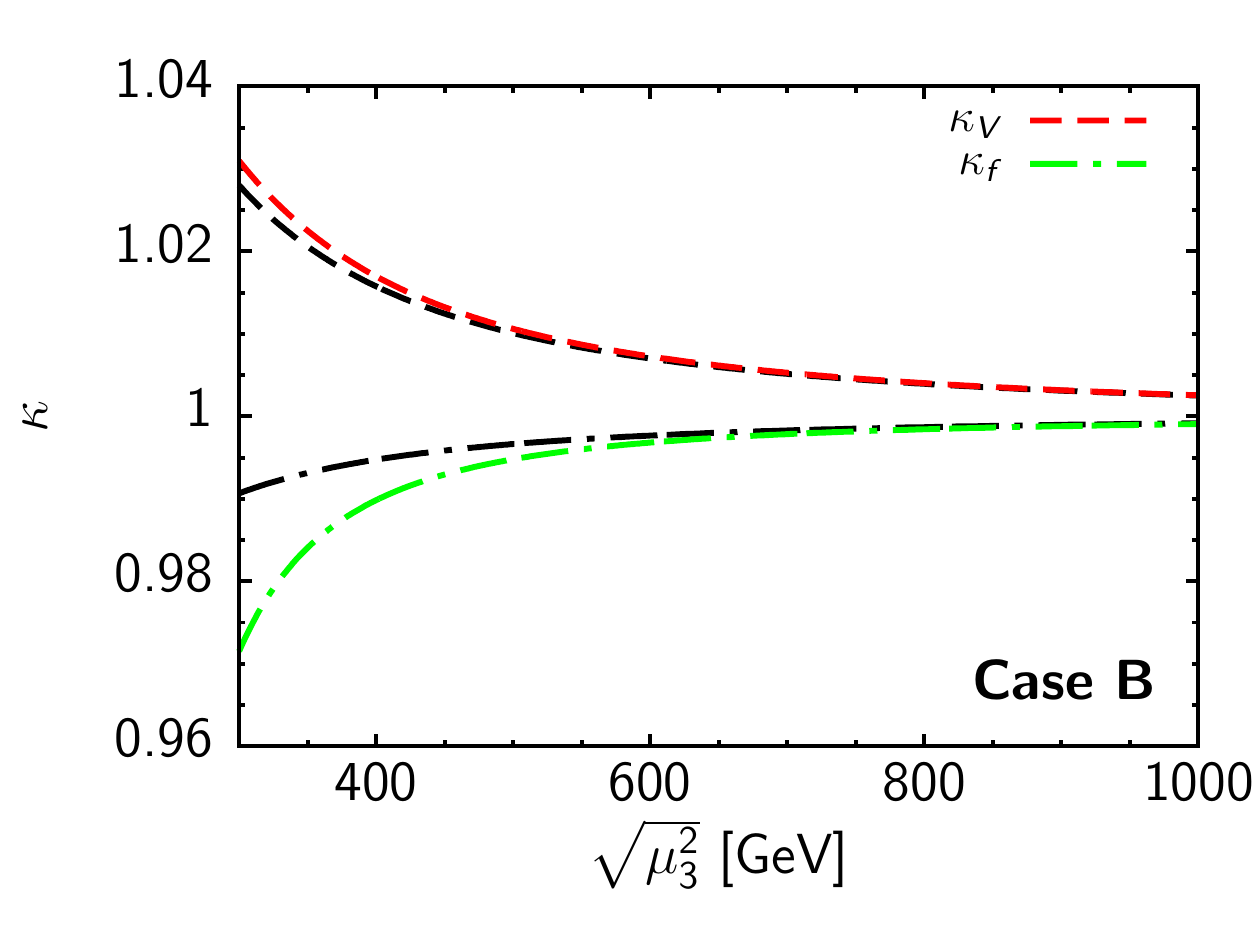}}
\caption{The light Higgs coupling modification factors $\kappa_V$ (upper dashed curves) and $\kappa_f$ (lower dot-dashed curves) as a function of $\mu_3$, for cases A (left) and B (right).  The colored (light) curves are the exact results while the black (dark) curves show the corresponding expansion formulas as in Eq.~(\ref{eq:hcoups}).}
\label{fig:kvf}
\end{center}
\end{figure}

The decoupling behavior of $\kappa_V$ and $\kappa_f$ is illustrated in Fig.~\ref{fig:kvf}.  We see that the expansion formula for $\kappa_V$ provides a very good approximation to the exact result, but the expansion formula for $\kappa_f$ deviates significantly from the exact result for small values of $\mu_3 \lesssim 400$~GeV, indicating that subleading terms become numerically relevant for these relatively low $\mu_3$ values.

We finally consider the decoupling behavior of the loop-induced couplings of $h$ to $\gamma\gamma$ and $Z\gamma$.\footnote{The loop-induced coupling of $h$ to $gg$ is modified by the same factor $\kappa_f$ that controls the $h$ couplings to SM fermions.}  Modifications to these couplings come from two sources: (i) the modifications of the fermion and $W$ boson loop contributions by factors of $\kappa_f$ and $\kappa_V$, respectively; and (ii) new contributions from the charged scalars $H_3^+$, $H_5^+$, and $H_5^{++}$ propagating in the loop.  Details of the calculation are given in Appendix~\ref{app:hgaga}.

The decoupling behavior of the charged scalar loop contributions to the $h \gamma\gamma$ and $h Z \gamma$ couplings can be understood by considering the relevant couplings of $h$ to charged scalars in the decoupling limit:
\begin{eqnarray}
	g_{hH_3^+H_3^{+*}} &\simeq& \left(4\lambda_2-\lambda_5\right) v 
	+ \frac{\left(M_1^2-3 M_1 M_2\right) v}{\mu_3^2}, \nonumber \\
	g_{hH_5^+H_5^{+*}} = g_{hH_5^{++}H_5^{++*}} 
	&\simeq& \left(4\lambda_2+\lambda_5\right) v + \frac{3 M_1 M_2 v}{\mu_3^2},
\end{eqnarray}
where we have kept only the leading term in the decoupling limit.  In particular, these triple-scalar couplings go to a constant of order $v$ in the decoupling limit.  Combined with the loop integral $\propto 1/m_i^2$, where $m_i$ is the mass of the charged scalar in the loop, we find that the contributions to the $h\gamma\gamma$ and $h Z \gamma$ amplitudes from charged scalars in the loop will decouple like $\mu_3^{-2}$.  In particular, we have
\begin{eqnarray}
	\Delta\kappa_{\gamma} &\simeq& -\frac{1}{F_1(M_W) + \frac{4}{3} F_{1/2}(m_t)}
	\frac{2 v^2}{3 \mu_3^2}\left[6\lambda_2+\lambda_5
	+ \frac{M_1^2+12M_1M_2}{4\mu_3^2}\right], \nonumber \\
	\Delta\kappa_{Z \gamma} &\simeq& \frac{1}{2(A_W+A_f)}\frac{1-2 s_W^2}{s_W c_W}
	\frac{2 v^2}{3\mu_3^2} \left[6\lambda_2+\lambda_5
	+ \frac{M_1^2+12M_1M_2}{4\mu_3^2}\right],
	\label{eq:scalarloops}
\end{eqnarray}
where $\Delta \kappa_{\gamma}$ and $\Delta \kappa_{Z\gamma}$ are the contributions to the effective $h\gamma\gamma$ and $h Z \gamma$ couplings due to the contributions of non-SM particles in the loop (see Appendix~\ref{app:hgaga} for details).  Here $F_1$ and $F_{1/2}$ ($A_W$ and $A_f$) represent the SM contributions from the $W$ boson and the top quark to $h\rightarrow\gamma\gamma$ ($h\rightarrow\gamma Z$).  Note that, near the decoupling limit, the charged scalar contributions to the $h \gamma\gamma$ and $h Z \gamma$ loop contributions have the same dependence on the GM model parameters, as given in the square brackets in Eq.~(\ref{eq:scalarloops}).

\begin{figure}
\begin{center}
\resizebox{0.49\textwidth}{!}{\includegraphics{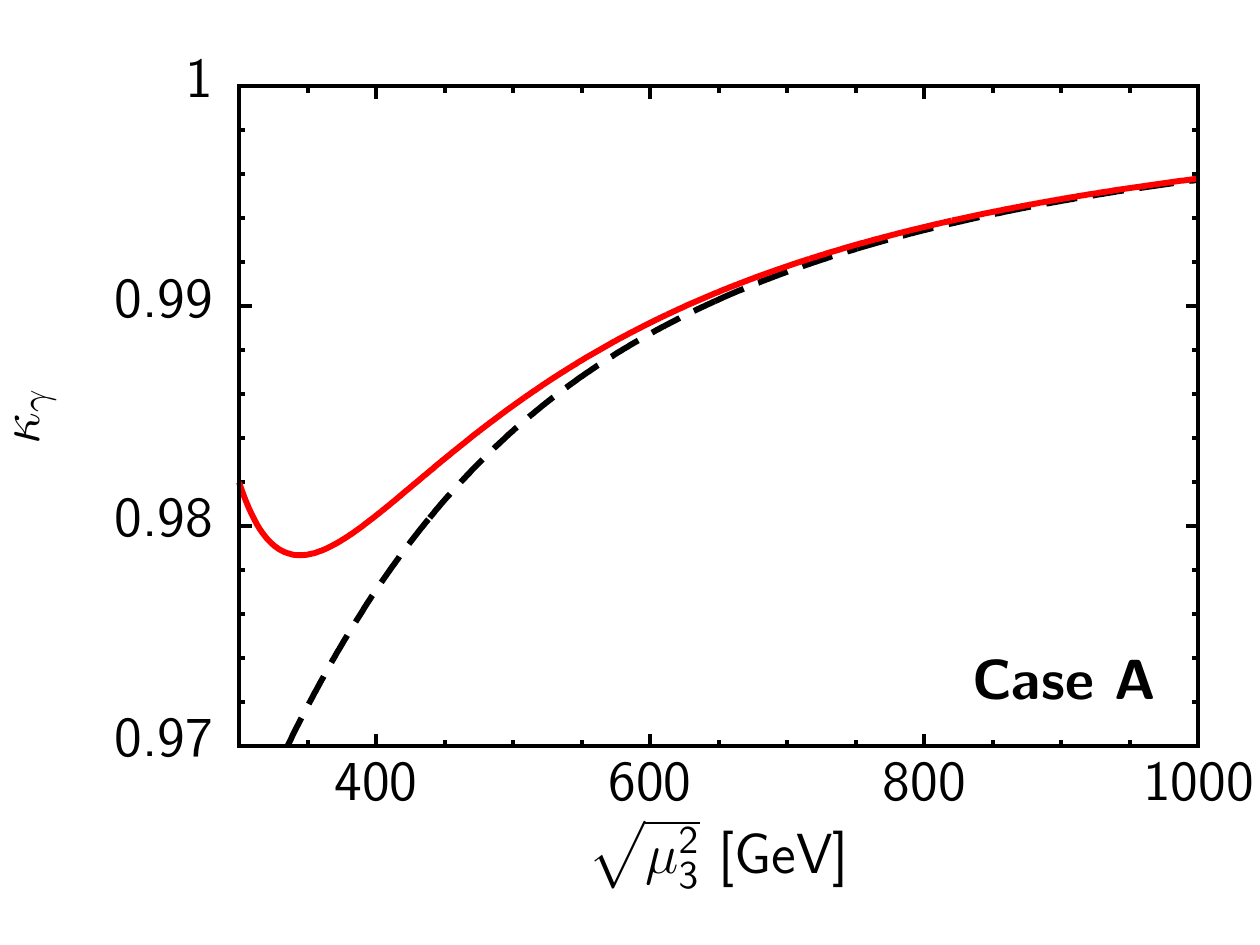}}
\resizebox{0.49\textwidth}{!}{\includegraphics{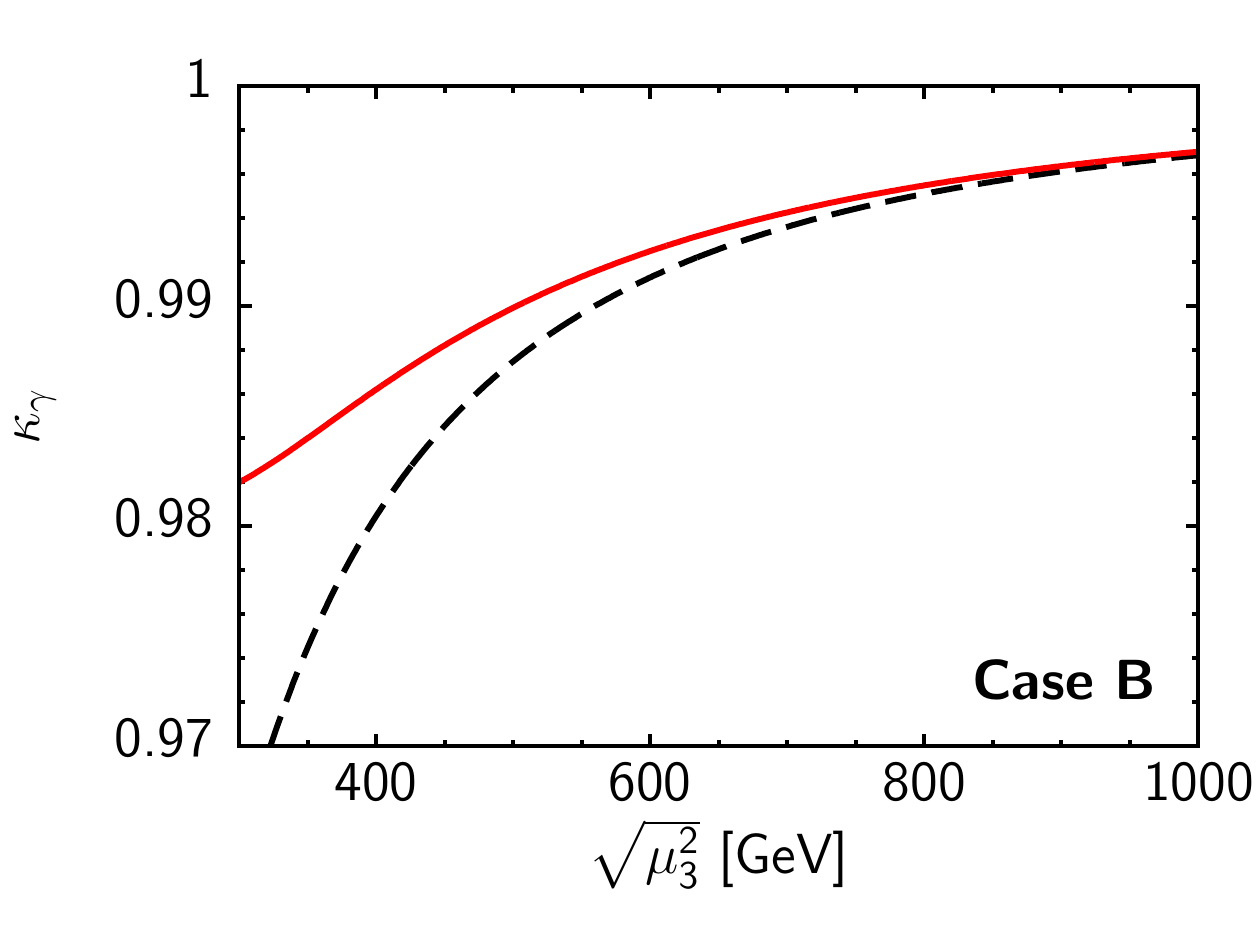}}
\resizebox{0.49\textwidth}{!}{\includegraphics{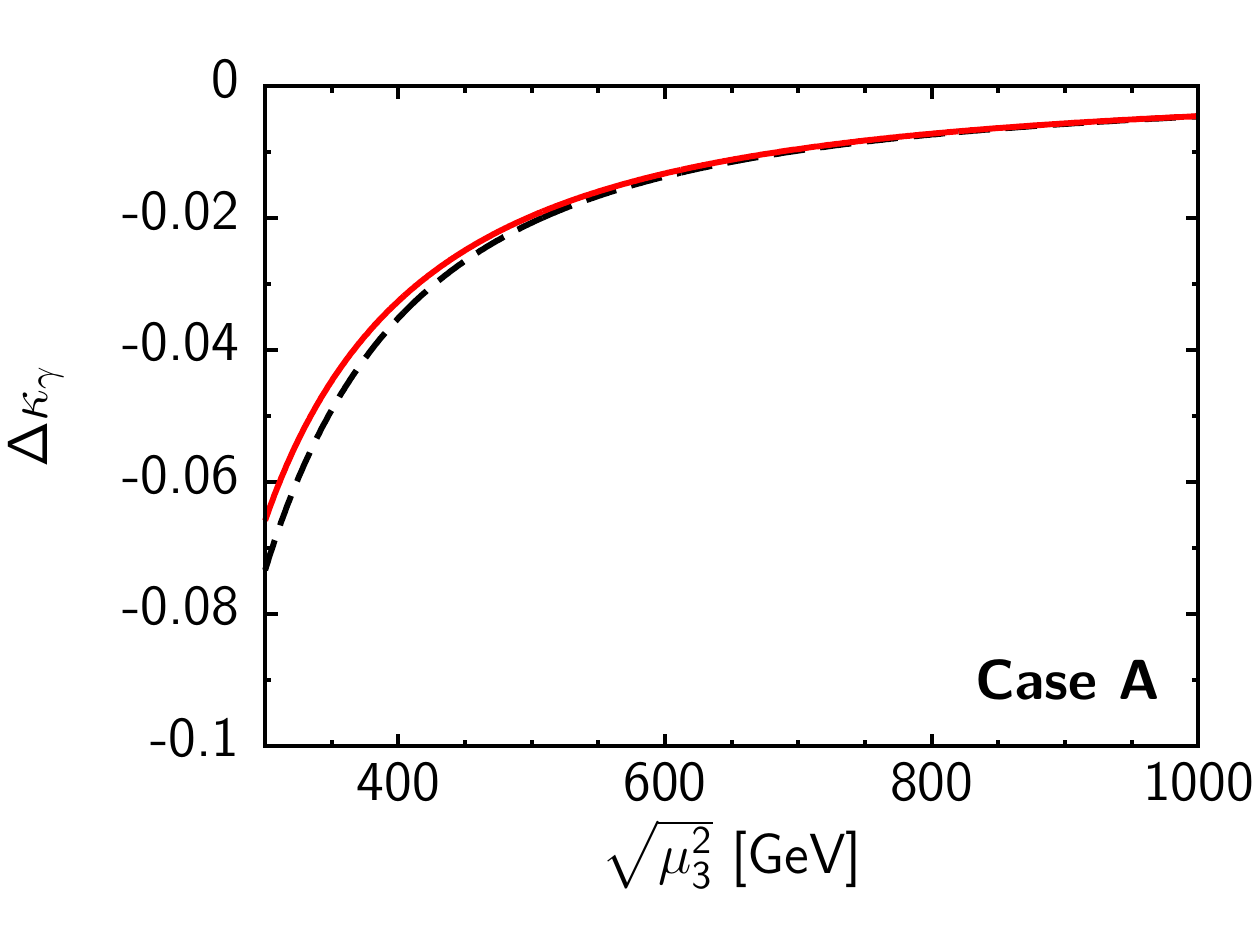}}
\resizebox{0.49\textwidth}{!}{\includegraphics{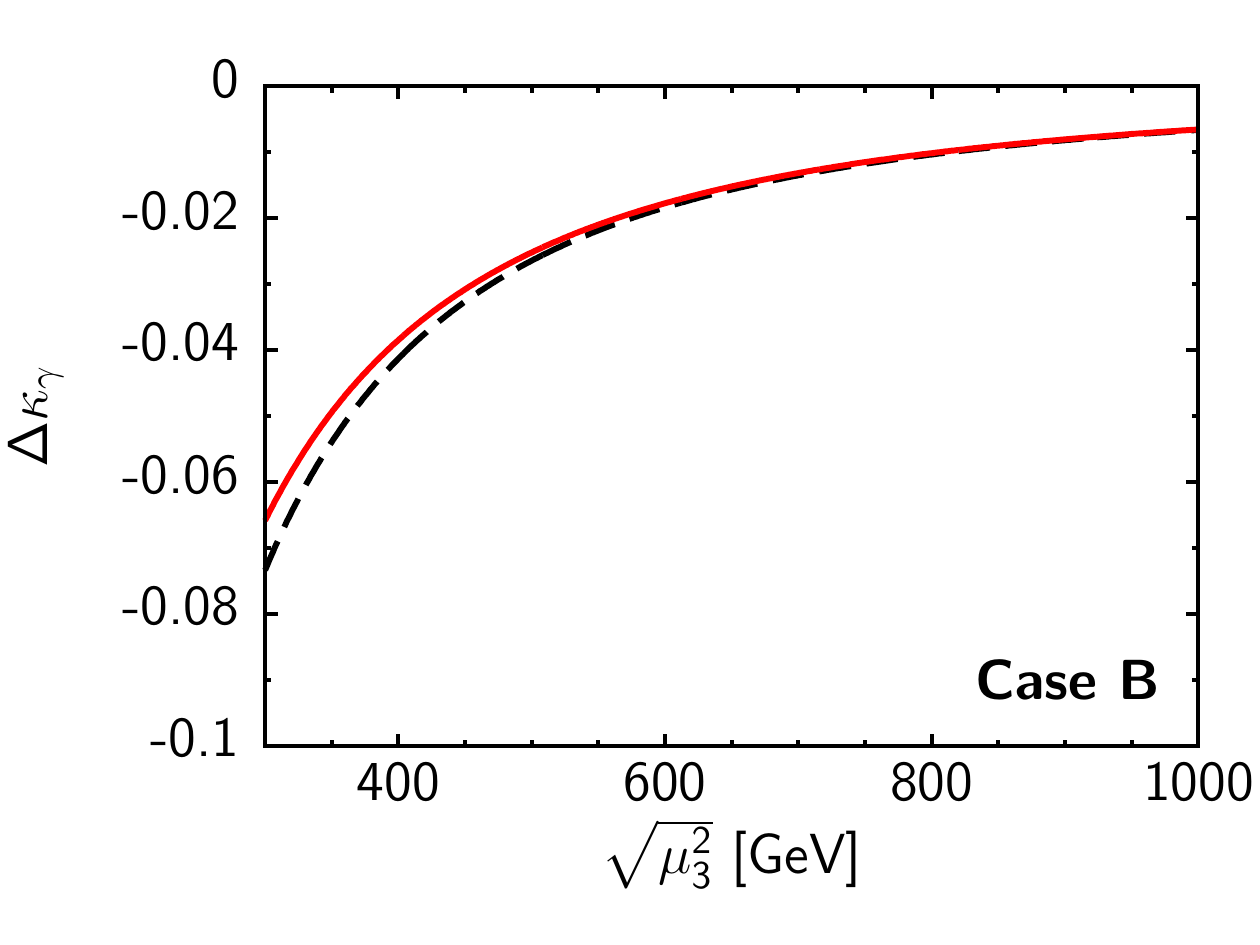}}
\caption{Top: The light Higgs coupling modification factor $\kappa_{\gamma}$ as a function of $\mu_3$, for cases A (left) and B (right). Bottom: The light Higgs coupling modification factor $\Delta\kappa_{\gamma}$, comprising only the contributions from the non-SM charged scalars in the loop, as a function of $\mu_3$.  In all plots the solid red (light) line shows the exact one-loop result, while the dashed black (dark) line is the expansion formula as discussed in the text.}
\label{fig:kgam}
\end{center}
\end{figure}

\begin{figure}
\begin{center}
\resizebox{0.49\textwidth}{!}{\includegraphics{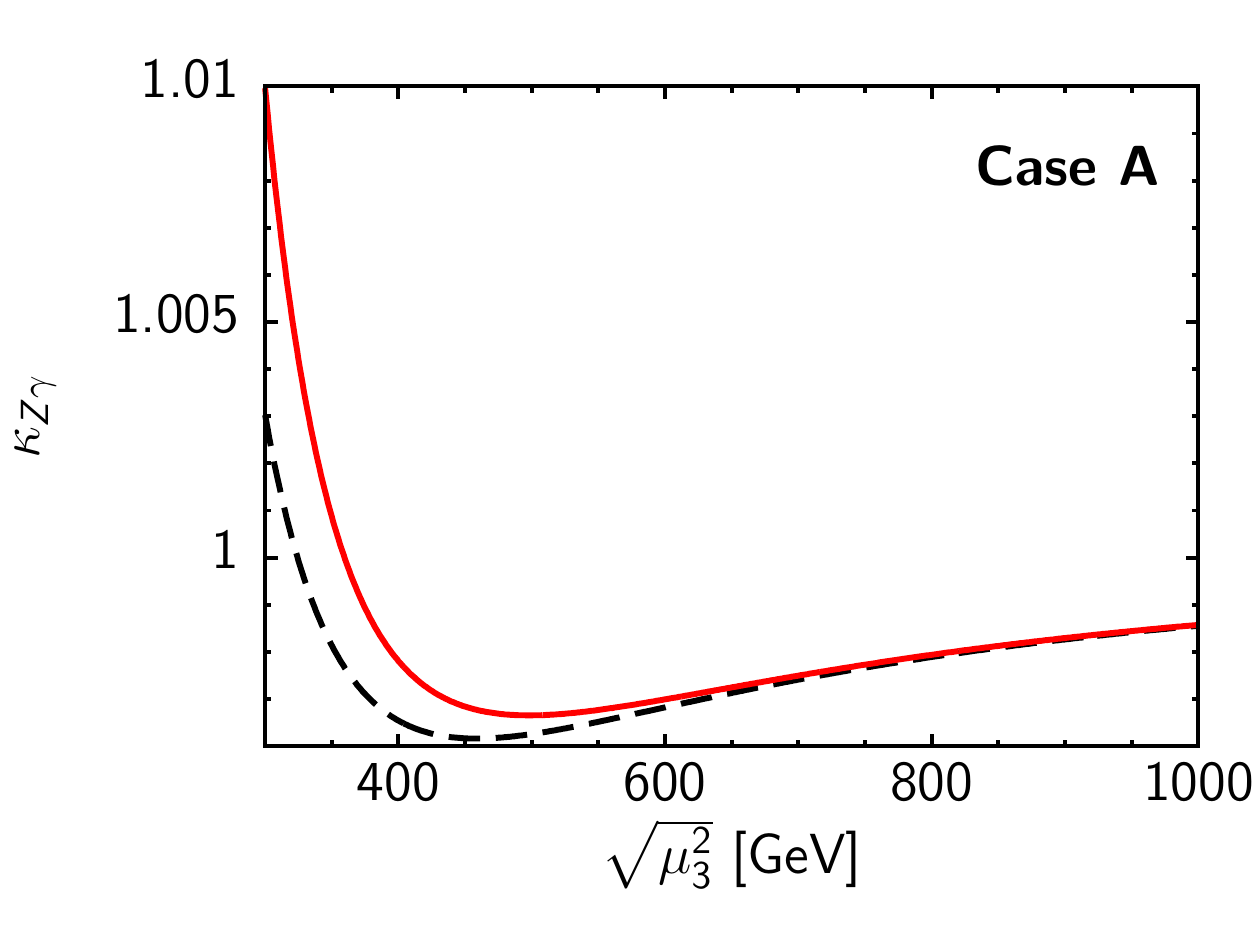}}
\resizebox{0.49\textwidth}{!}{\includegraphics{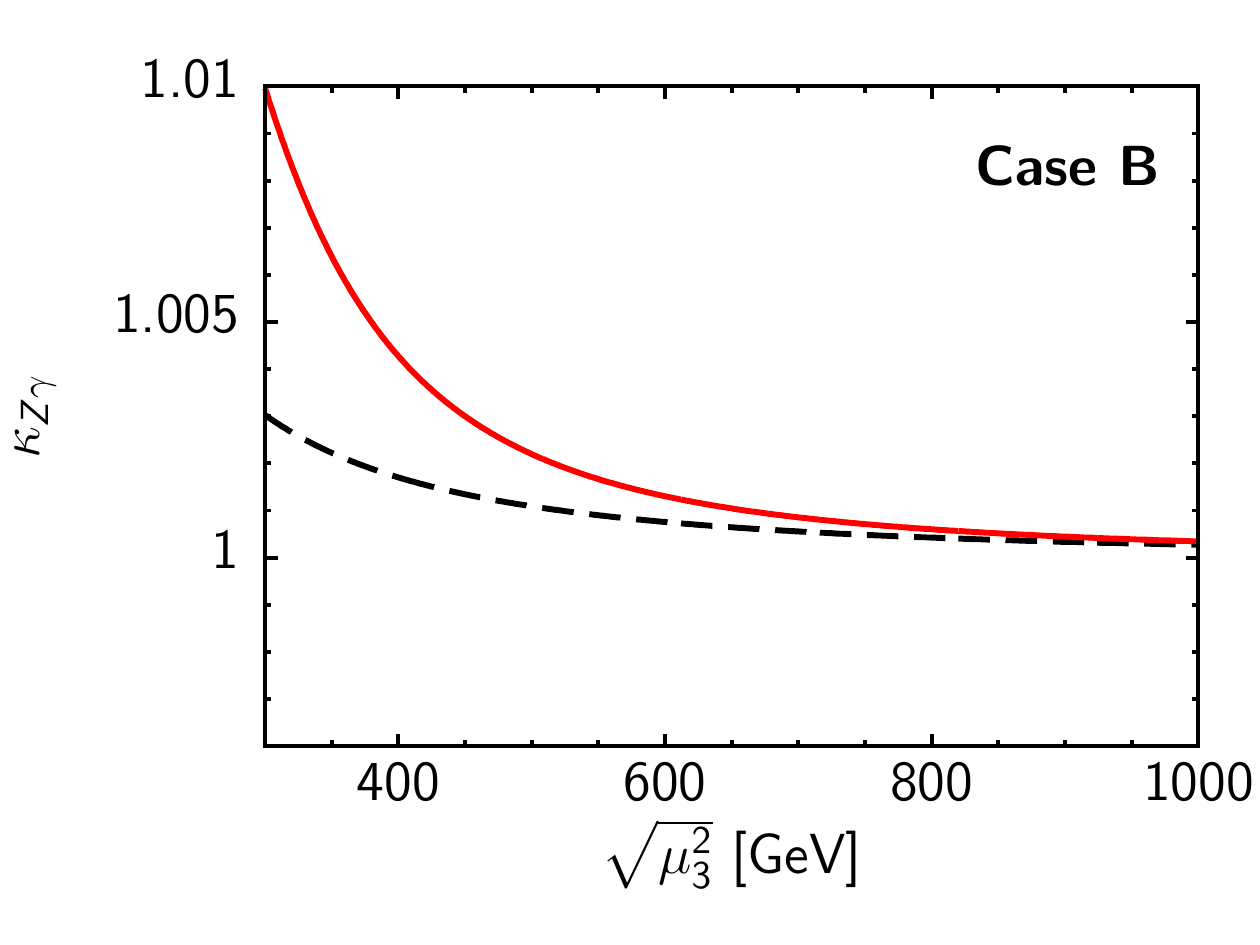}}
\resizebox{0.49\textwidth}{!}{\includegraphics{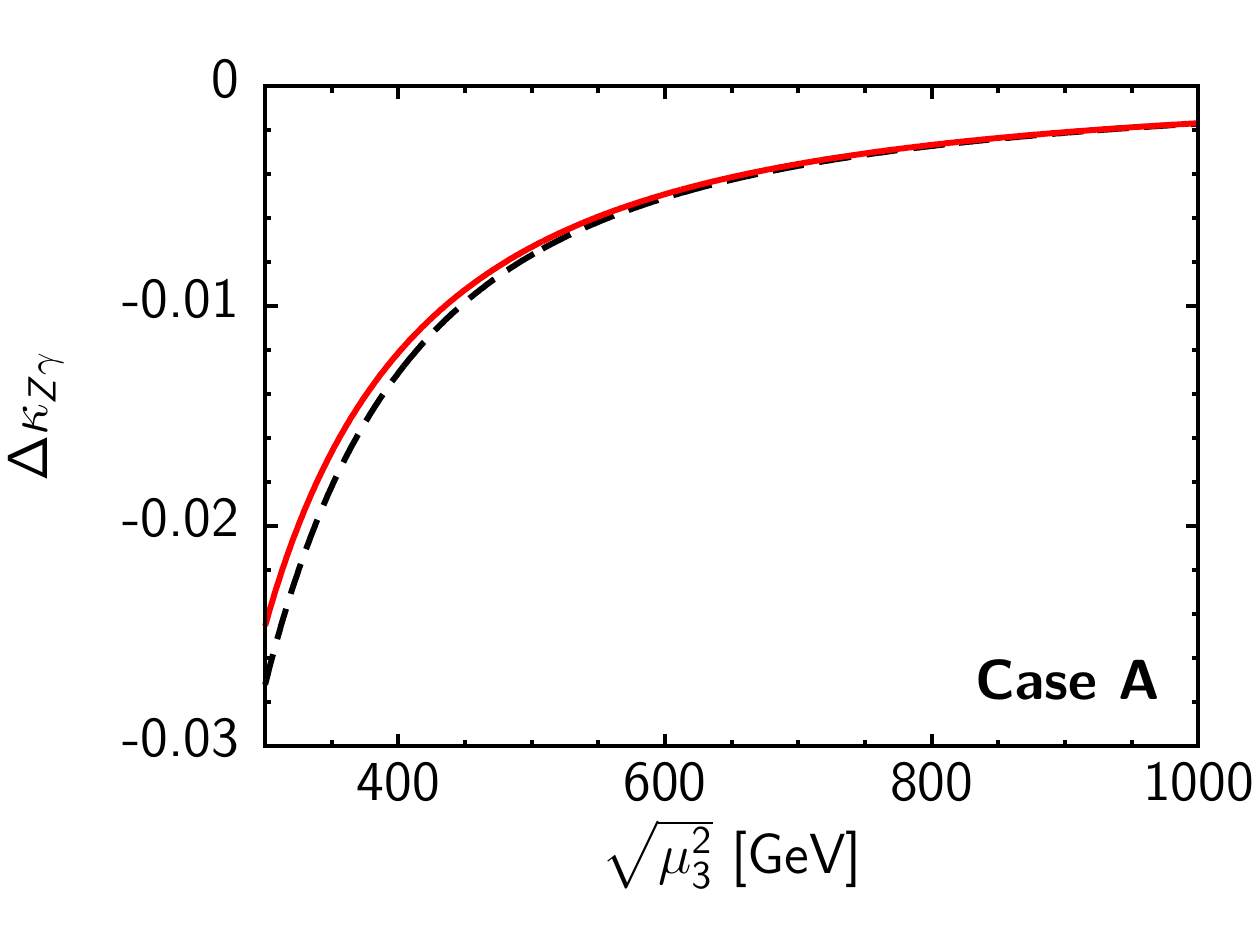}}
\resizebox{0.49\textwidth}{!}{\includegraphics{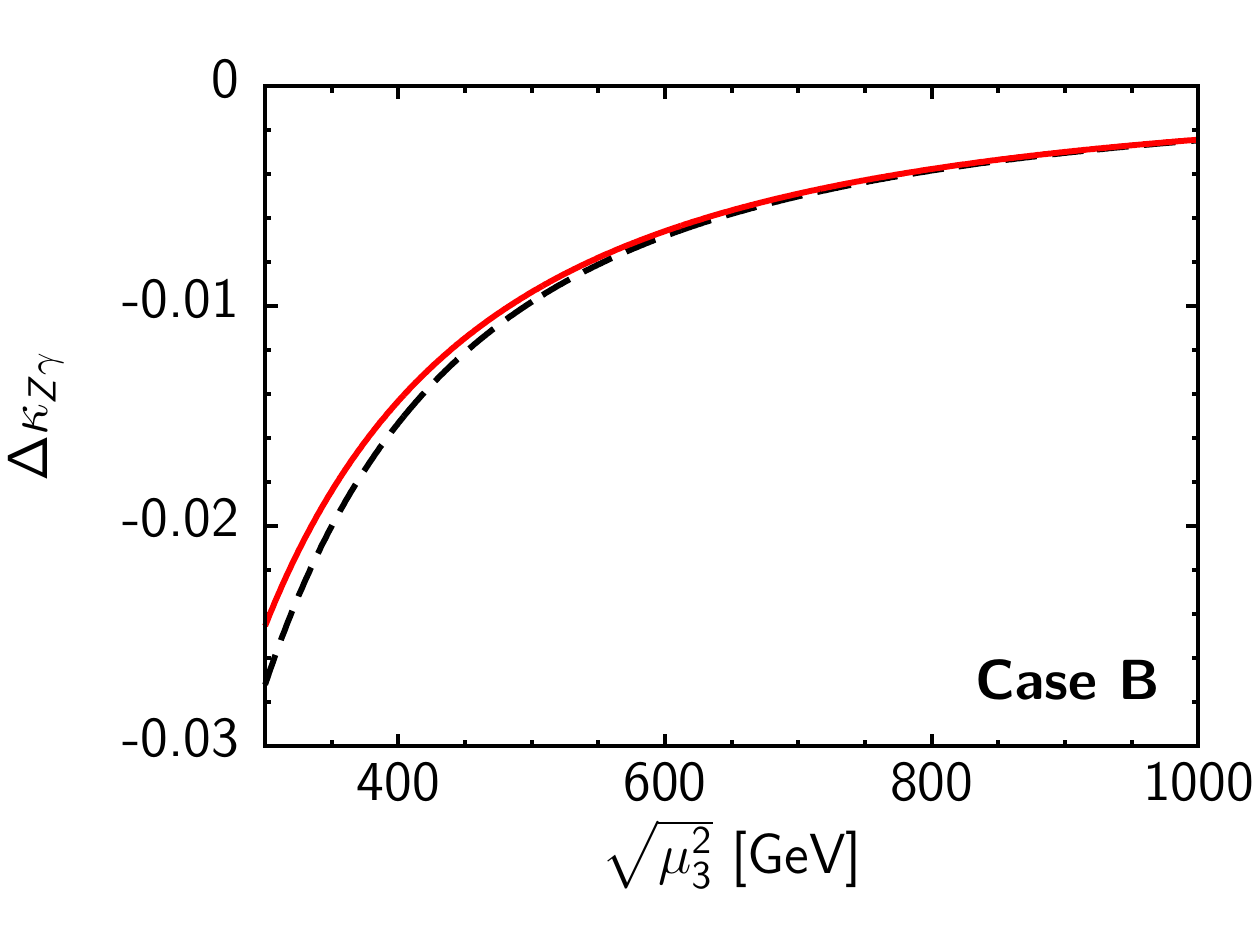}}
\caption{Top: The light Higgs coupling modification factor $\kappa_{Z\gamma}$ as a function of $\mu_3$, for cases A (left) and B (right).  Bottom: The light Higgs coupling modification factor $\Delta\kappa_{Z\gamma}$, comprising only the contributions from the non-SM charged scalars in the loop, as a function of $\mu_3$. In all plots the solid red (light) line shows the exact one-loop result, while the dashed black (dark) line is the expansion formula as discussed in the text.}
\label{fig:kgamZ}
\end{center}
\end{figure}

We illustrate the decoupling behavior of the loop-induced $h\gamma \gamma$ ($h Z \gamma$) effective coupling $\kappa_{\gamma}$ ($\kappa_{Z\gamma}$), as well as the contribution from only the new charged scalars $\Delta \kappa_{\gamma}$ ($\Delta \kappa_{Z\gamma}$), in Fig.~\ref{fig:kgam} (Fig.~\ref{fig:kgamZ}).  Of the SM fermion contributions, we include only the top quark loop.  The effective couplings $\kappa_{\gamma}$ and $\kappa_{Z\gamma}$ are defined normalized to the SM prediction.  We show the exact one-loop result (red solid lines) as well as that obtained using the expansion formulas of Eqs.~(\ref{eq:hcoups}) and (\ref{eq:scalarloops}).  The relatively large deviation at low $\mu_3$ between the exact result and the expansion formulas in the upper panels of Figs.~\ref{fig:kgam} and \ref{fig:kgamZ} follows from the relatively large deviation between the exact $\kappa_f$ and its expansion formula as shown in Fig.~\ref{fig:kvf}.

%%%%%%%%%%%%%%%%%%%%%%%%%%%%%%%%%%%%%%%%%%%%%%
\subsection{Comparison to decoupling in the two-Higgs-doublet model}
\label{sec:2HDM}

We now compare the decoupling behavior of the GM model to the well-studied case of the CP-conserving two-Higgs-doublet model (2HDM)~\cite{Gunion:2002zf}.  In particular, we examine how the couplings of the light custodial-singlet Higgs boson $h$ deviate from the SM limit as a function of the common mass scale of the heavy scalars.

The 2HDM contains five scalar states: two CP-even neutral scalars $h$ and $H$, a CP-odd scalar $A$, and two charged scalars $H^{\pm}$. As shown in Ref.~\cite{Gunion:2002zf}, the couplings of the light Higgs boson $h$ in the 2HDM behave as follows in the decoupling limit (we choose the Type-II structure for the fermion couplings),
\begin{eqnarray}
	\kappa_V^{\rm 2HDM} &\simeq& 1-\frac{\hat{\lambda}^2 v^4}{2\,m_A^4}, \nonumber \\
	\kappa_f^{\rm 2HDM} &\simeq& 1+\frac{\hat{\lambda} v^2}{m_A^2} 
	\times \left\{\begin{array}{l l}
		\cot\beta & {\rm for\ up\ type\ fermions} \\
		-\tan\beta & {\rm for\ down\ type\ fermions}, \\
	\end{array} \right. \nonumber \\
	g_{hhh}^{\rm 2HDM} &\simeq& \frac{3 m_h^2}{v} 
	\left[1-\frac{3\hat{\lambda}^2 v^2}{\lambda m_A^2}\right],
	\label{eq:2HDM}
\end{eqnarray}
where $\lambda$ and $\hat{\lambda}$ are linear combinations of the quartic couplings in the 2HDM, $m_A$ is the mass of the CP-odd scalar $A$, and the angle $\beta$ is defined as usual in terms of the ratio of the vevs of the two doublets, $\tan\beta=v_2/v_1$, where $v_1^2+v_2^2 = v^2 \simeq (246$~GeV)$^2$.  Values of $\tan\beta \sim 1$--50 are usually considered.  We also note that $g_{hhVV} = 1$ in the 2HDM.

Comparing the 2HDM couplings in the decoupling limit in Eq.~(\ref{eq:2HDM}) to those of the GM model in Eq.~(\ref{eq:hcoups}), we make the following observations:
\begin{itemize}
\item In case A, $\kappa_V$ decouples like $(v^4/M_{\rm new}^4)$ in both the GM model and the 2HDM, whereas $\kappa_f$ and $g_{hhh}$ decouple much faster in the GM model than in the 2HDM [like $(v^4/M_{\rm new}^4)$ in the GM model compared to $(v^2/M_{\rm new}^2)$ in the 2HDM].  
\item In case B, $\kappa_V$ decouples much more slowly in the GM model than in the 2HDM [like $(v^2/M_{\rm new}^2)$ in the GM model compared to $(v^4/M_{\rm new}^4)$ in the 2HDM], while $\kappa_f$ and $g_{hhh}$ decouple like $(v^2/M_{\rm new}^2)$ in both the GM model and the 2HDM.
\end{itemize}
As such, precision measurements of the Higgs couplings at a Higgs factory such as the International Linear Collider~\cite{Baer:2013cma} may be able to differentiate these models: relatively large deviations from the SM in all of the $h$ couplings would favor the GM model with large trilinear couplings as in case B, while large deviations in the fermion and trilinear Higgs couplings but SM-like vector boson couplings would favor the 2HDM. In the event that additional scalars are discovered at a relatively low mass scale, SM-like couplings of the light Higgs would tend to favor the GM model with small trilinear couplings as in case A compared to the 2HDM.

Another means of differentiating the 2HDM from the GM model can be found in the sign of the deviation of $\kappa_V$ from one.  In Higgs sectors containing only isospin doublets and singlets, the couplings of the light Higgs boson to $W$ or $Z$ boson pairs are always less than or equal to their SM values.  However, in the GM model near the decoupling limit, these couplings are always larger than in the SM, as seen in Eq.~(\ref{eq:hcoups}) for $\kappa_V$.  As such, a precision measurement of $\kappa_V$ that reveals a positive deviation relative to the SM prediction would provide definitive evidence for a non-negligible contribution to electroweak symmetry breaking from a scalar with isospin larger than 1/2.  Taken together with the stringent experimental constraint on the $\rho$ parameter, such an observation could only be accommodated in the GM model, one of its higher-isospin generalizations~\cite{Galison:1983qg}, or the extension of the SM by a scalar septet~\cite{Hisano:2013sn} (which also happens to preserve $\rho = 1$ at tree level).

%%%%%%%%%%%%%%%%%%%%%%%%%%%%%%%%%%%%%%%%%%%%%%
\subsection{Numerical scans}
\label{sec:scans}

\begin{figure}
\resizebox{0.5\textwidth}{!}{\includegraphics{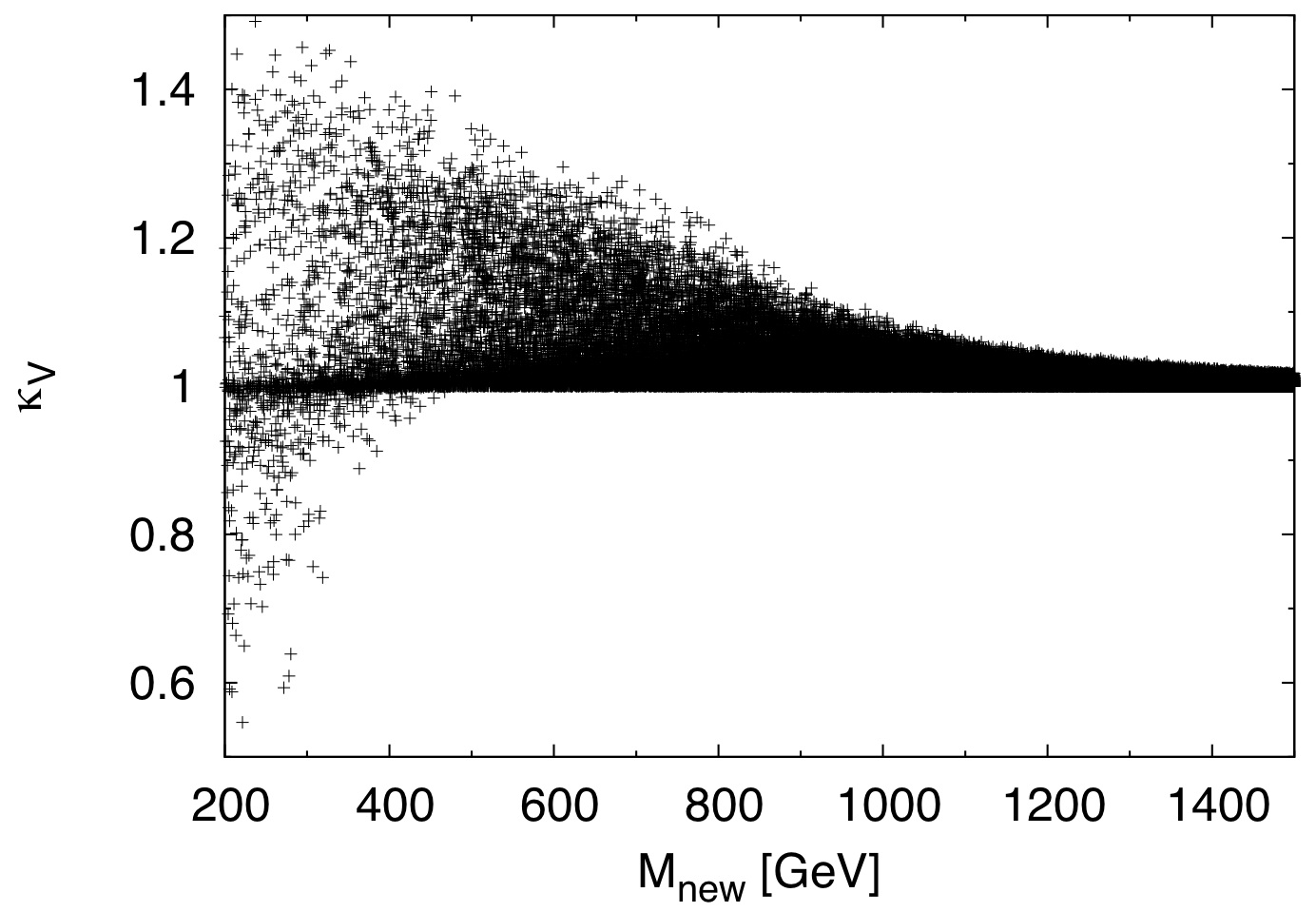}}\resizebox{0.5\textwidth}{!}{\includegraphics{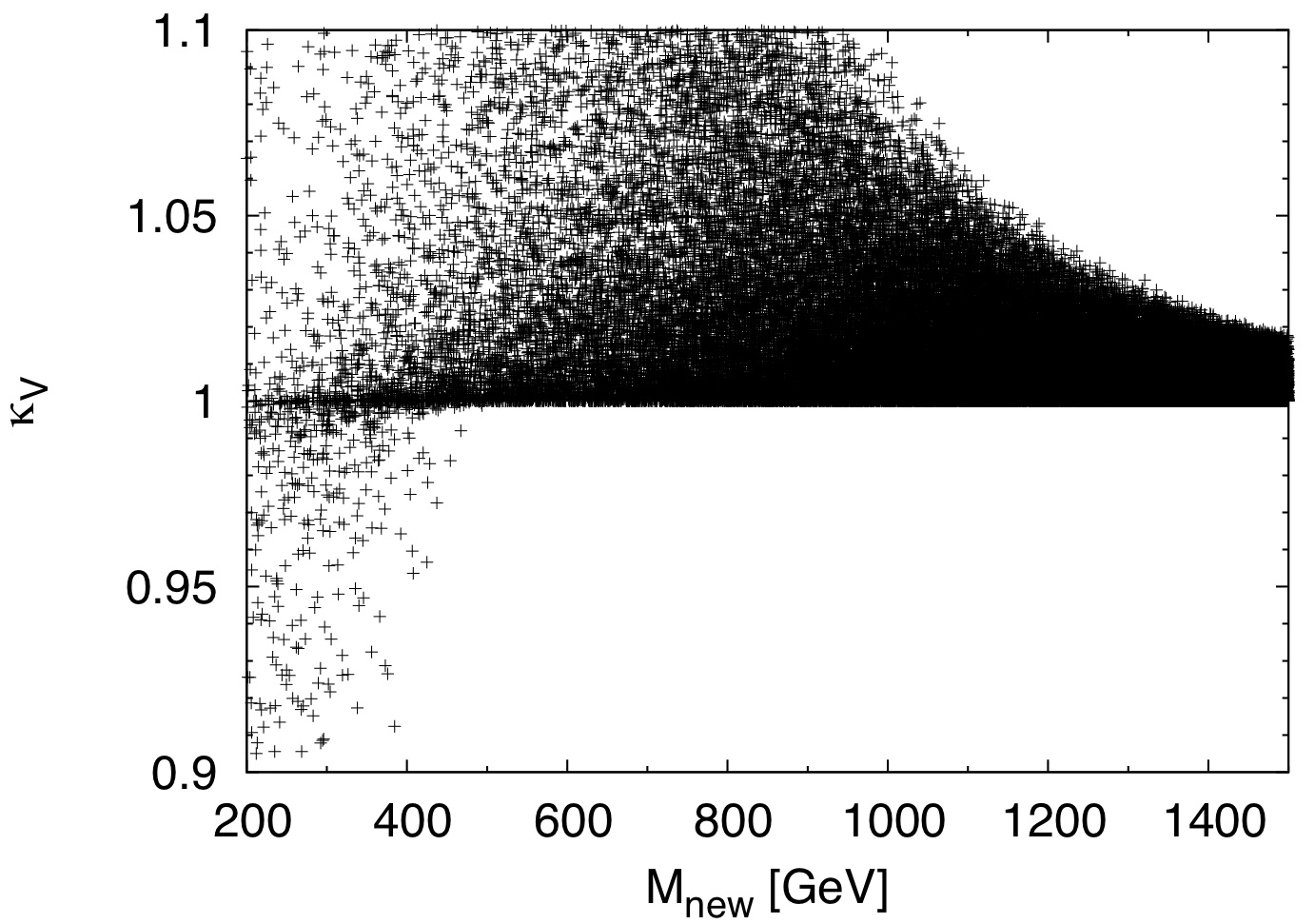}} \\
\resizebox{0.5\textwidth}{!}{\includegraphics{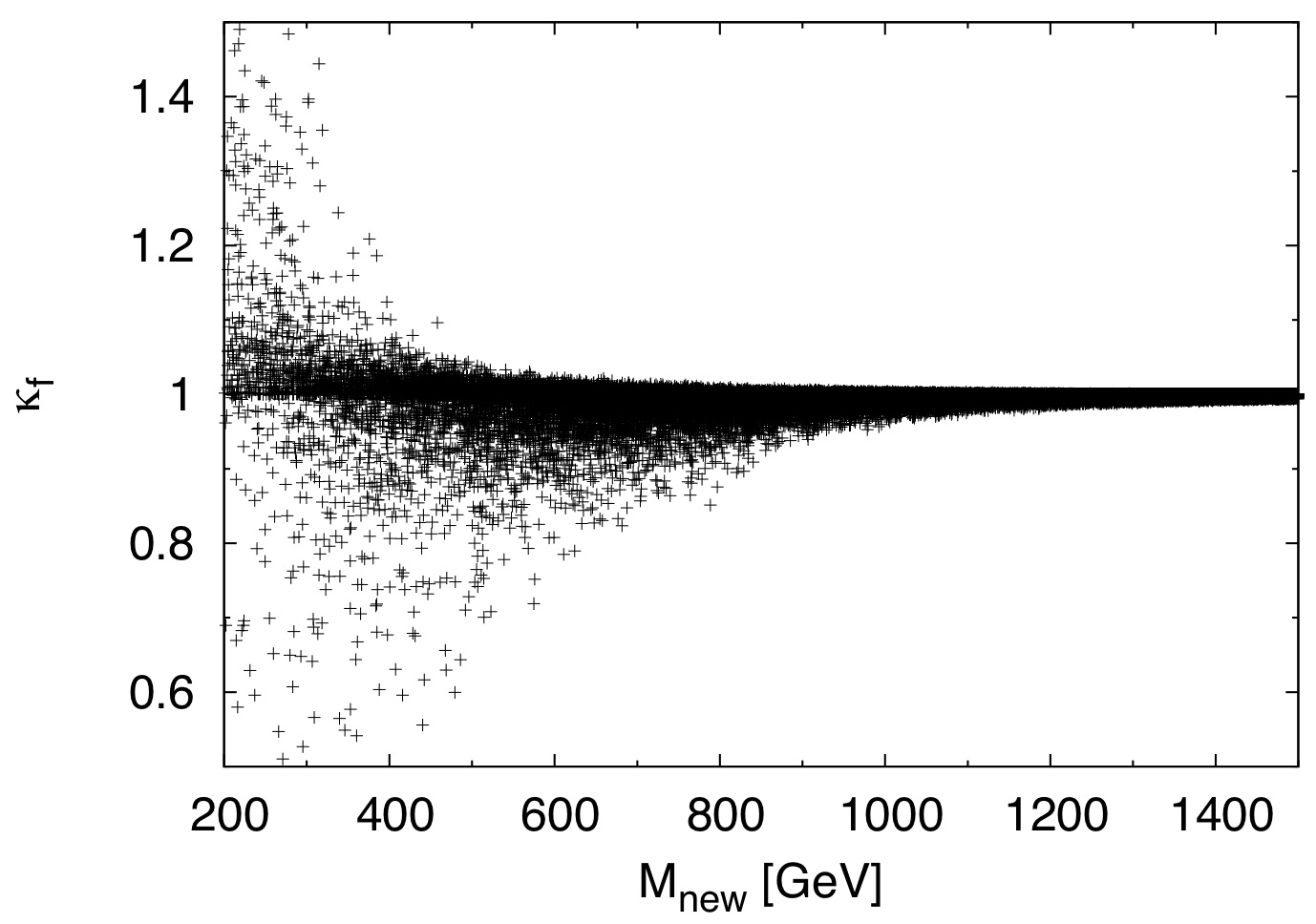}}\resizebox{0.5\textwidth}{!}{\includegraphics{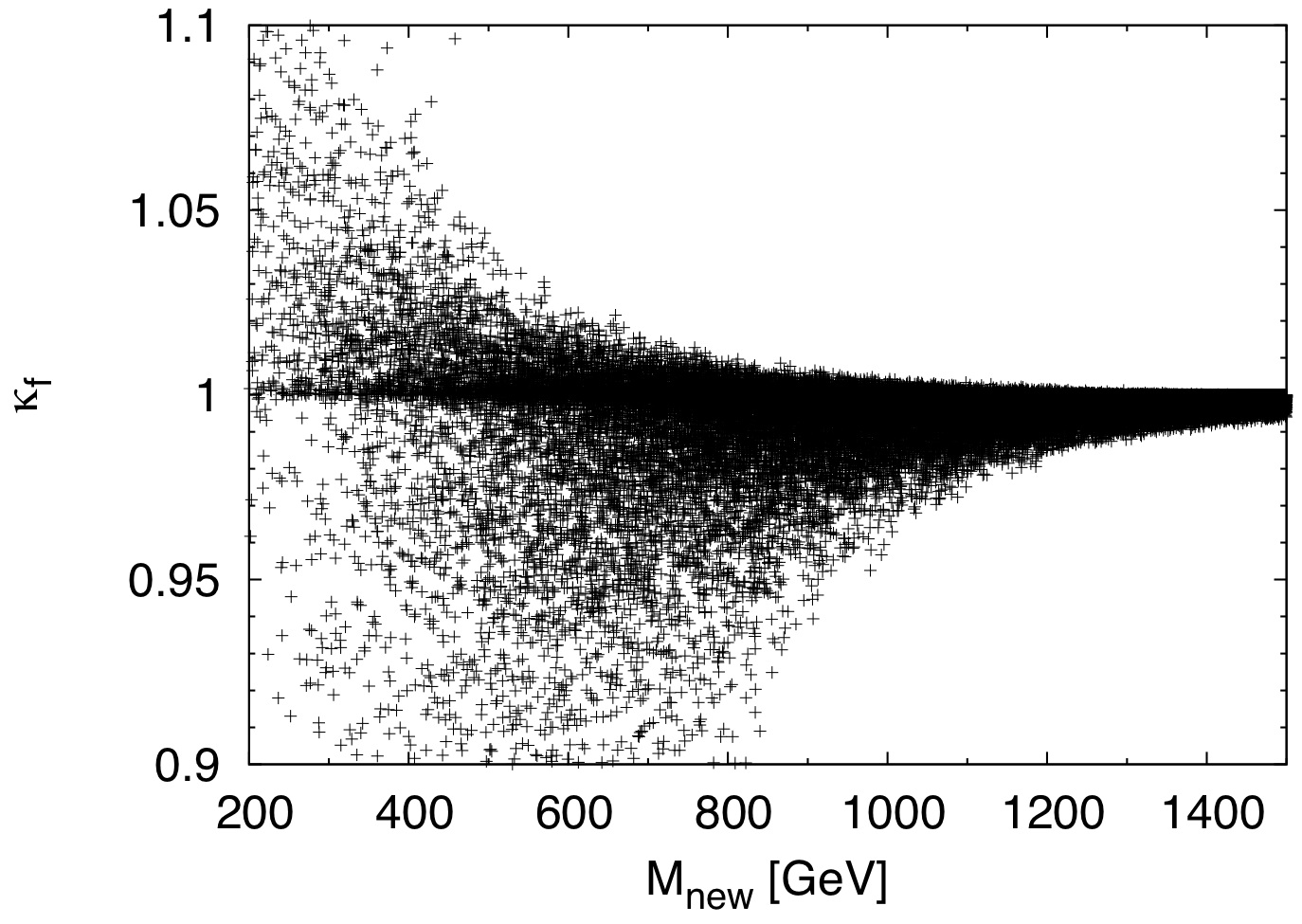}}
\caption{The light Higgs coupling modification factors $\kappa_V$ (upper) and $\kappa_f$ (lower) as a function of the mass of the lightest of the new scalars, $M_{\rm new} = \min(m_H, m_3, m_5)$.  The right panels shows a close-up of the region of small coupling deviations.}
\label{fig:kVscan}
\end{figure}

\begin{figure}
\resizebox{0.5\textwidth}{!}{\includegraphics{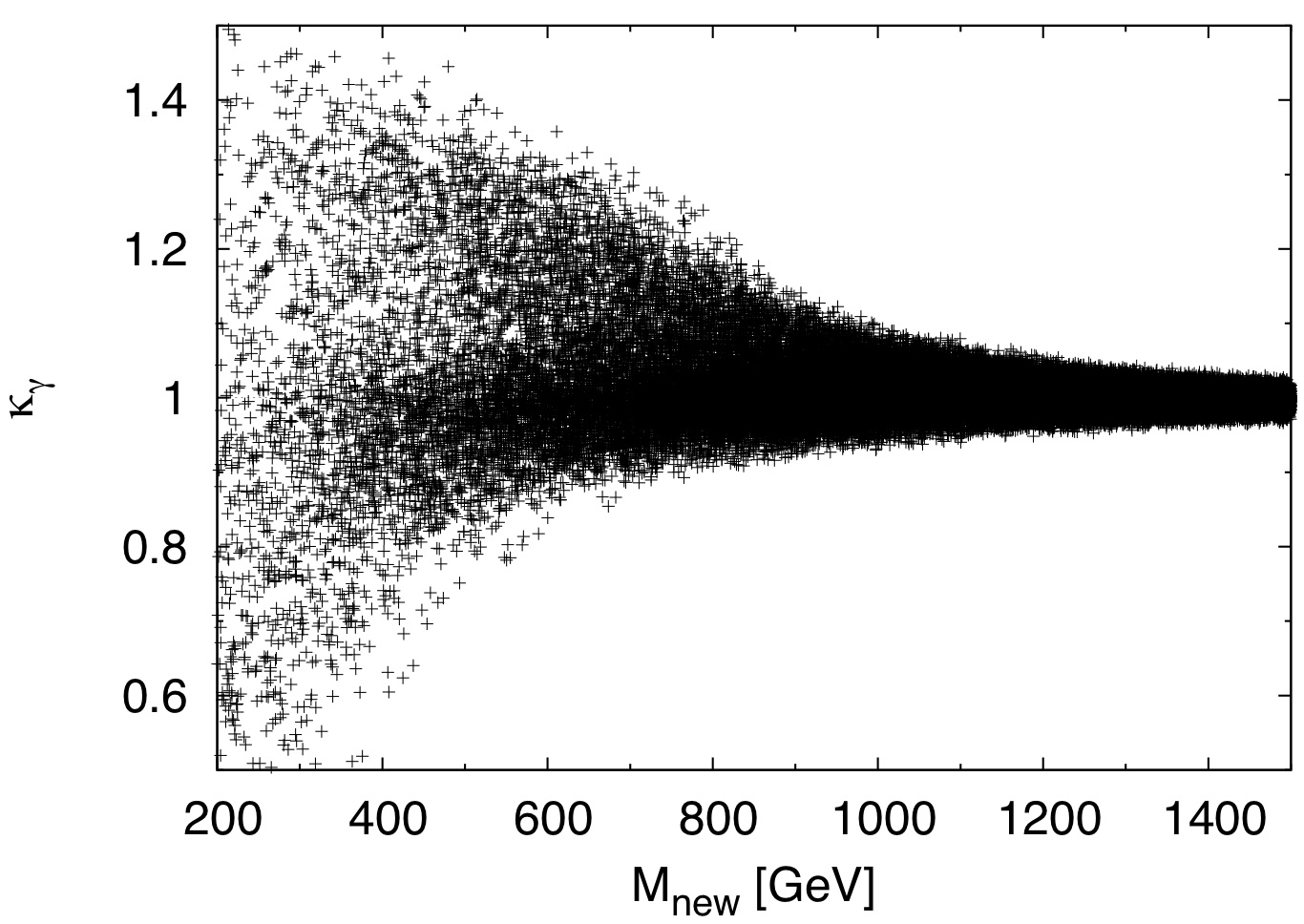}}\resizebox{0.5\textwidth}{!}{\includegraphics{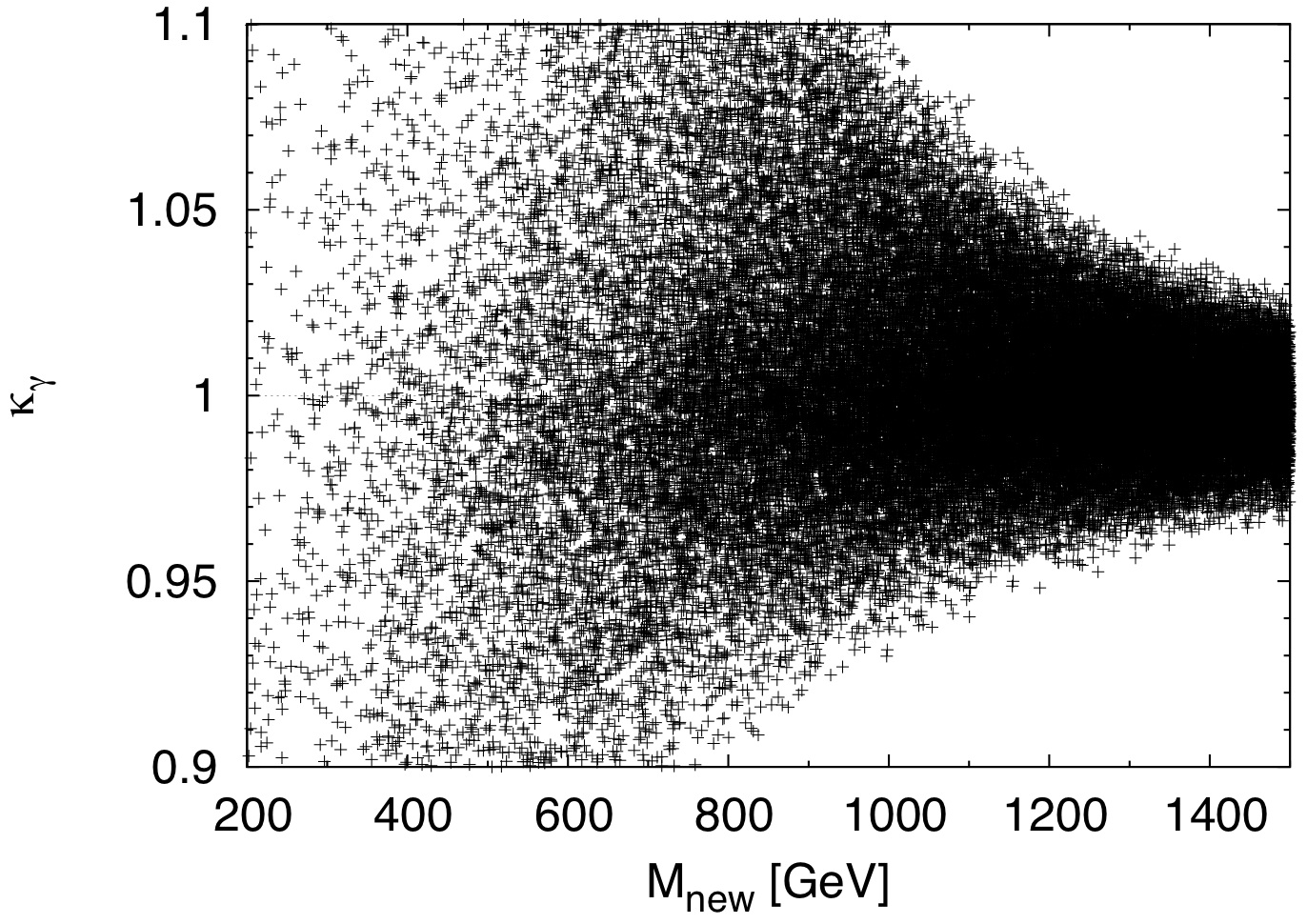}} \\
\resizebox{0.5\textwidth}{!}{\includegraphics{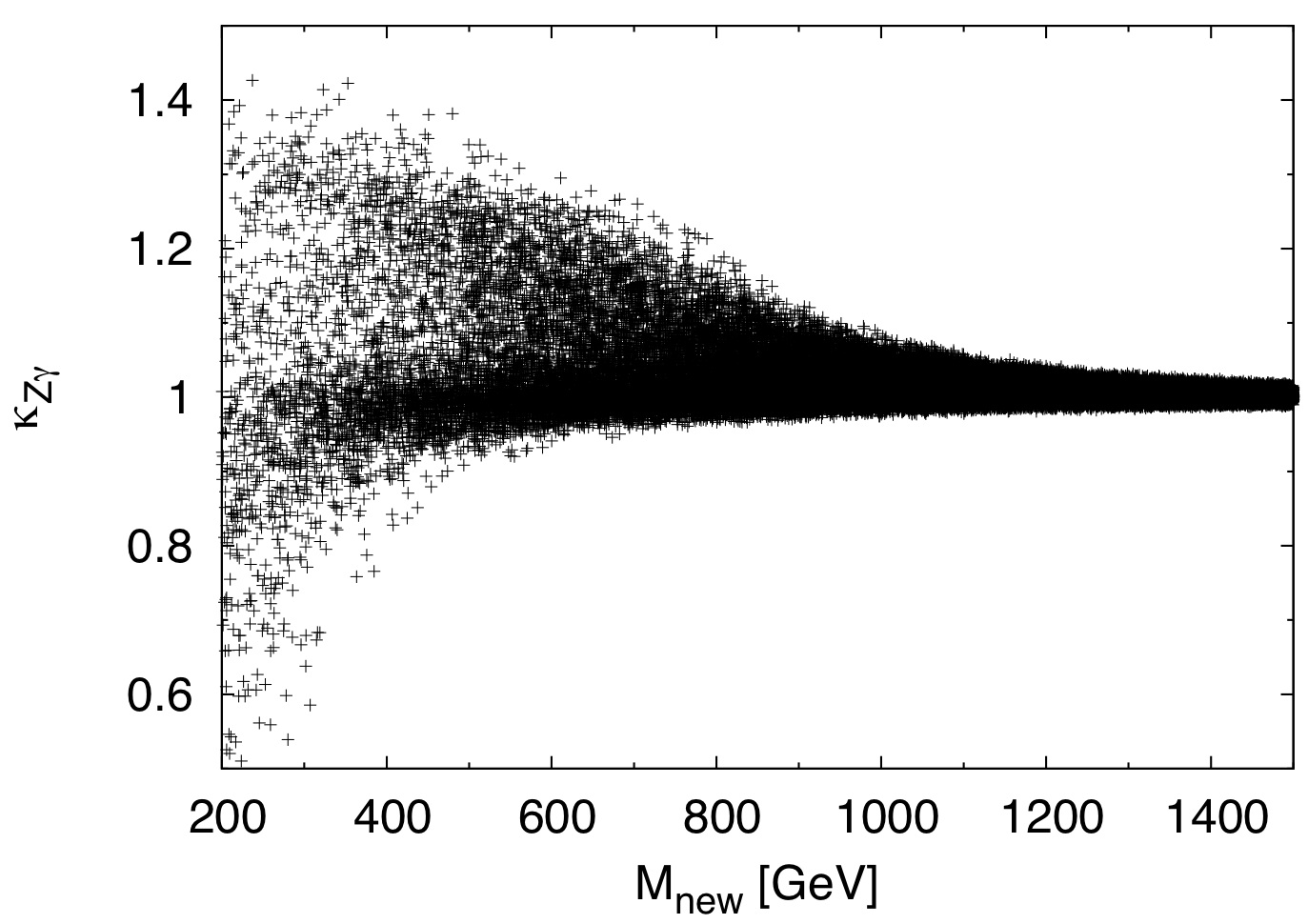}}\resizebox{0.5\textwidth}{!}{\includegraphics{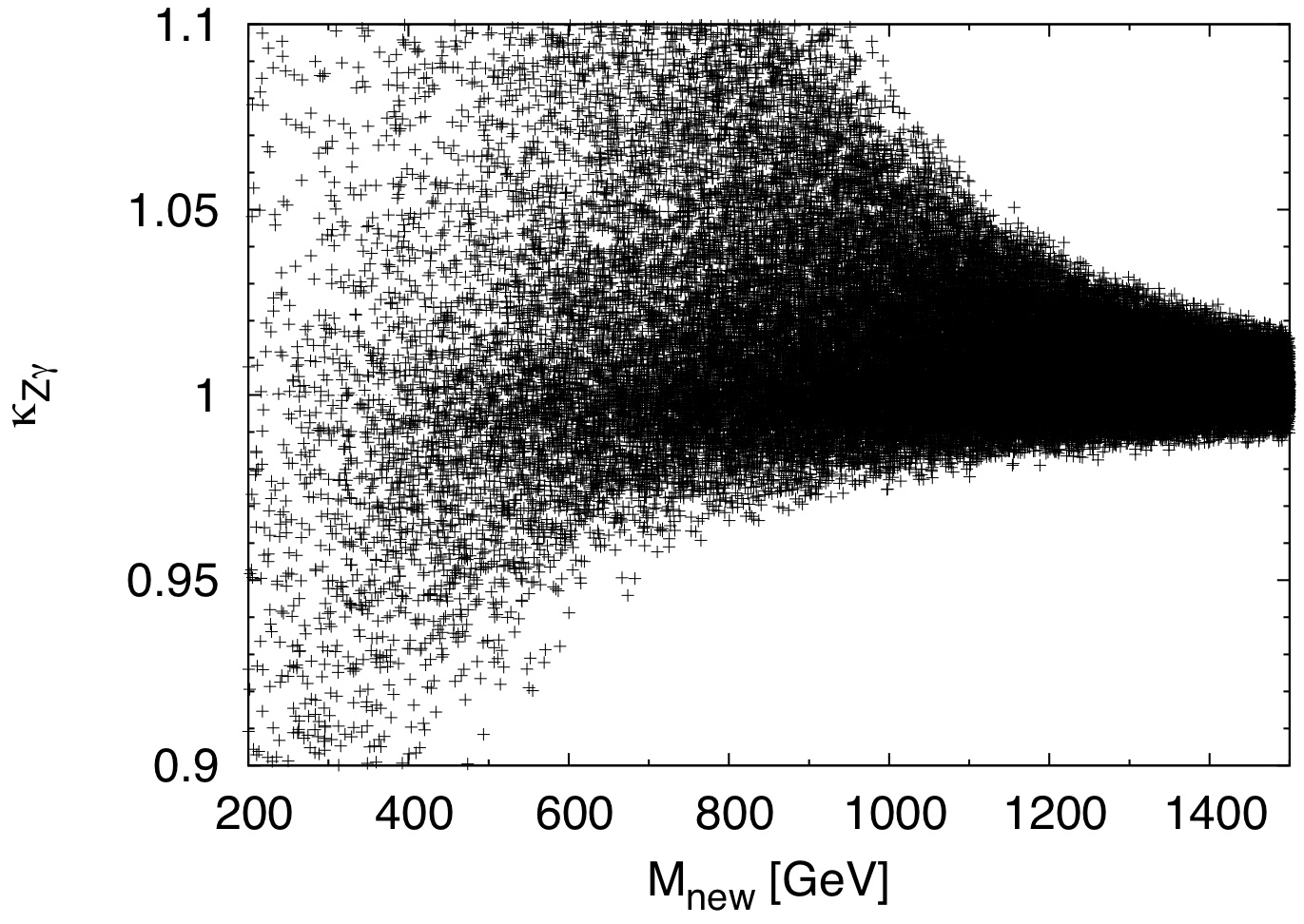}}
\caption{As in Fig.~\ref{fig:kVscan} but for $\kappa_{\gamma}$ (upper) and $\kappa_{Z\gamma}$ (lower).}
\label{fig:kgamscan}
\end{figure}

Finally we perform numerical scans over the parameter space of the GM model and examine the accessible range of couplings of the light Higgs $h$ to vector boson pairs, fermion pairs, photon pairs, and $Z\gamma$ compared to the corresponding couplings of the SM Higgs.  We set $m_h = 125$~GeV and allow $\mu_3^2$, $\lambda_{2-5}$, $M_1$, and $M_2$ to vary freely, imposing all the theoretical constraints of Sec.~\ref{sec:theoryconstraints}.  To avoid parameter regions in which the top quark Yukawa coupling becomes too large, we require that $\cot \theta_H \equiv v_{\phi}/2 \sqrt{2} v_{\chi} > 0.3$ in the spirit of Ref.~\cite{Barger:1989fj} ($\cot \theta_H$ plays the same role as is played by $\tan\beta$ in the Type-I two Higgs doublet model).  Numerical results are shown in Figs.~\ref{fig:kVscan}--\ref{fig:kgamscan} as a function of the mass of the lightest of the new scalars.  No experimental constraints from direct searches for the additional scalars or from their indirect effects on lower-energy observables such as $R_b$~\cite{Haber:1999zh} have been applied in Figs.~\ref{fig:kVscan}--\ref{fig:kgamscan}; we leave a proper study of these constraints to future work.

We see that the size of the deviations of the $h$ coupling scaling factors $\kappa_i$ from the SM prediction $\kappa_i = 1$ is comparable for all four scaling factors $\kappa_V$, $\kappa_f$, $\kappa_{\gamma}$ and $\kappa_{Z\gamma}$, and that the deviations can reach 10\% even for masses of the lightest new scalar around 800~GeV.  We also note that, as discussed before, $\kappa_V$ can be enhanced relative to its SM value by up to a few tens of percent; this is in contrast to the case of the 2HDM and models with additional electroweak-singlet scalars, in which $\kappa_V \leq 1$ at tree level.

Current LHC data collected at center-of-mass energies of 7 and 8~TeV already constrains $|\kappa_f| \sim 0.9 \pm 0.3$ and $\kappa_V \sim 1.1 \pm 0.15$ under the assumption that only these couplings are allowed to deviate from their SM values~\cite{ATLAS:2013sla}.  Meanwhile the photon coupling can be taken very roughly from the Higgs signal strength in the $\gamma\gamma$ channel; this is measured at ATLAS as $1.25 \pm 0.25$~\cite{ATLAS:2013sla} and at CMS as $0.78 \pm 0.27$ (multivariate analysis) or $1.11 \pm 0.31$ (cut-based analysis)~\cite{CMSgaga}.  These 1$\sigma$ allowed ranges can be fully populated in the GM model for heavy scalar masses below about 400--600~GeV, depending on the coupling considered.
Future measurements from the full LHC program (300~fb$^{-1}$ at 14~TeV) and the proposed LHC luminosity upgrade (3000~fb$^{-1}$ at 14~TeV) promise to reduce the uncertainties on these three couplings to a few percent, whereas a future high-luminosity $e^+e^-$ collider program could measure $\kappa_V$ and $\kappa_f$ with a precision well below the percent level~\cite{Dawson:2013bba}.   The coupling $\kappa_{Z\gamma}$ will be very difficult to constrain experimentally with a precision better than $\sim 50\%$ due to the low usable statistics of the $h \to Z\gamma$ decay mode~\cite{Dawson:2013bba}.

%%%%%%%%%%%%%%%%%%%%%%%%%%%%%%%%%%%%%%%%%%%%%%
\section{Conclusions}
\label{sec:conclusions}

The measured properties of the SM-like Higgs boson discovered at the LHC have so far been consistent with SM expectations.  Together with this, the fact that no additional new particles have yet been discovered at the LHC motivates the study of the decoupling limits of Higgs sector extensions, as they can lead to a 125~GeV resonance with very small deviations from SM Higgs couplings and heavier states that are out of reach with current data.

In this paper we studied the most general scalar potential of the GM model that preserves gauge invariance and the custodial SU(2) symmetry.  We started by collecting the theoretical constraints on the potential parameters required to satisfy tree-level unitarity in $2 \to 2$ scalar scattering amplitudes, ensure the potential is bounded from below, and avoid the existence of deeper custodial SU(2)-violating minima.  We then showed that the GM model with this most general scalar potential does possess a decoupling limit, and studied the phenomenological properties of the model as the decoupling limit is approached.  

We found that the mixing angle that controls the amount of triplet in the light Higgs state $h$, as well as the fraction of vev carried by the triplet $v_{\chi}/v$, both go to zero in the decoupling limit like $(v/M_{\rm new})$ or faster, while the fractional size of the mass splittings among the heavy scalars and the deviations of the couplings of the light Higgs boson from those of the SM Higgs go to zero like $(v^2 / M_{\rm new}^{2})$ or faster, where $M_{\rm new}$ is the mass scale of the heavy scalars.  The decoupling of the light Higgs boson couplings goes like $(v^2/M_{\rm new}^2)$ when the dimensionful trilinear couplings in the scalar potential grow with $M_{\rm new}$ as the decoupling limit is taken.  The decoupling is faster, like $(v^4/M_{\rm new}^4)$, when these trilinear couplings remain small (of order $v$) as the decoupling limit is taken.  

Compared to the decoupling limit of the 2HDM, the most notable difference is in the decoupling behavior of the $hVV$ coupling.  As described above, in the GM model this coupling can deviate from the SM prediction by an amount of order $(v^2/M_{\rm new}^2)$ or less, whereas in the 2HDM it can deviate at most by corrections of order $(v^4/M_{\rm new}^4)$---i.e., deviations in $\kappa_V$ can decouple much more slowly in the GM model than in the 2HDM.  

The $hVV$ coupling $\kappa_V$ also allows the GM model to be distinguished from any SM Higgs sector extension containing only SU(2)$_L$ doublets and singlets.  In the GM model, we find analytically that the leading modification to $\kappa_V$ in the decoupling limit is always positive; this is confirmed by a numerical scan (imposing all theoretical constraints) from which we find that $\kappa_V \geq 1$ for $M_{\rm new}  \gtrsim  500$~GeV.  For comparison, models containing only SU(2)$_L$-doublet and -singlet scalars always have $\kappa_V \leq 1$.
From our numerical scans we also found that the GM model can fully populate the current experimentally-allowed $1\sigma$ ranges of Higgs couplings to pairs of vector bosons, fermions, and photons when the new scalars are lighter than 400--600~GeV.  

We finally comment that the parameter space should be further constrained by direct searches for the heavy scalars as well as by indirect constraints from observables like $R_b$~\cite{Haber:1999zh,Chiang:2012cn}, $b \to s \gamma$, $B^0$--$\bar B^0$ mixing, and the electroweak oblique parameters~\cite{Gunion:1990dt,Kanemura:2013mc,Englert:2013zpa}.  We reserve a study of these constraints for future work.

%%%%%%%%%%%%%%%%%%%%%%%%%%%%%%%%%%%%%%%%%%%%%%
\begin{acknowledgments}
We thank Mayumi Aoki and Shinya Kanemura for checking our results for the unitarity constraints on the quartic scalar couplings.  We also thank Spencer Chang and Cheng-Wei Chiang for helpful discussions. 
This work was supported by the Natural Sciences and Engineering Research Council of Canada.  K.H.\ was also supported by the Government of Ontario through an Ontario Graduate Scholarship.
\end{acknowledgments}
%%%%%%%%%%%%%%%%%%%%%%%%%%%%%%%%%%%%%%%%%%%%%%

\appendix

%%%%%%%%%%%%%%%%%%%%%%%%%%%%%%%%%%%%%%%%%%%%%
\section{Feynman rules for scalar couplings}
\label{sec:feynrules}

\subsection{Triple-scalar couplings}
\subsubsection{Couplings involving $h$}
\label{sec:hfeynrules}

The Feynman rules  for three-scalar couplings involving $h$ are given by $-i g_{h s_1s_2}$, with all particles incoming and the couplings defined as follows:
\begin{eqnarray}
	g_{hhh} &=& 24\lambda_1 c_\alpha^3 v_\phi 
	- 6 s_\alpha c_\alpha \left(\sqrt{3} c_\alpha v_\chi - s_\alpha v_\phi\right)
	\left(2 \lambda_2 - \lambda_5\right) 
	- 8\sqrt{3} s_\alpha^3 v_\chi \left(\lambda_3+3\lambda_4\right) \nonumber \\
	&& + \frac{3\sqrt{3}}{2} M_1 c_\alpha^2 s_\alpha - 4\sqrt{3} M_2 s_\alpha^3   \, ,
	\nonumber \\
	g_{hhH} &=& 24\lambda_1 c_\alpha^2 s_\alpha v_\phi
	+ 2 \left[ \sqrt{3} c_\alpha v_\chi \left(3 c_\alpha^2-2\right) 
	+ s_\alpha v_{\phi} \left(1-3 c_\alpha^2\right) \right] \left(2\lambda_2-\lambda_5\right)
	\nonumber \\
	&& + 8\sqrt{3} c_\alpha s_\alpha^2 v_\chi \left(\lambda_3+3\lambda_4\right)
	- \frac{\sqrt{3}}{2} M_1 c_\alpha\left(3 c_\alpha^2-2\right)-4\sqrt{3} M_2 c_\alpha s_\alpha^2   \, , 
	\nonumber \\
	g_{hHH} &=& 24\lambda_1 c_\alpha s_\alpha^2 v_\phi 
	+ 2 \left[ \sqrt{3} s_\alpha v_\chi \left(3 c_\alpha^2-1\right)
	+ c_\alpha v_\phi \left(3 c_\alpha^2-2\right)\right] \left(2\lambda_2-\lambda_5\right)
	\nonumber \\
	&& - 8\sqrt{3} c_\alpha^2 s_\alpha v_\chi \left(\lambda_3+3\lambda_4\right)
	- \frac{\sqrt{3}}{2} M_1 s_\alpha\left(3 c_\alpha^2-1\right) 
	+ 4\sqrt{3} M_2 c_\alpha^2 s_\alpha    \, ,
	\nonumber \\
	g_{hH_3^0H_3^0} &=& g_{hH_3^+H_3^{+*}} 
	= 64\lambda_1 c_\alpha \frac{v_\chi^2 v_\phi}{v^2}
	- \frac{8}{\sqrt{3}} \frac{v_\phi^2 v_\chi}{v^2} s_\alpha \left(\lambda_3 + 3\lambda_4\right)
	- \frac{4}{\sqrt{3}} \frac{v_\chi M_1}{v^2}
	\left(s_\alpha v_\chi - \sqrt{3} c_\alpha v_\phi\right) \nonumber\\
	&&  -\frac{16}{\sqrt{3}} \frac{v_\chi^3}{v^2}s_\alpha
	\left(6\lambda_2 +\lambda_5\right) 
	- c_\alpha \frac{v_\phi^3}{v^2}\left(\lambda_5-4\lambda_2 \right)  
	+ \frac{2\sqrt{3}\,M_2 v_\phi^2}{v^2} s_\alpha
	\nonumber \\
	&& -\frac{8}{\sqrt{3}} \lambda_5 \frac{v_\chi v_\phi}{v^2}
	\left(s_\alpha v_\phi - \sqrt{3} c_\alpha v_\chi\right)      , 
	\nonumber \\
	g_{hH_3^0G^0} &=& g_{hH_3^+G^{+*}} 
	= \sqrt{2} (-16\lambda_1+8\lambda_2-3\lambda_5) \frac{v_\chi v_\phi^2}{v^2} c_\alpha
	- \frac{4\sqrt{2}}{\sqrt{3}}\frac{v_\chi^2 v_\phi}{v^2} s_\alpha
	\left(4\lambda_3+12\lambda_4-6\lambda_2+\lambda_5\right) \nonumber \\
	&& +\frac{M_1}{\sqrt{6} v^2}\left(2 v_\chi v_\phi s_\alpha 
	+ \sqrt{3}c_\alpha \left(8 v_\chi^2 -v_\phi^2\right)\right)
	+ 4\sqrt{6}\frac{v_\chi v_\phi}{v^2} M_2 s_\alpha \nonumber\\
	&& +\sqrt{\frac{2}{3}}\lambda_5\left(\frac{v_\phi^3}{v^2} s_\alpha 
	+ 8\sqrt{3} \frac{v_\chi^3}{v^2} c_\alpha\right)   , 
	\nonumber \\
	g_{hH_5^0H_5^0} &=& g_{hH_5^+H_5^{+*}} = g_{hH_5^{++}H_5^{++*}} 
	= -8\sqrt{3}\left(\lambda_3+\lambda_4\right) s_\alpha v_\chi 
	+ \left(4\lambda_2 + \lambda_5\right) c_\alpha v_\phi - 2\sqrt{3}\,M_2 s_\alpha   \,  ,
	\nonumber \\
	g_{hG^0G^0} &=& g_{hG^+G^{+*}} 
	= 8\lambda_1 \frac{v_\phi^3}{v^2} c_\alpha 
	- 2 \frac{v_\chi v_\phi}{v^2}\left(\sqrt{3} s_\alpha v_\phi - 8 c_\alpha v_\chi\right)
	\left(2\lambda_2 - \lambda_5\right) +16\sqrt{3} M_2 s_\alpha \frac{v_\chi^2}{v^2} 
	\nonumber \\
	&& - \frac{64}{\sqrt{3}} \frac{v_\chi^3}{v^2} s_\alpha \left(\lambda_3 + 3\lambda_4\right)
	- \frac{1}{2\sqrt{3}} M_1\frac{v_\phi}{v^2}\left( s_\alpha v_\phi 
	+ 8\sqrt{3} v_\chi c_\alpha \right).
	\label{eq:hhh} 
\end{eqnarray}
Here we use the shorthand notation $s_{\alpha} \equiv \sin\alpha$ and $c_{\alpha} \equiv \cos\alpha$. 

Accounting for the mixing $H_1^0, H_1^{0\prime} \rightarrow h, H$ and the difference in notation, the couplings involving physical scalars agree in the $M_1 = M_2 = 0$ limit with those of Ref.~\cite{Chang:2012gn} after correcting a small typo in the $H_1^0 H_3^+ H_3^-$ coupling~\cite{Spencer} (Ref.~\cite{Chang:2012gn} did not list the couplings involving Goldstone bosons).

%%%%%%%%%%%%%%%%%%%%%%%%%%%%%%%%%%%%%%%%%%%%%%

\subsubsection{Couplings involving $H$}
\label{sec:hhfeynrules}

The Feynman rules  for three-scalar couplings involving $H$ (but not $h$) are given by $-i g_{h s_1s_2}$, with all particles incoming and the couplings defined as follows:
\begin{eqnarray}
	g_{HHH} &=& 24\lambda_1 s_\alpha^3 v_\phi 
	+ 6 s_\alpha c_\alpha \left(\sqrt{3} s_\alpha v_\chi + c_\alpha v_\phi\right)
	\left(2 \lambda_2 - \lambda_5\right) 
	+ 8\sqrt{3} c_\alpha^3 v_\chi \left(\lambda_3+3\lambda_4\right)  \nonumber \\
	&& - \frac{3\sqrt{3}}{2} M_1 s_\alpha^2 c_\alpha - 4\sqrt{3} M_2 c_\alpha^3  \, ,
	\nonumber \\
	g_{HH_3^0H_3^0} &=& g_{HH_3^+H_3^{+*}} 
	= 64\lambda_1 s_\alpha \frac{v_\chi^2 v_\phi}{v^2}
	+ \frac{8}{\sqrt{3}} \frac{v_\phi^2 v_\chi}{v^2} c_\alpha \left(\lambda_3 + 3\lambda_4\right)
	+ \frac{4}{\sqrt{3}} \frac{v_\chi M_1}{v^2}
	\left(c_\alpha v_\chi +\sqrt{3} s_\alpha v_\phi\right) \nonumber\\
	&&  +\frac{16}{\sqrt{3}} \frac{v_\chi^3}{v^2}c_\alpha
	\left(6\lambda_2 +\lambda_5\right) 
	+ s_\alpha \frac{v_\phi^3}{v^2}\left(4\lambda_2 -\lambda_5\right)  
	- \frac{2\sqrt{3}\,M_2 v_\phi^2}{v^2} c_\alpha
	\nonumber \\
	&& +\frac{8}{\sqrt{3}} \lambda_5 \frac{v_\chi v_\phi}{v^2}
	\left(c_\alpha v_\phi + \sqrt{3} s_\alpha v_\chi\right)  , 
	\nonumber \\
	g_{HH_3^0G^0} &=& g_{HH_3^+G^{+*}} 
	= -\sqrt{2} \frac{v_\chi v_\phi^2}{v^2} s_\alpha (16\lambda_1-8\lambda_2+3\lambda_5)
	- \frac{4\sqrt{2}}{\sqrt{3}}\frac{v_\chi^2 v_\phi}{v^2} c_\alpha
	\left(4\lambda_3+12\lambda_4-6\lambda_2+\lambda_5\right) \nonumber \\
	&& -\frac{M_1}{\sqrt{6} v^2}\left(2 v_\chi v_\phi c_\alpha 
	+ \sqrt{3}s_\alpha \left(v_\phi^2-8 v_\chi^2\right)\right)
	- 4\sqrt{6}\frac{v_\chi v_\phi}{v^2} M_2 c_\alpha \nonumber\\
	&& -\sqrt{\frac{2}{3}} \lambda_5\left(\frac{v_\phi^3}{v^2} c_\alpha 
	- 8\sqrt{3} \frac{v_\chi^3}{v^2} s_\alpha\right)  , 
	\nonumber \\
	g_{HH_5^0H_5^0} &=& g_{HH_5^+H_5^{+*}} = g_{HH_5^{++}H_5^{++*}} 
	= 8\sqrt{3}\left(\lambda_3+\lambda_4\right) c_\alpha v_\chi 
	+ \left(4\lambda_2 + \lambda_5\right) s_\alpha v_\phi + 2\sqrt{3}\,M_2 c_\alpha  \,  , 
	\nonumber \\
	g_{HG^0G^0} &=& g_{HG^+G^{+*}} 
	= 8\lambda_1 \frac{v_\phi^3}{v^2} s_\alpha 
	+ 2 \frac{v_\chi v_\phi}{v^2}\left(\sqrt{3} c_\alpha v_\phi + 8 s_\alpha v_\chi\right)
	\left(2\lambda_2 - \lambda_5\right) -16\sqrt{3} M_2 c_\alpha \frac{v_\chi^2}{v^2}   
	\nonumber \\
	&& + \frac{64}{\sqrt{3}} \frac{v_\chi^3}{v^2} c_\alpha \left(\lambda_3 + 3\lambda_4\right)
	- \frac{1}{2\sqrt{3}} M_1\frac{v_\phi}{v^2}\left(8\sqrt{3} v_\chi s_\alpha  -c_\alpha v_\phi\right).
\end{eqnarray}

%%%%%%%%%%%%%%%%%%%%%%%%%%%%%%%%%%%%%%%%%%%%%%
\subsubsection{Couplings involving $H_3$ and $H_5$}
\label{sec:h3feynrules}

The Feynman rules  for three-scalar couplings involving scalars from the custodial triplet and fiveplet are given by $-i g_{s_1 s_2 s_3}$, with all particles incoming and the couplings defined as follows: 
\begin{eqnarray}
	g_{H_3^0H_3^0H_5^0} &=& -\frac{2\sqrt{2}}{\sqrt{3} v^2} 
	\left(-8\lambda_5 v_\chi^3 + 4 M_1 v_\chi^2 
	+ (-4\lambda_5+2\lambda_3)v_\phi^2 v_\chi+3 M_2 v_\phi^2\right), \nonumber \\
	g_{H_3^+H_3^{+*}H_5^0} &=& \frac{\sqrt{2}}{\sqrt{3} v^2}
	\left(-8\lambda_5 v_\chi^3+4 M_1 v_\chi^2
	+ (-4\lambda_5+2\lambda_3) v_\phi^2 v_\chi+3 M_2 v_\phi^2\right), \nonumber \\
	g_{H_3^0H_3^+H_5^{+*}} &=& -i\frac{\sqrt{2}}{v^2}
	\left(-8\lambda_5 v_\chi^3+4 M_1 v_\chi^2
	+ (-4\lambda_5+2\lambda_3) v_\phi^2 v_\chi+3 M_2 v_\phi^2\right), \nonumber \\
	g_{H_3^+H_3^+H_5^{++*}} &=& -\frac{2}{v^2} 
	\left(-8\lambda_5 v_\chi^3 + 4 M_1 v_\chi^2  
	+ (-4\lambda_5+2\lambda_3) v_\phi^2 v_{\chi} + 3 M_2 v_\phi^2  \right), \nonumber \\
	g_{H_5^0H_5^0H_5^0} &=& 2 \sqrt{6} \left( 2 \lambda_3 v_\chi - M_2 \right), \nonumber \\
	g_{H_5^+H_5^{+*}H_5^0} &=& \sqrt{6} \left( 2 \lambda_3 v_\chi - M_2 \right), \nonumber \\
	g_{H_5^+H_5^{+}H_5^{++*}} &=& -6 \left( 2 \lambda_3 v_\chi -  M_2 \right), \nonumber \\
	g_{H_5^{++}H_5^{++*}H_5^0} &=& -2 \sqrt{6} \left( 2 \lambda_3 v_\chi - M_2 \right). 
\end{eqnarray}

%%%%%%%%%%%%%%%%%%%%%%%%%%%%%%%%%%%%%%%%%%%%%%
\subsection{Scalar couplings to gauge bosons}
\label{app:sVVcoups}

The Feynman rules for the couplings of scalars to gauge bosons come from the gauge-kinetic terms in the Lagrangian,
\begin{equation}
	\mathcal{L} \supset \left( \mathcal{D}_{\mu} \phi \right)^{\dagger} \left( \mathcal{D}^{\mu} \phi \right)+ \frac{1}{2}\left( \mathcal{D}_{\mu} \xi \right)^{\dagger} \left( \mathcal{D}^{\mu} \xi \right) +\left( \mathcal{D}_{\mu} \chi \right)^{\dagger} \left( \mathcal{D}^{\mu} \chi \right) ,
	\label{eq:scalarkin}
\end{equation}
where $\phi = (\phi^+, \phi^0)^T$, $\xi = (\xi^+,\xi^0,-\xi^{+*})^T$, and $\chi=(\chi^{++},\chi^+,\chi^0)^T$, and the covariant derivative is given by
\begin{equation}
	\mathcal{D}_{\mu} = \partial_{\mu} - i \frac{g}{\sqrt{2}} \left( W_{\mu}^+ T^+ + W_{\mu}^- T^- \right) - i \frac{e}{s_W c_W} Z_{\mu} \left( T^3 - s_W^2 Q \right) - i e A_{\mu} Q.
\end{equation}

The gauge-kinetic Lagrangian can be equivalently expressed in terms of the matrix forms of $\Phi$ and $X$ given in Eq.~(\ref{eq:PX}) as
\begin{equation}
	\mathcal{L} \supset \frac{1}{2} {\rm Tr} \left[\left(D_\mu\Phi\right)^\dag\left(D_\mu\Phi\right)\right]
	+ \frac{1}{2} {\rm Tr}\left[\left(D_\mu X\right)^\dag\left(D_\mu X\right)\right],
\end{equation}
where the covariant derivative acting on each of the matrix forms is defined as 
\begin{eqnarray}
	D_{\mu} \Phi &=& \partial_\mu \Phi - i g W_{\mu}^a \tau^a \Phi 
	+ i g^{\prime} B_\mu \Phi \tau^3,    \nonumber  \\
	D_{\mu} X  &=& \partial_\mu X - i g W_{\mu}^a t^a X  
	+ i g^{\prime} B_\mu X t^3,
\end{eqnarray}
with $g = e/s_W$ and $g^{\prime} = e/c_W$.

We note that the couplings listed in the rest of this subsection are consistent with those in Ref.~\cite{HHG} after taking into account the different sign convention in the covariant derivative and the different sign in the definition of the neutral custodial triplet state $H_3^0$, equal to $-H_3^0$ in our notation.

\subsubsection{Couplings of one scalar to two gauge bosons}

The Feynman rules for the vertices involving a scalar and two gauge bosons are defined as $i g_{sVV} g^{\mu \nu}$. The couplings are given by
\begin{eqnarray}
	g_{hW^+W^{+*}}  &=& c_W^2 g_{hZZ} 
	= -\frac{e^2}{6 s_W^2} \left( 8\sqrt{3} s_\alpha v_\chi - 3 c_\alpha v_\phi\right),  \nonumber \\
	g_{HW^+W^{+*}}  &=&   c_W^2 g_{HZZ} 
	=   \frac{e^2}{6 s_W^2} \left( 8\sqrt{3} c_\alpha v_\chi + 3 s_\alpha v_\phi\right), \nonumber \\
	g_{H_5^0 W^+ W^{+*}} &=& \sqrt{\frac{2}{3}} \frac{e^2}{s_W^2} v_{\chi},  \nonumber \\
	g_{H_5^0 Z Z} &=& -\sqrt{\frac{8}{3}} \frac{e^2}{s_W^2 c_W^2} v_{\chi}, \nonumber \\
	g_{H_5^+ W^{+*} Z} &=&   -\frac{\sqrt{2} e^2 v_{\chi}}{c_W s_W^2}   , \nonumber \\
	g_{H_5^{++} W^{+*} W^{+*}} &=&   \frac{2 e^2 v_{\chi}}{s_W^2}. 
\end{eqnarray}

%%%%%%%%%%%%%%%%%%%%%%%%%%%%%%%%%%%%%%%%%%%%%%
\subsubsection{Couplings of two scalars to one photon}
\label{app:gacoups}

The Feynman rules for the vertices involving two scalars and a single photon $\gamma_\mu$ are defined as 
$i g_{\gamma s s^*} \left(p-p^*\right)_\mu$,
where $p$ ($p^*$) is the incoming momentum of incoming scalar $s$ ($s^*$).  The couplings are given by $g_{\gamma s s^*} = eQ_s$ with $Q_s$ being the electric charge of scalar $s$ in units of $e$.  All off-diagonal couplings and neutral scalar couplings to the photon are zero.

\subsubsection{Couplings of two scalars to one $Z$ boson}
\label{app:Zcoups}

The Feynman rules for the vertices involving two scalars and a single $Z$ boson are defined as
$i g_{Z s_1s_2} \left(p_1 - p_2\right)_\mu$,
where again $p_1$ ($p_2$) is the incoming momentum of incoming scalar $s_1$ ($s_2$).  The couplings are given by
\begin{eqnarray}
	g_{ZhH_3^0} &=& -i\sqrt{\frac{2}{3}}\frac{e}{s_Wc_W}\left(\sqrt{3}\frac{v_\chi}{v} c_\alpha + s_\alpha \frac{v_\phi}{v} \right), \nonumber \\
	g_{ZHH_3^0} &=& i\sqrt{\frac{2}{3}}\frac{e}{s_Wc_W}\left(c_\alpha \frac{v_\phi}{v} - \sqrt{3}\frac{v_\chi}{v} s_\alpha \right), \nonumber \\
	g_{ZH_3^0 H_5^0} &=& i \sqrt{\frac{1}{3}} \frac{e}{s_Wc_W}\frac{v_\phi}{v}, \nonumber \\
	g_{ZH_3^+H_3^{+*}} &=& g_{ZH_5^+H_5^{+*}} =g_{ZG^+G^{+*}} 
	= \frac{e}{2 s_Wc_W}\left(1 - 2s_W^2\right), \nonumber \\
	g_{ZH_5^{++}H_5^{++*}} &=& \frac{e}{s_Wc_W}\left(1 - 2 s_W^2\right), \nonumber \\
	g_{ZH_3^+H_5^{+*}} &=& -\frac{e}{2 s_Wc_W}\frac{v_\phi}{v}.
\end{eqnarray}
The couplings involving one Goldstone boson and one physical scalar are given by
\begin{eqnarray}
	g_{Z h G^0} &=& \frac{i e}{2\sqrt{3} s_Wc_W}\left(\sqrt{3} c_\alpha \frac{v_\phi}{v} -8 s_\alpha \frac{v_\chi}{v}\right), \nonumber \\
	g_{ZH G^0} &=& \frac{i e}{2\sqrt{3} s_Wc_W}\left(\sqrt{3} s_\alpha \frac{v_\phi}{v} +8 c_\alpha \frac{v_\chi}{v}\right), \nonumber \\
	g_{Z H_5^0 G^0} &=& -2 i \sqrt{\frac{2}{3}}\frac{e}{s_Wc_W} \frac{v_\chi}{v}, \nonumber \\
	g_{Z H_5^+G^{+*}} &=& -\frac{\sqrt{2} e}{s_Wc_W}\frac{v_\chi}{v}.
\end{eqnarray}
All others are zero. 

Note that all the diagonal couplings of $Z$ to charged scalars have the form
\begin{equation}
	g_{Zs s^*}=\frac{Q_s e}{2 s_Wc_W}\left(1 - 2 s_W^2\right),
	\label{eq:ssZ}
\end{equation}
where $Q_s$ is the electric charge of $s$.  This simple form is a consequence of custodial SU(2) symmetry.

\subsubsection{Couplings of two scalars to one $W$ boson}

The Feynman rules for the vertices involving two scalars and a single $W^+$ boson are defined as
$i g_{W^+ s_1s_2} \left(p_1 - p_2\right)_\mu$,
where again $p_1$ ($p_2$) is the incoming momentum of incoming scalar $s_1$ ($s_2$). The couplings are given by
\begin{eqnarray}
	g_{W^+ h H_3^{+*}}  &=&     - \sqrt{\frac{2}{3}} \frac{e}{  s_W} 
	\left( \sqrt{3} c_\alpha \frac{v_\chi}{v} +  s_\alpha  \frac{v_\phi}{v}    \right),      \nonumber  \\
	g_{W^+ H H_3^{+*}}  &=&     - \sqrt{\frac{2}{3}} \frac{e}{  s_W} 
	\left( \sqrt{3} s_\alpha \frac{v_\chi}{v} -  c_\alpha  \frac{v_\phi}{v}    \right),      \nonumber  \\
	g_{W^+  H_3^0 H_3^{+*}}   &=&  - \frac{i}{2} \frac{e}{s_W}	,	\nonumber \\
	g_{W^+ H_3^{+*} H_5^0 }   &=&   - \frac{1}{2 \sqrt{3}} \frac{e}{s_W} \frac{v_\phi}{v},		
	\nonumber \\
	g_{W^+ H_5^{+*} H_5^0}   &=&     \frac{\sqrt{3}}{2} \frac{e}{s_W}, 	\nonumber \\
	g_{W^+ H_3^0 H_5^{+*}}   &=&     \frac{i}{2} \frac{e}{s_W} \frac{v_\phi}{v},		\nonumber \\
	g_{W^+ H_3^+ H_5^{++*} }   &=&      \frac{1}{\sqrt{2}} \frac{e}{s_W} \frac{v_\phi}{v},
	\nonumber \\
	g_{W^+ H_5^+ H_5^{++*}}   &=&  	 \frac{1}{\sqrt{2}} \frac{e}{s_W}.	  
\end{eqnarray}

\subsubsection{Couplings of two scalars to two photons or $Z\gamma$}

The Feynman rules for the vertices involving two scalars and two photons $\gamma_\mu\gamma_\nu$ are defined as
$i g_{\gamma\gamma s s^*}g_{\mu\nu}$,
where the couplings are given by $g_{\gamma\gamma s s^*} = 2 e^2 Q_s^2$. 

The Feynman rules for the vertices involving two scalars, a $Z_{\mu}$ boson, and a photon $\gamma_\nu$ are defined as
$i g_{Z \gamma s_1s_2} g_{\mu\nu}$,
where the couplings are given by
\begin{eqnarray}
	g_{Z \gamma s s^*} &=& \frac{Q_s^2 e^2}{s_Wc_W} \left(1 - 2 s_W^2\right), \nonumber \\
	g_{Z \gamma H_3^+H_5^{+*}} &=& -\frac{e^2}{s_Wc_W}\frac{v_\phi}{v}, \nonumber \\
	g_{Z \gamma H_5^+G^{+*}} &=& -2\sqrt{2}\frac{e^2}{s_Wc_W}\frac{v_\chi}{v}.
\end{eqnarray}
We denote all the diagonal couplings to charged scalars and Goldstone bosons as $g_{Z \gamma s s^*}$. 
The simple form of these diagonal couplings is a consequence of custodial SU(2) symmetry.
 All other couplings are zero.

%%%%%%%%%%%%%%%%%%%%%%%%%%%%%%%%%%%%%%%%%%%%%%
\subsection{Scalar couplings to fermions}

The Feynman rules for the vertices involving a neutral scalar and two fermions are given as follows:
\begin{eqnarray}
	h \bar f f: &\quad& -i \frac{m_f}{v} \frac{\cos \alpha}{\cos \theta_H}, \nonumber \\
	H \bar f f: &\quad& -i \frac{m_f}{v} \frac{\sin \alpha}{\cos \theta_H}, \nonumber \\
	H_3^0 \bar u u: &\quad& \frac{m_u}{v} \tan \theta_H \gamma_5, \nonumber \\
	H_3^0 \bar d d: &\quad& -\frac{m_d}{v} \tan \theta_H \gamma_5.
\end{eqnarray}
Here $f$ stands for any charged fermion, $u$ stands for any up-type quark, and $d$ stands for any down-type quark or charged lepton.  

The Feynman rules for the vertices involving a charged scalar and two fermions are given as follows, with all particles incoming:
\begin{eqnarray}
	H_3^+ \bar u d: &\quad& -i \frac{\sqrt{2}}{v} V_{ud} \tan\theta_H
		\left( m_u P_L - m_d P_R \right), \nonumber \\
	H_3^{+*} \bar d u: &\quad& -i \frac{\sqrt{2}}{v} V_{ud}^* \tan\theta_H 
		\left( m_u P_R - m_d P_L \right), \nonumber \\
	H_3^+ \bar \nu \ell: &\quad& i \frac{\sqrt{2}}{v} \tan\theta_H m_{\ell} P_R, \nonumber \\
	H_3^{+*} \bar \ell \nu: &\quad& i \frac{\sqrt{2}}{v} \tan\theta_H m_{\ell} P_L.
\end{eqnarray}
Here $V_{ud}$ is the appropriate element of the CKM matrix and the projection operators are defined as $P_{R,L} = (1 \pm \gamma_5)/2$.  

The custodial five-plet states do not couple to fermions.

%%%%%%%%%%%%%%%%%%%%%%%%%%%%%%%%%%%%%%%%%%%%%%
\section{Formulas for $h\rightarrow\gamma\gamma$ and $h\rightarrow Z\gamma$}
\label{app:hgaga}

The matrix element for the loop-induced process $h \to \gamma\gamma$ is given by
\begin{equation}
	M_{\gamma} = \frac{\alpha_{EM}}{2 \pi v} \left[ \sum \mathcal{A}_{\gamma} \right]
	\epsilon^{\mu}(k_1) \epsilon^{\nu}(k_2) 
	\left[ \left( k_1 \cdot k_2 \right) g_{\mu\nu} - k_{2 \mu} k_{1 \nu} \right],
	\label{eq:Mgaga}
\end{equation}
where $\alpha_{EM}$ is the fine structure constant, $k_1$ and $k_2$ are the four-momenta of the photons, and $\left[ \sum \mathcal{A}_{\gamma} \right]$ represents the sum of the loop contributions from the $W$ boson, fermions (we include only the dominant top quark contribution), and scalars in the GM model:
\begin{equation}
	\left[ \sum \mathcal{A}_{\gamma} \right] 
	= \kappa_V F_1(\tau_W) + \kappa_f N_c Q_t^2 F_{1/2}(\tau_t)
	+ \sum_s \beta_s Q_s^2 F_0(\tau_s).
\end{equation}
Here $\kappa_V$ and $\kappa_f$ are the scaling factors for the couplings of $h$ to $W$ or $Z$ bosons and fermions, respectively, compared to those of the SM Higgs, $N_c Q_t^2 = 4/3$ for the top quark, $Q_s$ is the electric charge of scalar $s$ in units of $e$, and $\beta_s = g_{hss^*} v/2 m_s^2$.  The sum runs over all electrically charged scalars in the GM model.  The coupling $g_{hss^*}$ is defined in such a way that the corresponding interaction Lagrangian term is $\mathcal{L} \supset - g_{hss^*} h s s^*$ (this is consistent with the notation of Sec.~\ref{sec:hfeynrules}).

The loop factors are given in terms of the usual functions~\cite{HHG}, 
\begin{eqnarray}
	F_1(\tau) &=& 2+3\tau+3\tau(2-\tau) f(\tau), \nonumber \\
	F_{1/2}(\tau) &=& -2\tau[1+(1-\tau) f(\tau)], \nonumber \\
	F_0(\tau) &=& \tau[1-\tau f(\tau)],
\end{eqnarray}
where
\begin{equation}
	f(\tau) = \left\{ \begin{array}{l l}
	\left[\sin^{-1} \left(\sqrt{\frac{1}{\tau}}\right) \right]^2 & \quad  {\rm if} \ \tau \geq 1, \\
	-\frac{1}{4}\left[ \log \left(\frac{\eta_+}{\eta_-}\right) - i \pi \right]^2 & \quad  {\rm if} \ \tau < 1, \\
	\end{array} \right.
	\label{feq}
\end{equation}
with $\eta_{\pm} = 1 \pm \sqrt{1-\tau}$.  The argument is $\tau_i \equiv 4 m_i^2/ m_h^2$. 

The scaling factor $\kappa_{\gamma}$ of the loop-induced $h\rightarrow\gamma\gamma$ coupling in the GM model relative to that in the SM is then given by
\begin{equation}
	\kappa_{\gamma} = \frac{\kappa_V F_1(\tau_W) + \kappa_f N_c Q_t^2 F_{1/2}(\tau_t) + \sum_s \beta_s Q_s^2 F_0(\tau_s)}
	{F_1(\tau_W) + N_c Q_t^2 F_{1/2}(\tau_t)}.\label{kgam}
\end{equation}
We also define a factor $\Delta \kappa_{\gamma}$ which captures the new loop-induced contributions only,
\begin{equation}
	\Delta \kappa_{\gamma} = \frac{\sum_s \beta_s Q_s^2 F_0(\tau_s)}
	{F_1(\tau_W) + N_c Q_t^2 F_{1/2}(\tau_t)}.\label{dkgam}
\end{equation}

The matrix element for the loop-induced process $h \to Z \gamma$ is given by
\begin{equation}
	M_{Z\gamma} = \frac{\alpha_{EM}}{2 \pi v} \left[ \sum \mathcal{A}_{Z\gamma} \right] 
	\epsilon^\mu(k_Z) \epsilon^\nu(k_\gamma)
	\left[ k_{\gamma \mu} k_{Z \nu} - \left( k_Z \cdot k_\gamma \right) g_{\mu\nu} \right],
	\label{eq:MZga}
\end{equation}
where $k_Z$ and $k_{\gamma}$ are the four-momenta of the $Z$ and photon, respectively, and $\left[ \sum \mathcal{A}_{Z\gamma} \right]$ represents the sum of the loop contributions from the $W$ boson, fermions (we again include only the dominant top quark contribution), and scalars in the GM model:
\begin{equation}
	\left[ \sum \mathcal{A}_{Z\gamma} \right] = \kappa_V A_V + \kappa_f A_f + \frac{v}{2} A_S.
\end{equation}
The loop factors are given by~\cite{HHG,HGZ-Carena,HGZ-Chen}
\begin{eqnarray}
	A_W &=& -\cot\theta_W\left\{4\left(3-\tan^2\theta_W\right) I_2\left(\tau_W,\lambda_W\right)+\left[\left(1+\frac{2}{\tau_W}\right)\tan^2\theta_W-\left(5+\frac{2}{\tau_W}\right)\right] I_1\left(\tau_W,\lambda_W\right)\right\}, \nonumber \\
	A_f &=& \sum_f N_{cf} 
	\frac{-2 Q_f \left(T^{3L}_f - 2 Q_f \sin^2\theta_W\right)}{\sin\theta_W\cos\theta_W}
	\left[ I_1(\tau_f,\lambda_f)-I_2(\tau_f,\lambda_f) \right], \nonumber \\
	A_S &=& \sum_s 2 \frac{g_{hss^*}\,C_{Zss^*}\,Q_s}{m_s^2}
	I_1\left( \tau_s, \lambda_s \right),
	\label{eq:Zgaamps}
\end{eqnarray}
where for the top quark $N_{cf} = 3$, $Q_f = 2/3$, and $T^{3L}_f = 1/2$.  The scalar amplitude depends on the coupling $C_{Zss^*} \equiv g_{Zss^*}/e$ of the scalar to the $Z$ boson, defined in  such a way that the corresponding coupling of the scalar to the photon is $C_{\gamma s s^*} \equiv g_{\gamma s s^*}/e = Q_s$ (this is consistent with the notation of Secs.~\ref{app:gacoups} and \ref{app:Zcoups}).  The sum over scalars runs over $H_3^+$, $H_5^+$, and $H_5^{++}$.  Note that, even though the off-diagonal couplings $Z H_3^+ H_5^{+*}$ and $Z W^+ H_5^{+*}$ are nonzero, there are no corresponding $h H_3^+ H_5^{+*}$ or $h W^+ H_5^{+*}$ couplings, so that there are no diagrams contributing to the $h \to Z \gamma$ amplitude that involve more than one type of particle in the loop.

The loop factors are given in terms of the functions~\cite{HHG}
\begin{eqnarray}
	 I_1(a,b) &=& \frac{ab}{2(a-b)} + \frac{a^2b^2}{2(a-b)^2} \left[f(a) - f(b)\right]
	 + \frac{a^2b}{(a-b)^2} \left[g(a) - g(b)\right], \nonumber \\
	 I_2(a,b) &=& -\frac{ab}{2(a-b)} \left[f(a) - f(b)\right],
\end{eqnarray}
where the function $f(\tau)$ was given in Eq.~(\ref{feq}) and
\begin{equation}
	g(\tau) = \left\{ \begin{array}{l l}
	\sqrt{\tau-1} \sin^{-1} \left(\sqrt{\frac{1}{\tau}}\right) & \quad  {\rm if} \ \tau \geq 1, \\
	\frac{1}{2} \sqrt{1-\tau} \left[ \log \left(\frac{\eta_+}{\eta_-}\right) - i \pi \right] 
		& \quad  {\rm if} \ \tau < 1,
	\end{array} \right.
	\label{geq}
\end{equation}
with $\eta_{\pm}$ defined as for $f(\tau)$.  The arguments of the functions are $\tau_i \equiv 4 m_i^2/m_h^2$ as before and $\lambda_i \equiv 4 m_i^2/M_Z^2$.

The scaling factor $\kappa_{Z\gamma}$ of the loop-induced $h \to Z\gamma$ coupling in the GM model relative to that in the SM is then given by
\begin{equation}
	\kappa_{Z \gamma} = \frac{\kappa_V A_W + \kappa_f A_f + \frac{v}{2} A_S}{A_W + A_f}. \label{kZgam}
\end{equation}
We also define a factor $\Delta \kappa_{Z\gamma}$ which captures the new loop-induced contributions only,
\begin{equation}
	\Delta \kappa_{Z\gamma} = \frac{\frac{v}{2} A_S}{A_W + A_f}. \label{dkZgam}
\end{equation}

We finally note that an informative check of the $h \to Z \gamma$ amplitudes can be made by comparing to the corresponding $h \to \gamma\gamma$ amplitudes after taking $M_Z \to 0$ in the kinematics and replacing the $Z$ boson coupling to the particle in the loop with the corresponding photon coupling.  In this limit, the loop functions $I_1$ and $I_2$ become
\begin{eqnarray}
	\lim_{b \to \infty} I_1(a,b) &=& - \frac{a}{2} + \frac{a^2}{2} f(a), \nonumber \\
	\lim_{b \to \infty} I_2(a,b) &=& \frac{a}{2} f(a),
\end{eqnarray}
so that the combinations appearing in the scalar and fermion loops reduce in this limit to
\begin{eqnarray}
	\lim_{\lambda \to \infty} I_1(\tau,\lambda) &=& - \frac{1}{2} F_0(\tau), \nonumber \\
	\lim_{\lambda \to \infty} \left[ I_1(\tau,\lambda) - I_2(\tau,\lambda) \right] 
	&=& \frac{1}{4} F_{1/2}(\tau).
\end{eqnarray}

Taking $M_Z \to 0$ and replacing $C_{Zss^*}$ with $C_{\gamma ss^*} = Q_s$, the $h \to Z \gamma$ amplitude from loops of scalars becomes
\begin{equation}
	\frac{v}{2} A_S \rightarrow - \sum_s \frac{g_{hss^*} v}{2 m_s^2} Q_s^2 F_0(\tau_s),
\end{equation}
which is exactly minus the scalar loop amplitude that enters $h \to \gamma\gamma$ (the relative minus sign is due to the opposite sign convention for the Lorentz structure in Eqs.~(\ref{eq:Mgaga}) and (\ref{eq:MZga})).

Similarly, taking $M_Z \to 0$ and replacing the vectorial part of the $Zf \bar f$ coupling, 
$(\frac{1}{2} T^{3L}_f - Q_f s_W^2)/s_W c_W$, with the corresponding photon coupling $Q_f$, the $h \to Z \gamma$ amplitude from a fermion loop becomes
\begin{equation}
	A_f \rightarrow - N_{cf} Q_f^2 F_{1/2}(\tau_f).
\end{equation}
Again this is exactly minus the fermion loop amplitude that enters $h \to \gamma\gamma$.

To understand the structure of the $W$ loop amplitude, we first rewrite the expression for $A_W$ in Eq.~(\ref{eq:Zgaamps}) in a form that clearly separates the kinematic dependence of the loop diagram on $M_Z$ from its dependence on the $WWZ$ and $WWZ\gamma$ couplings.  In that expression, the first factor of $\cot\theta_W$ comes from the $WWZ$ (or $WWZ\gamma$) coupling, while the factors of $\tan^2\theta_W = M_Z^2/M_W^2 - 1 = 4/\lambda_W - 1$ inside the curly brackets come from the kinematics.  Expressing the kinematic $M_Z$ dependence entirely in terms of $\tau_W$ and $\lambda_W$, we find,
\begin{equation}
	A_W = -\cot\theta_W \left\{ \left[ 8 - \frac{16}{\lambda_W} \right] 
	I_2\left(\tau_W,\lambda_W\right)
	+ \left[\frac{4}{\lambda_W} \left(1+\frac{2}{\tau_W}\right) 
	- \left(6 + \frac{4}{\tau_W} \right) \right] I_1\left(\tau_W,\lambda_W\right)\right\}.
\end{equation}
Taking $M_Z \to 0$ in the kinematics and replacing the $WWZ$ and $WWZ\gamma$ couplings with the corresponding $WW\gamma$ and $WW\gamma\gamma$ couplings, the $h \to Z\gamma$ amplitude from the $W$ boson loop becomes
\begin{equation}
	A_W \rightarrow - F_1(\tau_W),
\end{equation}
which is exactly minus the $W$ loop amplitude that enters $h \to \gamma\gamma$.

%%%%%%%%%%%%%%%%%%%%%%%%%%%%%%%%%%%%%%%%%%%%%%
\section{Translation to the notations of other papers}
\label{app:translations}

In Table~\ref{tab:translations} we give the translations of our parameterization of the scalar potential in Eq.~(\ref{eq:potential}) to others in the literature.  The first explicit scalar potential for the GM model was written down by Chanowitz and Golden (CG)~\cite{Chanowitz:1985ug}, imposing a discrete symmetry $X \to -X$ to eliminate the cubic terms.  This parameterization of the potential was also used by Gunion, Vega and Wudka~\cite{Gunion:1989ci}, Chang et al.~\cite{Chang:2012gn}, and Efrati and Nir~\cite{Efrati:2014uta}.  A second parameterization was introduced by Aoki and Kanemura (AK)~\cite{Aoki:2007ah}, who were the first to write down the cubic terms.  The analysis by Chiang and Yagyu (CY)~\cite{Chiang:2012cn} used a slight variation on this potential, which was also used in the follow-up paper Ref.~\cite{Chiang:2013rua}.  These are the only other analyses that we know of that take into account nonzero values of the cubic terms.  A third parameterization, preserving the $X \to -X$ symmetry, was introduced by Englert, Re and Spannowsky (ERS)~\cite{Englert:2013zpa,Englert:2013wga}.  Our parameterization of the quartic couplings matches this last one (with the addition of the cubic terms).

\begin{table}
\begin{center}
\begin{tabular}{ccccc}
\hline \hline
This paper & CG~\cite{Chanowitz:1985ug} & AK~\cite{Aoki:2007ah} & CY~\cite{Chiang:2012cn} & ERS~\cite{Englert:2013zpa} \\
\hline
$\lambda_1$ & $\lambda_1 + \lambda_3$ & $\lambda_1$ & $\lambda_1$ & $\lambda_1$ \\
$\lambda_2$ & $2 \lambda_3 + \lambda_4$ & $\lambda_3$ & $\lambda_4$ & $\lambda_2$ \\
$\lambda_3$ & $3 \lambda_5$ & $\lambda_4$ & $\lambda_3$ & $\lambda_3$ \\
$\lambda_4$ & $\lambda_2 + \lambda_3 - \lambda_5$ & $\lambda_2$ & $\lambda_2$ & $\lambda_4$ \\
$\lambda_5$ & $2 \lambda_4$ & $-\lambda_5$ & $-\lambda_5$ & $\lambda_5$ \\
$M_1$ & 0 & $-\mu_1$ & $-\mu_1$ & 0 \\
$M_2$ & 0 & $-\mu_2$ & $-\mu_2$ & 0 \\
\hline\hline
\end{tabular}
\end{center}
\caption{Translation table for the various scalar potential parameterizations of the Georgi-Machacek model that appear in the literature.  See text for details.}
\label{tab:translations}
\end{table}

%%%%%%%%%%%%%%%%%%%%%%%%%%%%%%%%%%%%%%%%%%%%%%

\end{document}